\documentclass[rmp,11pt,,amsmath,amssymb,floatfix]{revtex4}
\usepackage{graphicx}
\usepackage{bm}
\usepackage{endnotes}

\def \be{\begin{equation}}
\def \ee{\end{equation}}
\def \ba{\begin{array}}
\def \ea{\end{array}}
\def \beq{\begin{eqnarray}}
\def \eeq{\end{eqnarray}}

\begin{document}
\title{Phase space representation of quantum dynamics.}

\author{Anatoli Polkovnikov$^1$}
\affiliation {$^1$Department of Physics, Boston University, Boston, MA 02215}

\begin{abstract}

 We discuss a phase space representation of quantum dynamics of systems with many degrees of freedom. This representation is based on a perturbative expansion in quantum fluctuations around one of the classical limits. We explicitly analyze expansions around three such limits: (i) corpuscular or Newtonian limit in the coordinate-momentum representation, (ii) wave or Gross-Pitaevskii limit for interacting bosons in the coherent state representation, and (iii) Bloch limit for the spin systems. We discuss both the semiclassical (truncated Wigner) approximation and further quantum corrections appearing in the form of either stochastic quantum jumps along the classical trajectories or the nonlinear response to such jumps. We also discuss how quantum jumps naturally emerge in the analysis of non-equal time correlation functions. This representation of quantum dynamics is closely related to the phase space methods based on the Wigner-Weyl quantization and to the Keldysh technique. We show how such concepts as the Wigner function, Weyl symbol, Moyal product, Bopp operators, and others automatically emerge from the Feynmann's path integral representation of the evolution in the Heisenberg representation. We illustrate the applicability of this expansion with various examples mostly in the context of cold atom systems including sine-Gordon model, one- and two-dimensional Bose Hubbard model, Dicke model and others.

\end{abstract}
\maketitle

\tableofcontents
\section{Introduction}

Representing quantum mechanics using entirely the language of phase space variables, like in classical physics, attracted significant theoretical attention since its early days. It was clear from the beginning that such a description should be closely connected to statistical physics, where quantum fluctuations should be captured, at least partially, by statistical fluctuations. A very important development in addressing this question came from the work of Moyal~\cite{moyal}, who showed that the Liouville equation for the Wigner function~\cite{wigner} (Wigner transform of the density matrix) assumes the form identical to the Liouville equation for the density matrix in the classical limit if one generalizes the notion of Poisson brackets to Moyal brackets (see Ref.~\cite{hillery} for review). This formalism closely relies on the correspondence between quantum operators and classical functions of phase space variables first suggested by Weyl~\cite{weyl}: the expectation value of any operator is equal to the average of the Weyl symbol of this operator weighted with the Wigner function~\cite{hillery}. The Wigner function just plays the role of the probability distribution of phase space variables. Because the Wigner function is not positively defined, it is often referred to as the Wigner quasi-probability distribution.

Phase space methods were also developed and found many applications in quantum optics (see e.g. Refs.~\cite{walls-milburn, gardiner-zoller}). There it is convenient to work with the quantum coherent states first introduced by Schr\"odinger~\cite{schrodinger} and then reintroduced into the quantum optics and termed as ``coherent states'' by Glauber~\cite{glauber}. Quantum classical correspondence is then found by associating creation and annihilation operators of photons with complex amplitudes, which also define the phase space variables. The Liouville equation for the coherent state Wigner transform of the density matrix takes a form very similar to the equation for the Wigner function in the coordinate momentum representation. More recently phase space methods in quantum optics were adopted to atomic systems~\cite{steel, sinatra} and already found numerous applications in the field of cold atoms (see e.g. Ref.~\cite{blakie_08} for review).

The Wigner function and the corresponding equations of motion give a natural framework for the representation of quantum dynamics using entirely the language of phase space variables. The time evolution in this language can be then represented through deterministic motion of phase space degrees of freedom accompanied by the quantum jumps. The main goal of this work is to discuss this representation in detail. Within this approach one naturally recovers the classical limit and the structure of quantum corrections. There are many classical limits we are familiar with from the elementary physics, each providing a natural choice of the phase space variables: i) corpuscular, particle, or Newtonian limit, where coordinates and momenta of particles constitute the phase space; ii) wave limit, where number and phase of the fields (or equivalently complex amplitudes corresponding to creation and annihilation operators) form the phase space. This limit naturally appears in optics, wave mechanics, or more recently in the physics of weakly interacting degenerate Bose gases; iii) mixed limit where some degrees of freedom are treated as particles and some as waves. This limit naturally appears in gauge theories, most notably in electromagnetism; iv) classical spin limit, where spins are treated as rotators, and so on. The feature which unifies all classical descriptions is that the time evolution of the corresponding phase space variables is defined by the unique solution of deterministic equations of motion for given initial conditions. Very often even in classical systems the initial conditions are different from one experiment to another. Then one has to perform an additional averaging over the probability distribution of the initial conditions. Because the number of classical equations of motion scales linearly with the number of degrees of freedom, it is usually possible to simulate classical dynamics in relatively large systems. Contrary exact quantum simulations are usually limited to very small systems because of exponential dependence of the Hilbert space size on the number of degrees of freedom.

In this work we will be interested in approaching non-equilibrium dynamics in quantum systems, where we start from a well defined initial state either pure or mixed, either equilibrium or not. Then we change the parameters of the Hamiltonian in a certain way, and follow the time evolution of the observables of interest describing the system. We assume that our system is closed (i.e. there is no external heat bath) and that there is a well defined Hamiltonian, possibly time dependent, which determines the time evolution.  One of the very important questions, which needs to be addressed, is how the exponential complexity of quantum dynamics kicks in if we gradually increase $\hbar$. More specifically one needs to address the following questions: i) what we should do with equations of motion, ii) what we should do with quantum observables, and iii) what we should do with initial conditions as we increase $\hbar$ starting from zero. We would like to note that the Planck's constant $\hbar$ plays the role of the saddle-point parameter such that in the limit $\hbar\to 0$ a unique classical theory is recovered. In practice, there can be other saddle point parameters which play the same role like the (inverse) mode occupation in bosonic systems, the (inverse) spin size in the spin systems and so on. So by $\hbar$ we effectively understand some parameter which governs the strength of quantum fluctuations. As we go along it will become clear what plays the role of this parameter near particular classical limits.

These questions were partially addressed before. In particular, it was realized that the Wigner transform of the density matrix is especially suitable for analyzing dynamics near the classical limit when the quantum fluctuations are small~\cite{walls-milburn, steel}. It was also understood that in the leading order quantum fluctuations appear only through the (Wigner) distribution of the initial conditions while they do not affect the equations of motion themselves~\footnote{It should be noted that this statement is only true if the quantum Hamiltonian is written in the symmetric Weyl form. Otherwise there can be non-vanishing corrections to the equations of motion, which remain, however, fully deterministic.}. The resulting semiclassical or truncated Wigner approximation (TWA) was successfully applied to various non-equilibrium problems with bosonic cold atom systems (see Ref.~\cite{blakie_08} for review). There were also developed schemes based on other representations of the density matrix in the coherent state basis, in particular, based on the positive P-representation~\cite{walls-milburn, gardiner-zoller, steel, drummond_07}. Somewhat related method allowed one to map problem of interacting lattice bosons into stochastic evolution using the basis of Fock (number) states~\cite{carusotto}. In Ref.~\cite{deuar_09} a mixed representation of the density matrix, which smoothly interpolates between positive P- and Wigner representations,  was suggested. Such schemes, however, have their own issues and do not generally allow for systematic expansion in quantum fluctuations. In Refs.~\cite{ap_twa, ap_cat} it was shown that TWA naturally appears in the coherent state path integral representation in bosonic systems as the semiclassical approximation incorporating quantum fluctuation in the leading order to the classical (Gross-Pitaevskii) dynamics. Furthermore in Ref.~\cite{ap_twa} it was shown that the consequent quantum corrections to the expectation value of a particular observable appear in the form of the nonlinear response of this observable to the infinitesimal changes in the classical fields across the trajectories. Recently TWA was also extended to other situations going beyond traditional domain of interacting spinless Bose systems. E.g. in Ref.~\cite{agkp} the TWA was used to analyze dynamics of a particular spin-boson system described by the Dicke model. In Ref.~\cite{cohen} TWA was first applied to the spinor condensates. In Ref.~\cite{plimak_new} for the first time TWA was extended to tackle multi-time correlation functions. In Ref.~\cite{ludwig1} TWA was applied to analyze non-equilibrium vortex formation due to quantum fluctuations in a two-dimensional quantum rotor model.

The same semiclassical approximation (though not termed as TWA) was developed in the context of the quantum dynamics in the coordinate-momentum representation using the language of the Moyal brackets~\cite{hillery, zurek_rmp}. In the leading order in $\hbar$, where the Moyal brackets reduce to the Poisson brackets, one finds an analogue of TWA, where quantum fluctuations enter only through the initial conditions while the equations of motion are affected only in higher orders in $\hbar$. This approach was extensively used for studying quantum chaos and decoherence in single-particle systems (see e.g. Refs.~\cite{zurek_rmp, zurek_98, zurek_02}), particularly for a single particle system coupled to a heat bath. It was argued that the decoherence in chaotic systems was responsible for extending the validity of the semiclassical approach by eliminating effects of interference between single particle trajectories. In the single-particle context TWA was also applied to analyze collisions of wave packets with one-dimensional barriers~\cite{scully_83}.

In driven many-particle systems the most commonly used approach to the non-equilibrium phenomena is the Keldysh diagrammatic  technique~\cite{keldysh, kamenev, kamenev2}. This technique is often used to derive the quantum kinetic equation and obtain single-particle distribution functions in non-equilibrium systems~\cite{kamenev2}. Usually this technique is applied for finding steady states in driven systems (see E.g. Refs.~\cite{yb1, yb2, ehud_1/f}), but it can be also used for analyzing dynamics in situations, where initial conditions are important~\cite{ag, agkp}. While formally the Keldysh technique and phase space methods are equivalent, there is an important difference between them~\footnote{There is some ambiguity in defining what the Keldysh techniques is. Sometimes this techniques is understood in the broader sense of the functional representation of the time evolution on the Schwinger-Keldysh contour. See e.g. Ref.~\cite{plimak1}}. The elementary objects in the phase space methods are the Wigner function and phase space variables. In the Keldysh technique the elementary objects are the single particle's Green's functions. Closely related to the Keldysh diagrammatic technique there are also non-equilibrium approaches based on the functional integral representation of evolution~\cite{plimak1, plimak2, plimak3}, which can be further accompanied with various approximations like effective actions~\cite{gasenzer, rey, dias1, dias2, braunss}. These methods are now actively developed.

In this paper we will discuss how quantum dynamics can be formulated entirely in the phase space using the language of deterministic (classical) trajectories and stochastic quantum jumps.  We will review both the earlier results and present the new ones. A very important goal of this work is to show that the formalism is basically identical in various phase space representations (coordinate-momentum, coherent state, and spin coherent state). While some details between these representations are different the structures of the description are very similar suggesting generality of this approach. We will review how all these results can be derived using the Feynman path integral formulation of the time evolution. This derivation resembles the one routinely used in the Keldysh technique~\cite{kamenev, kamenev1,kamenev2} with the only difference that instead of working in the Schr\"odinger representation, where the density matrix evolves in time, we will work in the Heisenberg representation, where the operators describing observables change in time. In the coherent state representation the methods described in this work are also related to the functional integral approach used by Plimak {\em et al.}~\cite{plimak1}. There the density matrix only determines the initial state. The main difference of that work with the approach discussed here is that we will not attempt to find the evolution of the whole density matrix, which is an exponentially complex object in many-particle systems and contains tremendous amount of unmeasurable information. Instead we will be interested in finding expectation values of particular observables, like the energy or its fluctuations, various equal or nonequal time correlation functions, etc. We will show that such concepts like the Wigner function, Weyl ordering, TWA, and quantum corrections naturally appear from the Feynmann's path integral without need to make any assumptions. We will show that quantum corrections to TWA can be written either in the form of the nonlinear response to infinitesimal quantum jumps or in the form of stochastic quantum jumps with non-positive quasi-probability weight. In both cases each jump carries a factor of the effective Planck's constant squared. The representation of the quantum operators through the phase space variables also naturally emerges from the path integral formalism in the form first suggested by Bopp~\cite{bopp, hillery} and generalized here to time dependent operators:
\beq
\hat x(t)\to x(t)+ {i\hbar \over 2}{\partial\over\partial p(t)},\quad
\hat p(t)\to p(t)- {i\hbar \over 2}{\partial\over\partial x(t)}
\label{eq1}
\eeq
or
\beq
\hat x(t)\to x(t)- {i\hbar \over 2}{\overleftarrow\partial\over\partial p(t)},\quad
\hat p(t)\to p(t)+ {i\hbar \over 2}{\overleftarrow{\partial}\over\partial x(t)},
\label{eq1a}
\eeq
where the partial derivatives are understood as the response to the infinitesimal quantum jumps occurring at time $t$ and the choice of the sign is dictated by casuality of the evolution (see the next section for details). This Bopp representation will play the key role in the whole formalism. Throughout this paper we will reserve ``hat''-notations to denote quantum operators and ``non-hat''-notations for the functions defined in the phase space . There is a representation similar to Eqs.~(\ref{eq1}) and (\ref{eq1a}) for the creation and annihilation operators (see Eqs.~(\ref{eq:pa_dag}) and (\ref{eq:pa})). Without further details it is clear that the representation (\ref{eq1}) of the quantum coordinate and momentum operators above immediately reproduces the classical limit at $\hbar\to 0$.  It is interesting to note that the requirements of the casuality of the representation of the quantum dynamics through the phase space trajectories dictates that only certain multi-time correlation functions are allowed (see Sec.~\ref{sec:der}). These turn out to be precisely the same correlation functions which appear in the theory of measurements~\cite{aash_rmp,nazarov}.

Let us note that in three (and higher) dimensional continuous systems with short range interactions TWA suffers from ultra-violet divergencies~\cite{deuar_07, blakie_08}. This indicates that the expansion of dynamics around the classical limit can be ill defined and likely requires some short-distance renormalization of the bare parameters of the Hamiltonian. In the context of cold atoms certain ways to deal with these divergencies by effectively truncating the Hilbert space were suggested in the literature~\cite{blakie_08}. However, comprehensive understanding of these issues is far from being complete and is beyond the scope of this work. It is not even understood whether these ultra-violet divergencies are intrinsic property of three-dimensional systems or they are simply related to the singular nature of short-range potentials in three dimensions, which is well known from equilibrium physics~\cite{fetter_walecka}. A simple way to avoid these divergencies,  as we illustrate in Sec.~\ref{sec:applications}, is to work with lattice systems.

The paper is organized as follows. First in Sec.~\ref{sec:Moyal} we give a brief introduction into the phase space representation of the density matrix and operators. Starting from the coordinate-momentum representation we describe such concepts as the Wigner function, Weyl symbol, Moyal product and Moyal brackets, and Bopp operators. We then discuss similar concepts in the coherent state picture. In particular, we introduce coherent state analogues of the Poisson and Moyal brackets as well as of the Moyal product. In the end of this section we discuss close analogy between the coordinate-momentum and coherent state representations. Then in Sec.~\ref{sec:main} we give a detailed overview of the formalism discussed in this paper. We tried to highlight all the features of the formalism important for practical implementation of TWA and quantum corrections to particular problems skipping the details of the derivations of these results. We split this section into three parts discussing first dynamics in the coordinate-momentum representation, then in the coherent state representation, and finally quantum dynamics of spin systems. Then in Sec.~\ref{sec:applications} we discuss the implementation of the formalism to particular problems. We start from the simplest possible example of a harmonic oscillator suddenly driven from equilibrium gradually increasing the complexity of the systems and describing dynamics in the sine-Gordon model, one and two dimensional Bose-Hubbard models, Dicke model and others. For small systems we give comparison of the results obtained within this approach with the exact results and discuss in which situations the quantum expansion works well and in which it fails. Our goal was not to describe the corresponding physics of these models in detail, but rather to focus on the applicability of the method. Then in Sec.~\ref{sec:der} we guide the reader through the derivation of the results using the path integral approach. This section is rather technical and can be skipped unless the reader is interested in understanding these details. One of the main purposes of this section is to show that the Weyl symbol, Wigner function, quantum corrections, casual representation of the dynamics, Bopp operators, etc. naturally appear from the path integral approach without need to make any assumptions or to justify this choice {\em a-posteriori}. Finally in Sec.~\ref{sec:liouv_wig} we discuss connections of the path integral method used here and the other approaches based on the von Neumann's equation for the density matrix. In Appendix~\ref{sec:app_implement} we illustrate the detailed implementation of the first quantum correction to the TWA for the sine-Gordon and Bose-Hubbard models. In Appendix~\ref{appendix_hamilt_ordering} we discuss a somewhat subtle issue of emergence of the Weyl ordering of the Hamiltonian in classical equations of motion within the path integral representation of the evolution (which usually  requires the normal ordering of the Hamiltonian~\cite{kamenev2}).

\section{Brief overview of the phase space representation of quantum systems.}
\label{sec:Moyal}

In this section we will briefly introduce the basic tools, which will be essential in the next sections when we will talk about dynamics. In particular, here we will introduce such concepts as the Wigner function, Weyl symbol, Moyal bracket, and Bopp operators. The Moyal brackets will also allow us to write down equations of motion for the density matrix in the phase space representation and discuss their close connection to the classical Liouville's equations (see Sec.~\ref{sec:liouv_wig}). We will discuss and contrast the two basic phase space representations: coordinate-momentum and the bosonic coherent state. The first representation is naturally connected to the corpuscular classical limit while the second representation corresponds to the wave classical limit. In this work we will not talk about the Grassmann's number representation of the fermionic phase space.

There are several reviews available in literature describing the phase space representation of quantum systems.In particular Ref.~\cite{hillery} gives an excellent overview of such representation in the coordinate-momentum phase space. Details of the coherent state representation of bosonic systems can be found in Refs.~\cite{blakie_08, gardiner-zoller, walls-milburn}. We will thus not attempt to give a comprehensive review of these  representations and skip the proofs, which can be found in the above mentioned texts. We will only introduce the tools necessary for understanding the consequent sections describing dynamics. We will also contrast these two phase space representations. By introducing coherent state Poisson and Moyal brackets we will show that these two descriptions of quantum systems identically map to each other under the change of the phase space variables. The reader familiar with these concepts can skip this section and directly proceed to the next one.

Let us note that the phase space representation of quantum systems is not unique~\cite{hillery, gardiner-zoller, walls-milburn}. In this work we will focus only on the Wigner-Weyl representation because, as we will show later, it automatically emerges from the path integral description of quantum dynamics and gives the most natural connection between classical and quantum dynamics. This representation also treats conjugate phase space operators symmetrically.

\subsection{Coordinate-momentum representation}.

In the context of ordinary canonical variables like coordinates and momenta the phase space approach to quantum mechanics was pioneered by Moyal~\cite{moyal}. It is based on the Weyl ordering of operators and the Wigner distribution function (see e.g. Refs.~\cite{hillery, connell}).

{\em Weyl symbol and Wigner function.} The key ingredient in the phase space description of quantum systems is the so called Weyl symbol, which gives a one to one map between quantum operators and the ordinary functions defined in the phase space. For Hermitian operators this map is real. Thus for an arbitrary operator $\hat \Omega(\hat {\bf x},\hat {\bf p})$  the Weyl symbol $\Omega_W({\bf x},{\bf p})$ is formally defined as
\be
\Omega_W({\bf x},{\bf p})=\int d\bm{\xi} \left<{\bf x}-{\bm{\xi}\over 2}\right|\hat \Omega(\hat {\bf x},\hat {\bf p})\left|{\bf x}+{\bm {\xi}\over 2}\right>\exp\left[{i\over \hbar} {\bf p}\cdot\bm{\xi}\right].
\label{w_weyl}
\ee
We use the vector notations to highlight that we are dealing with general d-dimensional multi-particle phase space of the dimension 2D, where the factor of two reflects that for each degree of freedom we are dealing with pairs of conjugate variables. In particular, for N-particle system in d-dimensions we have $D=Nd$. If the operator $\hat\Omega(\hat {\bf x},\hat {\bf p})$ is written in the symmetrized form then it is straightforward to see that the Weyl symbol $\Omega_W$ is obtained by simple substitution $\hat {\bf x}\to {\bf x}$ and $\hat{\bf p}\to {\bf p}$. In particular, this is true for all operators of the form $\hat \Omega(\hat{\bf x}, \hat{\bf p})=\hat A(\hat{\bf x})+\hat B(\hat{\bf p})$.

If the ordering in $\hat\Omega(\hat {\bf x},\hat {\bf p})$ is such that the coordinate operators appear on the left of momentum operators then the Weyl symbol (\ref{w_weyl}) can be also written as
\be
\Omega_{W}({\bf x}, {\bf p})=\int {d \bm{\xi} d \bm{\eta}\over (4\pi\hbar)^D}\, \Omega\left({\bf x}-{\bm{\xi}\over 2}, {\bf p}+{\bm {\eta}\over 2}\right)\mathrm e^{-i\bm {\xi \eta}/2\hbar}.
\label{weyl}
\ee
Here $\Omega({\bf x},{\bf p})$ is the function obtained from the operator $\hat\Omega$ by direct substitution $\hat {\bf x}\to {\bf x}$ and $\hat {\bf p}\to {\bf p}$. The equivalence of Eqs.~(\ref{w_weyl}) and (\ref{weyl}) can be established by inserting the identity resolution $\hat {\bf I}=1/2\int d\bm{\eta}\, |{\bf p}+\bm{\eta}/2\rangle\langle {\bf p}+\bm{\eta}/2|$ into Eq.~(\ref{w_weyl}).

The second key ingredient of phase space methods is the Wigner function which is defined as the Weyl symbol of the density matrix $\hat \rho$:
\be
W({\bf x},{\bf p})=\int d\bm{\xi} \left< {\bf x}-{\bm{\xi}\over 2}\right|\, \hat\rho\,  \left|\, {\bf x}+{\bm{\xi}\over 2}\right> \mathrm e^{i{\bf p}\bm{\xi}/\hbar}\equiv\int d\bm{\xi}\, \rho({\bf x}-\bm{\xi}/2,{\bf x}+\bm{\xi}/2)\,\mathrm e^{i{\bf p}\bm{\xi}/\hbar}.
\label{wig_xp}
\ee
For any proper density matrix the Wigner function is normalized such that
\be
\int {d{\bf x}d{\bf p}\over (2\pi\hbar)^D} W({\bf x},{\bf p})=1.
\ee

The Wigner function together with Weyl symbols of various operators gives complete description of a given system. In particular, the expectation value of any operator $\hat\Omega(\hat {\bf x},\hat {\bf p})$ is found by averaging the Weyl symbol of this operator over the phase space with the Wigner function playing the role of the probability distribution:
\be
\langle \hat{\Omega}(\hat {\bf x},\hat {\bf p})\rangle= \int\int {d {\bf x} d {\bf p}\over (2\pi\hbar)^D} W({\bf x},{\bf p})\Omega_{W}({\bf x},{\bf p}).
\label{eq:omega2_0}
\ee
Because the Wigner function is not positively defined it is often referred to as the Wigner quasi-probability distribution.

{\em Moyal product.} Weyl symbols of operators satisfy important Moyal product relation~\cite{hillery} which defines the Weyl symbol of the product of two operators $\hat\Omega_1\hat\Omega_2$ through the Weyl symbols of the individual operators:
\be
(\Omega_1\Omega_2)_W({\bf x},{\bf p})=\Omega_{1,W}({\bf x},{\bf p})\exp\left[-{i\hbar\over 2}\Lambda\right]\Omega_{2,W}({\bf x},{\bf p}),
\label{moyal_prod}
\ee
where
\be
\Lambda={\overleftarrow\partial\over\partial {\bf p}}\cdot{\overrightarrow\partial\over\partial {\bf x}}-{\overleftarrow\partial\over\partial {\bf x}}\cdot{\overrightarrow\partial\over\partial {\bf p}}=\sum_j {\overleftarrow\partial\over\partial p_j}{\overrightarrow\partial\over\partial x_j}-{\overleftarrow\partial\over\partial x_j}{\overrightarrow\partial\over\partial p_j}
\label{lambda_def}
\ee
is the so called simplectic operator. It is straightforward to verify the validity of Eq.~(\ref{moyal_prod}) using the definition of the Weyl symbol~(\ref{weyl}). For any two functions $A({\bf x},{\bf p})$ and $B({\bf x},{\bf p})$ the product $-A\Lambda B=\left\{A,B\right\}$ defines the Poisson bracket between these functions:
\be
\left\{A,B\right\}= \sum_{j} {\partial A\over\partial x_j}{\partial B\over\partial p_j}-{\partial A\over\partial p_j}{\partial B\over\partial x_j}= \left\{A,B\right\}.
\label{eq:PB}
\ee
Note there are different sign conventions in defining Poisson brackets; e.g. compare refs.~\cite{LLI} and~\cite{hillery}. In Eq.~(\ref{eq:PB}) we adopted the choice of Ref.~\cite{hillery}. So the Moyal product of the two operators (\ref{moyal_prod}) can be also expressed through the exponent of the Poisson bracket:
\be
(\Omega_1\Omega_2)_W({\bf x},{\bf p})=\Omega_{1,W}({\bf x},{\bf p})\exp\left[{i\hbar\over 2}\left\{\dots\right\}\right]\Omega_{2,W}({\bf x},{\bf p}).
\ee
From Eq.~(\ref{moyal_prod}) it follows that the Weyl symbol of the commutator of the two operators $\hat\Omega=\left[\hat\Omega_1,\hat\Omega_2\right]$ can be expressed as
\be
\Omega_W=-2i\Omega_{1,W}\sin\left[{\hbar\over 2}\Lambda\right]\Omega_{2,W}=2i\Omega_{1,W}\sin\left[{\hbar\over 2}\left\{\dots\right\}\right]\Omega_{2,W}\equiv i\hbar\left\{\Omega_1,\Omega_2\right\}_{MB},
\label{moyal_bracket}
\ee
where $\left\{\dots\right\}_{MB}$ defines the Moyal bracket:
\be
\left\{\dots\right\}_{MB}={2\over \hbar}\sin\left[{\hbar \over 2}\left\{\dots\right\}\right]
\ee
In the limit $\hbar\to 0$ the Weyl symbol of the commutator of any two operators reduces to the Poisson bracket between the corresponding classical functions multiplied by $i\hbar$. This is of course a well known manifestation of the correspondence principle~\cite{LL3}.

{\em Bopp operators.} Wigner-Weyl quantization is intrinsically connected with the symmetric representation of the phase space operators first suggested by Bopp~\cite{bopp, kubo}:
\be
\hat{\bf x}= {\bf x}+{i\hbar\over 2}{\partial\over {\partial} {\bf p}},\quad
\hat{\bf p}= {\bf p}-{i\hbar\over 2}{\partial\over \partial {\bf x}}.\label{bopp_rp}
\ee
It is clear that this representation satisfies the canonical commutation relations. There is an equivalent Bopp representation based on the left derivatives:
\be
\hat{\bf x}= {\bf x}-{i\hbar\over 2}{\overleftarrow\partial\over {\partial} {\bf p}},\quad
\hat{\bf p}= {\bf p}+{i\hbar\over 2}{\overleftarrow\partial\over \partial {\bf x}},\label{bopp_rp1}
\ee
where the left arrow implies that the derivative acts on the operator on the left. Loosely speaking the representation (\ref{bopp_rp1}) is obtained from Eq.~(\ref{bopp_rp}) by integrating by parts. The formal proof that the Bopp operators reproduce the Weyl symbol can be found in Refs.~\cite{hillery, osborn}. We will also prove these statement in Sec.~\ref{sec:der} when we will analyze non-equal time correlation functions.

Let us illustrate how the Bopp operators can be used in practice. Suppose we are interested in a Weyl symbol of the operator $\hat x\hat p$. For simplicity we choose a single particle in one dimension. Then using the Bopp operators we can immediately construct the Weyl symbol of this product:
\be
\hat{x}\hat{p}=\hat{x}\hat{p}\, 1\to \left(x+{i\hbar\over 2}{\partial\over {\partial} p}\right)\left(p-{i\hbar\over 2}{\partial\over {\partial} x}\right)1=xp+{i\hbar\over 2}.
\label{omega_rp}
\ee
Thus we conclude that $(xp)_W=xp+i\hbar/2$. The same result can be obtained using the right derivatives:
\be
\hat{x}\hat{p}=1\,\hat{x}\hat{p}\to 1\left(x-{i\hbar\over 2}{\overleftarrow{\partial}\over {\partial} p}\right)\left(p+{i\hbar\over 2}{\overleftarrow{\partial}\over {\partial} x}\right)=xp+{i\hbar\over 2}.
\label{omega_rp_a}
\ee
We inserted unity into Eqs.~(\ref{omega_rp}) and (\ref{omega_rp_a}) to illustrate that the are no other operators on the right of $\hat p$ and left of $\hat x$. The same result can be obtained directly from Eq.~(\ref{weyl}) as well as from the symmetrization
\be
\hat x\hat p={\hat x \hat p+\hat p\hat x\over 2}+{1\over 2}[\hat x,\hat p]\to xp+{i\hbar \over 2}
\ee
and from the Moyal product expansion:
\be
(xp)_W=x\exp\left[-{i\hbar \over 2}\Lambda\right]p=xp-{i\hbar \over 2}x\Lambda p=xp+{i\hbar\over 2}.
\ee
which is exactly equal to the result obtained from Eq.~(\ref{weyl}). As we will see below in the Heisenberg picture the derivative terms in Eqs.~(\ref{bopp_rp}) and (\ref{bopp_rp1}) get an additional interpretation as a response to infinitesimal quantum jumps at a particular moment of time. Then the choice of the left or right derivative is dictated by the casuality of the evolution.

\subsection{Coherent state representation.}

All the results above immediately generalize to the coherent state representation of the phase space. Bosonic coherent states $|\psi\rangle_c$ are defined as eigenstates of bosonic annihilation operators $\hat\psi$ such that $\hat\psi|\psi\rangle_c=\psi|\psi\rangle_c$. Explicitly the coherent states read:
\be
|\psi\rangle_c=\exp[\psi\,\hat\psi^\dagger]|0\rangle=\sum_n {\psi^n\over \sqrt{n!}}|n\rangle,
\label{coh_st}
\ee
where $|0\rangle$ is the vacuum state and $|n\rangle=(\hat\psi^\dagger)^n/\sqrt{n!}\,|0\rangle$ is the n-particle Fock state. It is straightforward to check that
\be
\hat\psi^\dagger|\psi\rangle_c={\partial\over\partial \psi} |\psi\rangle_c.
\ee
 Obviously there is a direct analogy between the action of the coordinate $\hat x$ and momentum $\hat p$ operators on the coordinate basis ($\hat x |x\rangle=x|x\rangle$, and $\hat p|x\rangle=-i\hbar \partial_x|x\rangle$) and the action of the annihilation $\hat\psi$ and creation $\hat\psi^\dagger$ operators on the coherent state basis. As we will see this analogy will persist throughout the paper. The coherent states (\ref{coh_st}) are not orthogonal and not normalized. In particular
\be
\phantom{x}_c\langle\psi|\psi'\rangle_c=\exp[\psi^\star\psi'].
\ee
Because of non-orthogonality, the coherent state basis is over-complete. Using these states one can write the resolution of the identity
\be
\hat 1=\int d\psi d\psi^\star \mathrm e^{-|\psi|^2} |\psi\rangle_{c\,c}\langle \psi|,
\ee
where the integration measure is defined as $d\psi d\psi^\star=\Re\psi d\Im\psi/\pi$.

{\em Weyl symbol and Wigner function.} Using coherent states one can define the Weyl symbol of and arbitrary operator $\hat\Omega(\hat{\bm{\psi}},\hat{\bm{\psi}}^\dagger)$ as
\be
\Omega_W(\bm{\psi,\psi^\star})={1\over 2^M}\int\int d\bm{\eta}^\star d\bm{\eta} \left<
\bm{\psi - {\eta\over 2}}\right|\hat\Omega(\hat{\bm\psi},\hat{\bm\psi}^\dagger) \left| \bm{\psi+{\eta\over
2}}\right>\mathrm e^{-|\bm\psi|^2-{1\over
4}|\bm{\eta}|^2}\,\mathrm e^{{1\over
2}(\bm{\eta^\star\psi-\eta\psi^\star})}.
\label{weyl_coherent_def}
\ee
Here the vector ${\bm \psi}=\{\psi_j\}$ consists of complex amplitudes corresponding to different single-particle eigenstates, $M$ is the Hilbert space size. So $j$ can denote coordinates, momenta, different internal spin states, etc. As in the coordinate-momentum representation the Weyl symbol of $\hat\Omega$ is the partial Fourier transform of the matrix elements of this operator between different coherent states. The additional gaussian factors come from the normalization of the coherent states. Again as in the coordinate momentum representation the Weyl symbol of a symmetrically ordered operator can be obtained by simple substitution $\hat{\bm \psi}\to\bm{\psi}$ and $\hat{\bm\psi}^\dagger\to\bm{\psi}^\star$. For normally ordered operators Eq.~(\ref{weyl_coherent_def}) gives
\be
\Omega_{W}(\bm{\psi},\bm{\psi}^\star)={1\over 2^M}\int\int d\bm{\eta} d\bm{\eta}^\star \Omega\left(\bm{\psi}^\star-{\bm{\eta}^\star/
2},\bm{\psi}+{\bm{\eta}/ 2}\right)\mathrm e^{-|\bm{\eta}|^2/2}.
\label{Weyl_coherent}
\ee

Similarly to the coordinate-momentum representation the Wigner function is defined as the Weyl symbol of the density matrix:
\be
W(\bm{\psi,\psi^\star})={1\over 2^M}\int\int d\bm{\eta}^\star d\bm{\eta} \left<
\bm{\psi - {\eta\over 2}}\right|\hat\rho \left| \bm{\psi+{\eta\over
2}}\right>\mathrm e^{-|\bm\psi|^2-{1\over
4}|\bm{\eta}|^2}\,\mathrm e^{{1\over
2}(\bm{\eta^\star\psi-\eta\psi^\star})}.
\label{wig_coherent}
\ee
The expectation value of any operator, by analogy with Eq.~(\ref{eq:omega2_0}), is given by the average of the Weyl symbol with the weight given by the Wigner function:
\be
\langle \hat{\Omega}(\hat{\bm\psi},{\hat{\bm \psi}^\dagger},t)\rangle=\int\int d {\bm \psi} d \bm{\psi}^\star\, W({\bm \psi},\bm{\psi}^\star)\Omega_{W}(\bm{\psi},\bm{\psi}^\star).
\label{eq:omega5_0}
\ee
So the Wigner function again plays the role of the quasi-probability distribution of the complex amplitudes.

{\em Moyal product and Bopp operators.} The Weyl symbol of the product of two operators can be written by analogy to Eq.~(\ref{moyal_prod}):
\be
(\Omega_1\Omega_2)_W=\Omega_{1,W} \exp\left[\Lambda_c\over 2\right]\Omega_{2,W},
\label{moyal_prod_coh}
\ee
where we introduced the symplectic coherent state operator (cf. Eq.~(\ref{lambda_def})):
\be
\Lambda_c=\sum_j {\overleftarrow{\partial}\over\partial\psi_j} {\overrightarrow{\partial}\over\partial\psi_j^\star}- {\overleftarrow{\partial}\over\partial\psi_j^\star} {\overrightarrow{\partial}\over\partial\psi_j}.
\ee
Similarly to the coordinate-momentum representation it is convenient to introduce the coherent state Poisson bracket between two arbitrary functions $A$ and $B$ as
\be
\left\{A,B\right\}_c=A\Lambda_c B =\sum_j {\partial A\over \partial\psi_j}{\partial B\over\partial \psi_j^\star}- {\partial A\over \partial\psi_j^\star}{\partial A\over\partial \psi_j}.
\label{coh_pois_brack}
\ee
The coherent state Poisson bracket is a classical analogue of the commutator. In particular, $\{\psi_i,\psi_j^\star\}_c=\delta_{i,j}$ is a classical analogue of $[\hat\psi_i,\hat\psi_j^\dagger]=\delta_{i,j}$. In general it is easy to see that in the classical limit of large occupation numbers $[\hat A,\hat B]\to \{A,B\}_c$. Note that the coherent state Poisson bracket defined above~(Eq.~(\ref{coh_pois_brack})) directly corresponds to the commutator without need to multiply by $i\hbar$. This choice is dictated by the convenience so that the classical Hamiltonian equations of motion in the wave limit read $i\hbar\partial_t\psi_j=\{\psi_j, H\}_c$. It is easy to check that for bosonic systems with two-body interactions this equation becomes identical to the Gross-Pitaevskii equation (see Sec.~\ref{sec:liouv_wig} for details).

Using the coherent state Moyal product rule~(\ref{moyal_prod}) and the symplectic structure of the operator $\Lambda_c$ we can find the expression for the Weyl symbol of the commutator of the two operators $\hat\Omega=[\hat\Omega_1,\hat\Omega_2]$
\be
\Omega_W=2\Omega_{1,W}\sinh\left[\Lambda_c\over 2\right]\Omega_{2,W}=\left\{\Omega_{1,W},\Omega_{2,W}\right\}_{MBC},
\label{commut_coh}
\ee
where we introduced the coherent state Moyal bracket by analogy with standard Moyal bracket~(\ref{moyal_bracket}):
\be
\left\{\dots\right\}_{MBC}=2\sinh\left[{1\over 2}\left\{\dots\right\}_c\right].
\label{coh_moyal_bracket}
\ee

Wigner Weyl quantization is intrinsically associated with the coherent state Bopp operators which can be introduced by analogy with Eqs.~(\ref{bopp_rp}):
\beq
&&\hat\psi_j^\dagger\to\psi_j^\star-{1\over 2}{{\partial}\over\partial\psi_j}=\psi_j^\star+{1\over 2}{\overleftarrow{\partial}\over\partial\psi_j},
\label{bopp_psi_dag}\\
&&\hat\psi_j\to\psi_j+{1\over 2}{{\partial}\over\partial\psi_j^\star},=\psi_j-{1\over 2}{\overleftarrow{\partial}\over\partial\psi_j^\star}.
\label{bopp_psi}
\eeq
The choice of the representation with the conventional (right) derivatives and the one with left derivatives is arbitrary. As we will see below for time dependent problems it is dictated by casuality. This representation of creation and annihilation operators is clearly symmetric and preserves the correct commutation relations. It also automatically reproduces the Weyl symbol of any operator (the proof of this statement will be given in Sec.~\ref{sec:der_coherent}). To illustrate this let us show a couple of examples of computing Weyl symbols of $\hat\psi^\dagger\hat\psi$ and $\hat\psi^\dagger\hat\psi^\dagger\hat\psi\hat\psi$ first using the Moyal product rule and then using the Bopp operators. Thus using Eq.~(\ref{moyal_prod_coh}) we find
\be
(\hat\psi^\dagger\hat\psi)_W=\psi^\star \left(1-{1\over 2} {\overleftarrow{\partial}\over\partial\psi^\star} {\overrightarrow{\partial}\over\partial\psi}\right)\psi=|\psi^2|-{1\over 2}.
\label{weyl_example_coh1}
\ee
The same expression can be obtained using the Bopp operators
\be
\hat\psi^\dagger\hat\psi=\hat\psi^\dagger\hat\psi 1\to\left(\psi^\star-{1\over 2}{{\partial}\over\partial\psi}\right) \left(\psi+{1\over 2}{{\partial}\over\partial\psi^\star}\right)1=|\psi|^2-{1\over 2}.
\ee
Note that the Weyl symbol for the number operator can be obtained also by symmetrizing it first $\hat\psi^\dagger\hat\psi={1/2}[\hat\psi^\dagger\hat\psi+\hat\psi\hat\psi^\dagger]-1/2$ and then substituting $\hat\psi\to\psi$ and $\hat\psi^\dagger\to\psi^\star$ in the symmetrized form. This Weyl symbol can also be obtained by direct application of Eq.~(\ref{Weyl_coherent}) to the operator $\hat\Omega=\hat\psi^\dagger\hat\psi$.

Let us also briefly consider the other example first using the Moyal product:
\be
(\hat\psi^\dagger\hat\psi^\dagger\hat\psi\hat\psi)_W=(\psi^\star)^2 \left(1-{1\over 2} {\overleftarrow{\partial}\over\partial\psi^\star} {\overrightarrow{\partial}\over\partial\psi}+{1\over 8} {\overleftarrow{\partial}^2\over\partial^2\psi^\star} {\overrightarrow{\partial}^2\over\partial\psi^2}\right)\psi^2=|\psi^2|^2-2|\psi|^2+{1\over 2}
\ee
and then using Bopp operators
\be
\hat\psi^\dagger\hat\psi^\dagger\hat\psi\hat\psi\to \left(\psi^\star-{1\over 2}{{\partial}\over\partial\psi}\right)^2 \left(\psi+{1\over 2}{{\partial}\over\partial\psi^\star}\right)^2 1=|\psi^2|^2-2|\psi|^2+{1\over 2}.
\ee
Again both methods give the same result which can be also verified using the direct integration in Eq.~(\ref{Weyl_coherent}).

\subsection{Coordinate-momentum versus coherent state representations.}

To summarize the discussion above we will contrast the two phase space pictures in Table~\ref{table1}.
\begin{table}[h]
\caption{Coherent state versus coordinate momentum phase space}
\centering
\begin{tabular}{| c | c | c |}
\hline\hline
Representation & coordinate-momentum & coherent  \\
 \hline
Phase space variables & ${\bf x}, {\bf p}$ & $\bm{\psi},\bm{\psi}^\star$\\ \hline
Quantum operators & $\hat {\bf x},\hat {\bf p}$ & $\hat{\bm{\psi}},\hat{\bm{\psi}}^\dagger$\\ \hline
Standard representation &
\begin{tabular}{c}
$\hat {\bf x}\to {\bf x},\;\hat {\bf p}\to -i\hbar\partial_{\bf x}$\\ (coordinate basis)
\end{tabular}
& \begin{tabular}{c}
$\hat{\bm{\psi}}\to \bm{\psi}$, $\hat{\bm{\psi}}^\dagger\to \partial_{\bm{\psi}}$\\
(coherent state basis)
\end{tabular}\\ [2ex]\hline
\begin{tabular}{c}
Canonical\\ commutation relations
\end{tabular}&
\begin{tabular}{c}
$[\hat x_\alpha,\hat p_\beta]=i\hbar\delta_{\alpha,\beta} $\\
($\alpha,\beta$ refer to different particles)
\end{tabular}
& \begin{tabular}{c} $[\hat \psi_i,\hat \psi_j^\dagger]=\delta_{ij} $\\
($i,j$ refer to single-particle states)
\end{tabular}\\ [2ex]\hline
\begin{tabular}{c}
Quantum-classical \\
correspondence \end{tabular}&
\begin{tabular}{c}
$\hat {\bf x}\to {\bf x},\, \hat {\bf p}\to {\bf p},\, [\hat A,\hat B]\to i\hbar\{A,B\}$\\
$\{A,B\}=\sum_\alpha {\partial A\over \partial x_\alpha}{\partial B\over \partial p_\alpha}
-{\partial A\over \partial p_\alpha}{\partial B\over \partial x_\alpha}$
\end{tabular}&
\begin{tabular}{c}
$\hat{\bm{\psi}}\to \bm{\psi},\,\hat{\bm{\psi}}^\dagger\to\bm{\psi}^\star,\, [\hat A,\hat B]\to \{A,B\}_c$\\
$\{A,B\}_c=\sum_j {\partial A\over \partial \psi_j}{\partial B\over \partial \psi_j^\star}
-{\partial A\over \partial \psi_j^\star}{\partial B\over \partial \psi_j}$\\
\end{tabular}\\[2ex] \hline
Wigner function &
$W({\bf x},{\bf p})=\int\! d\bm{\xi} \left< {\bf x}-{\bm{\xi}\over 2}\right|\, \hat\rho\,  \left|\, {\bf x}+{\bm{\xi}\over 2}\right> \mathrm e^{i{\bf p}\bm{\xi}/\hbar}$ &
\begin{tabular}{c}
$W(\bm{\psi,\psi}^\star)={1\over 2}\int\!\int\! d\bm{\eta}^\star d\bm{\eta} \left< \bm{\psi - {\eta\over 2}}\right|\hat\rho \left| \bm{\psi+{\eta\over 2}}\right>$\\
$~~~~~~~~~~\times\mathrm e^{-|\bm\psi|^2-{1\over 4}|\bm{\eta}|^2}\,\mathrm e^{{1\over
2}(\bm{\eta^\star\psi-\eta\psi^\star})}$
\end{tabular}\\[2ex] \hline
Weyl symbol &
$\Omega_W({\bf x},{\bf p})=\int\! d\bm{\xi} \left< {\bf x}-{\bm{\xi}\over 2}\right|\, \hat\Omega\,  \left|\, {\bf x}+{\bm{\xi}\over 2}\right> \mathrm e^{i{\bf p}\bm{\xi}/\hbar}$ &
\begin{tabular}{c}
$\Omega_W(\bm{\psi,\psi}^\star)\!={1\over 2}\!\int\!\int\! d\bm{\eta}^\star d\bm{\eta} \left< \bm{\psi - {\eta\over 2}}\right|\hat\Omega \left| \bm{\psi+{\eta\over 2}}\right>$\\
$~~~~~~~~~~\times\mathrm e^{-|\bm\psi|^2-{1\over 4}|\bm{\eta}|^2}\,\mathrm e^{{1\over
2}(\bm{\eta^\star\psi-\eta\psi^\star})}$
\end{tabular}\\[2ex] \hline
Moyal product &
\begin{tabular}{c}
$(\Omega_1\Omega_2)_W=\Omega_{1,W}\exp\left[-{i\hbar\over 2}\Lambda\right]\Omega_{2,W},$\\
$\Lambda=\sum_\alpha {\overleftarrow\partial\over\partial p_\alpha}{\overrightarrow\partial\over\partial x_\alpha}-{\overleftarrow\partial\over\partial x_\alpha}{\overrightarrow\partial\over\partial p_\alpha}$
\end{tabular} &
\begin{tabular}{c}
$(\Omega_1\Omega_2)_W=\Omega_{1,W} \exp\left[\Lambda_c\over 2\right]\Omega_{2,W}$,\\
$\Lambda_c=\sum_j {\overleftarrow{\partial}\over\partial\psi_j} {\overrightarrow{\partial}\over\partial\psi_j^\star}- {\overleftarrow{\partial}\over\partial\psi_j^\star} {\overrightarrow{\partial}\over\partial\psi_j}$
\end{tabular}\\[2ex] \hline
Moyal bracket &
$\left\{\Omega_1,\Omega_2\right\}_{MB}=-{2\over \hbar}\Omega_1\sin\left[{\hbar \over 2}\Lambda\right]\Omega_2$ &
$\left\{\Omega_1,\Omega_2\right\}_{MBC}=2\Omega_1\sinh\left[{1\over 2}\Lambda_c\right]\Omega_2$ \\[2ex] \hline
Bopp operators &
\begin{tabular}{c}
$\hat{\bf x}= {\bf x}+{i\hbar\over 2}{\partial\over {\partial} {\bf p}}={\bf x}-{i\hbar\over 2}{\overleftarrow\partial\over {\partial} {\bf p}},$\\[2ex]
$\hat{\bf p}= {\bf p}-{i\hbar\over 2}{\partial\over \partial {\bf x}}= {\bf p}+{i\hbar\over 2}{\overleftarrow\partial\over \partial {\bf x}}$
\end{tabular} &
\begin{tabular}{c}
$\hat{\bm{\psi}}^\dagger=\bm{\psi}^\star-{1\over 2}{{\partial}\over\partial\bm{\psi}}=\bm{\psi}^\star+{1\over 2}{\overleftarrow{\partial}\over\partial\bm{\psi}}$,\\ [2ex]
$\hat{\bm \psi}=\bm{\psi}+{1\over 2}{{\partial}\over\partial\bm{\psi}^\star},=\bm{\psi}-{1\over 2}{\overleftarrow{\partial}\over\partial\bm{\psi}^\star}$
\end{tabular} \\[2ex] \hline
\end{tabular}
\label{table1}
\end{table}
The main purpose of doing this is to emphasize the similarity between the coordinate-momentum and coherent state representations. Normally the latter is introduced in the context of the second quantization. This table shows that there is no secondary aspect to the quantization in the coherent state representation. These two representations correspond to two dual descriptions of the phase space: corpuscular and wave. Note that this dual description is only available for bosonic systems. For fermionic coherent state phase space description requires use of Grassmann variables~\cite{negele_orland}, which do not have a natural classical interpretation. While one can formally define the Wigner-Weyl quantization for the fermionic systems, its extension to dynamics is not understood by now and will be a subject of the future research.

\section{Quantum dynamics in phase space.}
\label{sec:main}

In this section we will discuss the representation of quantum dynamics in the phase space providing sufficient details for further exploration of this formalism and for its practical applications. We will skip the details of the derivation of these results, which will be discussed later in Sec.~\ref{sec:der}. First in Sec.~\ref{sec:coord-moment} we will overview the situation in the coordinate-momentum representation, where the phase space is represented by the coordinates and momenta of particles. In the corresponding classical limit the particles move according to deterministic trajectories in this phase space defined by the microscopic Newton's (or more generally Hamiltonian's) equations of motion. Next in Sec.~\ref{sec:main_coherent} we will describe dynamics of interacting bosons in the coherent state phase space. There the classical (Gross-Pitaevskii) limit corresponds to the (matter) waves. And finally in Sec.~\ref{sec:spin} we will discuss spin systems. Using the Schwinger mapping of spins to bosons and then using the results of Sec.~\ref{sec:main_coherent} we will avoid the need of working with rather complicated spin-coherent states. As we already mentioned in this paper we will not talk about the wave limit for fermions. In many situations dynamics of fermions is described by bosonic collective excitations and thus can be analyzed within the methods described here. We will mention an example of such mapping of fermions to bosons and the application of the phase space formalism to the dynamics when we discuss the Dicke model in Sec.~\ref{sec:dicke}.

Before proceeding let us note that there is a quite common misconception that the Gross-Pitaevskii equations are only applicable to the condensates and describe only the mean-field dynamics, i.e. that they apply only to situations when the condensate is described by a single wave. This statement is equivalent to saying that the Maxwell's equations in electromagnetism are only applicable to lasers (coherent states of photons) or that the Newton's equations are only applicable to rigid bodies when the motion of all particles can be described by a single center of mass degree of freedom. In reality the Gross-Pitaevskii equations are simply classical wave equations of motion similar to the Newton's equations, Maxwell's equations, or Bloch equations. This point perhaps becomes clearer as we go along and see that the Gross-Pitaevskii dynamics can be expressed through the standard Hamiltonian formalism~\cite{LLI} using the language of coherent state Poisson brackets introduced earlier in Sec.~\ref{sec:Moyal}.

\subsection{Coordinate-momentum representation}
\label{sec:coord-moment}

\subsubsection{Semiclassical dynamics of a single particle in an external potential. Truncated Wigner Approximation.}

 Let us start the discussion from considering a single particle in one dimension moving in an external, generally time dependent, potential $V(x,t)$. In quantum mechanics the time evolution of a wave function (density matrix) is described by the Schr\"odinger (von Neumann) equation:
\be
i\hbar {\partial\Psi(x,t)\over \partial t}=\hat{ \mathcal H}(t)\Psi(x,t),
\ee
or
\be
i\hbar {\partial \rho(x,x',t)\over \partial t}=\left[\hat{\mathcal H}(t),\rho(x,x',t)\right],
\ee
where $\Psi(x,t)$ ($\rho(x,x',t)$) stand for the wave function (density matrix) of the particle; square brackets denote the commutator and
\be
\hat{\mathcal H}(t)={\hat p^2\over 2m}+V(x,t)
\label{hamilt}
\ee
is the Hamiltonian. We use the capital $\Psi$ for the wave function to avoid confusion with coherent state complex amplitudes. The corresponding classical Hamiltonian equations of motion are given by the standard formulae:
\be
{dx\over d\tau}=\left\{x,\mathcal H\right\},\;{d p\over d\tau}=\left\{p,\mathcal H\right\},
\label{hamilt_eqs}
\ee
where
\begin{displaymath}
\left\{A,B\right\}=\partial_x A\partial_p B-\partial_p A\partial_x B
\end{displaymath}
is the Poisson bracket. The classical equations of motion should be supplemented by the initial conditions. Together they determine a unique classical trajectory. However, it is important to emphasize that in principle there can be some statistical uncertainty in knowing the initial conditions. This is especially true in many-particle systems. In this case one needs to average the observable of interest over these initial conditions. For example, the average of some classical observable $\Omega_{\rm cl}(x,p,t)$ at some moment of time $t$ is given by
\be
\langle \Omega_{\rm cl}(x,p,t)\rangle =\int dx_0 dp_0 W_{\rm cl}(x_0,p_0)\Omega_{\rm cl} (x(t),p(t),t),
\label{cl_averaging}
\ee
where $W_{\rm cl}(x_0,p_0)$ is the probability distribution of the initial coordinate and momentum. In the fully deterministic case $x_0=\tilde x$ and $p_0=\tilde p$ we have $W_{\rm cl}(x_0,p_0)=\delta(x_0-\tilde x)\delta(p_0-\tilde p)$ and we are back to the unique value of $\Omega$ with no averaging needed.

In quantum mechanics the coordinate and the momentum of a particle can not be simultaneously defined due to the uncertainty principle. Therefore it is not possible to fully localize the particle in the phase space and thus the distribution function should always have nonzero width. A natural object which substitutes the classical probability distribution $\Omega_{\rm cl}(x,p)$ in the quantum case is the Wigner function or equivalently the Wigner transform of the initial density matrix introduced in Sec.~\ref{sec:Moyal} (see Eq.~(\ref{wig_xp})). In the purely classical limit $\hbar\to 0$ the Wigner function reduces to the classical probability distribution. In particular, for the equilibrium density matrix at finite temperature the classical limit of the Wigner function is the Boltzmann's distribution. Similarly the object which replaces the classical observable is the Weyl symbol of the quantum operator (see Eq.~(\ref{w_weyl})).

Interestingly in the leading order in quantum fluctuations Eq.~(\ref{cl_averaging}), which expresses the expectation value of any operator as an average of the Weyl symbol weighted with the Wigner function, immediately generalizes to the time dependent problems. Namely, up to the second order in $\hbar$ quantum fluctuations do not affect the classical equations of motion, which now play the role of the characteristics along which the Wigner function is conserved (see also Sec.~\ref{sec:liouv_wig}). This is equivalent to the classical picture, where the probability distribution $W_{\rm cl}(x(t), p(t),t)$ is conserved along the classical trajectories. Thus in the leading order in quantum fluctuations Eq.~(\ref{cl_averaging}) still remains valid if we replace $W_{\rm cl}(x_0,p_0)\to W(x_0, p_0)$ and $\Omega_{\rm cl}(x(t),p(t),t)\to \Omega_{W}(x(t),p(t),t)$:
\be
\langle \hat{\Omega}(\hat{x},\hat{p},t)\rangle\approx \int\int d x_0 d p_0 W_0(x_0,p_0)\Omega_{W}(x_{cl}(t),p_{cl}(t),t),
\label{eq:omega2}
\ee
Comparison of Eqs.~(\ref{cl_averaging}) and (\ref{eq:omega2}) highlights a close connection between the quantum and the classical statistical averaging. Since, as we noted earlier, the Wigner function is generally non-positive its direct interpretation as a probability distribution becomes somewhat problematic. For equilibrium initial ensembles at high temperatures the Wigner function coincides with the Boltzmann's function and thus becomes positive. However, as the temperature is lowered the difference between classical and quantum distributions becomes stronger: the classical distribution gets narrower both in terms of coordinates and momenta while the quantum Wigner distribution always satisfies the minimum uncertainty principle. The semiclassical approximation (\ref{eq:omega2}) to the dynamics is very important because it gives the first step in going from classical to quantum description of the dynamics. A similar approximation in the coherent state phase space (see below) was termed as truncated Wigner approximation (TWA)~\cite{walls-milburn} for the reasons which will be clear in Sec.~\ref{sec:liouv_wig}. We will stick to this terminology also in the coordinate-momentum representation, thought it is not commonly used there, because this is essentially the same approximation.

\subsubsection{Nonequal time correlation functions.}
\label{sec:neq_time}

The semiclassical (TWA) approximation can be extended to finding non-equal time correlation functions like
\be
\langle \hat\Omega_1(\hat x,\hat p,t_1)\hat \Omega_2(\hat x,\hat p, t_2)\rangle.
\ee
In the classical limit such correlations do not depend on the ordering of the operators. However, quantum mechanically the ordering is important. Within the phase space approach (see also Ref.~\cite{plimak_new}) this non-commutativity can be elegantly absorbed into the language of quantum jumps, which naturally emerge extending the notion of Bopp operators~(\ref{bopp_rp}) and (\ref{bopp_rp1}) to the Heisenberg representation:
\beq
\hat{x}(t)\to x(t)+{i\hbar\over 2}{{\partial}\over {\partial} p(t)}=x(t)-{i\hbar\over 2}{\overleftarrow{\partial}\over{\partial} p(t)},
\label{omega_r}\\
\hat{p}(t)\to p(t)-{i\hbar\over 2}{\partial\over \partial x(t)}=p(t)+{i\hbar\over 2}{\overleftarrow\partial\over \partial x(t)}.
\label{omega_p}
\eeq
Now the derivatives above are understood as quantum jumps, i.e. if we are interested in measuring e.g. $\langle \hat {x}(t_1)\hat{x}(t_2)\rangle$ with $t_1<t_2$ then at the moment $t_1$ the conjugate momentum undergoes an infinitesimal jump: $p(t_1)\to p(t_1)+\delta p(t_1)$. After that the system continues to evolve deterministically (within TWA) and at the moment $t_2$ in addition to the expected term $x(t_1) x(t_2)$ there is an extra contribution equal to the response of $x(t_2)$ to this jump $\delta p(t_1)$ multiplied by $i\hbar/2$. I.e.
\be
\langle \hat {x}(t_1)\hat{x}(t_2)\rangle\approx \int\int d x_0 d p_0 W_0(x_0,p_0) x(t_1)x (t_2)+{i\hbar \over 2} \int\int d x_0 d p_0 W_0(x_0,p_0) {\delta x (t_2)\over \delta p(t_1)},
\ee
where in the last term the limit $\delta p(t_1)\to 0$ is implied. The expression above is approximate only to the extent of TWA being approximate. If $t_1>t_2$ then one can still use the same procedure introducing the jump at $t_1$. But then in order to evaluate the response of $x(t_2)$ it becomes necessary to propagate the equations of motion backwards in time. This is inconvenient and unphysical, since it violates the casuality of the description of the dynamics. Instead one can use the left derivative rule, introduce the jump in $p(t_2)$ and evaluate the response of $x(t_1)$ with respect to this jump multiplied by $-i\hbar/2$:
 \be
\langle \hat {x}(t_1)\hat{x}(t_2)\rangle\approx \int\int d x_0 d p_0 W_0(x_0,p_0) x(t_1)x (t_2)-{i\hbar \over 2} \int\int d x_0 d p_0 W_0(x_0,p_0) {\delta x (t_1)\over \delta p(t_2)}.
\ee
 In this way we clearly preserve the casuality of the evolution. It is interesting to note the higher order correlation functions including more than two different times can be computed in a casual way only for special orderings  of the operators (see also the discussion in Ref.~\cite{plimak1}). We will return to this point in Sec.~(\ref{sec:der}) when we discuss the non-equal time correlation functions in detail.

From Eqs.~(\ref{omega_r}) and (\ref{omega_p}) it is obvious that if we are interested in finding the expectation value of the symmetric combination of say $\hat{x}(t_1)$ and $\hat{x}(t_2)$ or $\hat{x}(t_1)$ and $\hat{p}(t_2)$ then the quantum jumps cancel each other and we can use classical substitutions $\hat{x}\to x$ and $\hat p\to p$ to find such averages. For example,
\be
{\hat{x}(t_1)\hat{x}(t_2)+\hat{x}(t_2)\hat{x}(t_1)\over 2}\to x(t_1)x(t_2).
\ee
This is consistent with general expectations that symmetric correlation functions are classical~\cite{aash_rmp} which are measured by ideal classical detectors~\cite{nazarov}. In general for multi-time averages purely classical substitutions occur if we are interested in time-symmetric ordered correlation functions (see Sec.~\ref{sec:der} and Ref.~\cite{plimak_new}).  Conversely the antisymmetric correlation function has only the contribution coming from quantum jumps:
\be
 \hat{x}(t_1)\hat{x}(t_2)-\hat{x}(t_2)\hat{x}(t_1)\to i\hbar {\partial\over \partial p(t_1)}\; x(t_2).
\ee
Clearly in this case extending Bopp operators to Heisenberg representation according to Eqs. (\ref{omega_r}) and (\ref{omega_p}) also gives a natural extension of the commutation relations to non-equal time operators through the response to quantum jumps. In fact this statement is exact and is not limited to the TWA.

\subsubsection{Beyond the truncated Wigner approximation: nonlinear response and stochastic quantum jumps.}

The truncated Wigner approximation is formally exact if we are dealing with harmonic systems. For a single particle this means that the potential $V(x)$ should be quadratic. If this is not the case then TWA is only approximate. There are two equivalent ways of representing quantum corrections to TWA both using the language of quantum jumps. In the first way the corrections are represented as the nonlinear response (of the observable of interest) to these jumps and in the second way these jumps are stochastically distributed. The second representation is somewhat reminiscent to the Langevin noise~\cite{walls-milburn, gardiner-zoller}, however, with important differences. The representation of quantum dynamics through the noise also appears in quantum optics if one deals with two-body potentials and writes the Liouville equation for the density matrix in the P- or positive P-representation~\cite{gardiner-zoller, walls-milburn,drummond_07}. However, the jumps in the P-representation do not give a systematic expansion in quantum fluctuations. In the formalism discussed here based on the Wigner-Weyl quantization each jump carries an extra factor of $\hbar^2$ and thus the expansion in the number of jumps is equivalent to the expansion in the powers of the Planck's constant. As we show next the jumps have several unusual properties and only superficially resemble standard noise.

For simplicity we will consider again a single-time expectation value of some observable $\hat\Omega$. This simplification is not really important. In a more general case one can combine Eqs.~(\ref{omega_r}) and (\ref{omega_p}) with the results presented below to compute quantum corrections to multi-time correlation functions. Rather than giving a complete general formula we will show the first few terms in the quantum expansion:
\beq
&&\langle \hat{\Omega}(\hat{x},\hat{p},t)\rangle= \int\int d x_0 d p_0 W_0(p_0,x_0)\nonumber\\
&&\biggl(
1-\int_0^t d\tau {1\over 3!\,2^2}{\hbar^2\over i^2}{\partial^3 V(x)\over \partial x(\tau)^3}{\partial^3\over\partial p(\tau)^3} -\int_0^t d\tau {1\over 5!\, 2^4}{\hbar^4\over i^4}{\partial^5
V(x)\over \partial x(\tau)^5} {\partial^5\over\partial p(\tau)^5}\nonumber\\
&&+\int_0^t d\tau_1\int_{\tau_1}^t d\tau_2 {\hbar^4\over (3!\,2^2)^2 i^4}{\partial^3 V(x)\over \partial x(\tau_1)^3} {\partial^3\over\partial p(\tau_1)^3}\,{\partial^3 V(x)\over \partial x(\tau_2)^3} {\partial^3\over\partial p(\tau_2)^3}+\dots \biggr) \Omega_{W}(x(t),p(t),t).
\label{corr2}
\eeq
The expansion above is clearly in powers of $\hbar^2$. In the leading (zeroth) order one simply recovers TWA. In the next order in $\hbar^2$ one needs to introduce a quantum jump in the momentum: $p(\tau)\to p(\tau)+\delta p(\tau)$, calculate the third order nonlinear response of the observable to this jump, multiply this response by the third derivative of the potential with respect to $x(\tau)$ as well as by other factors and integrate over $\tau$. In the next order of $\hbar^4$ one can either have two jumps with the third order response or a single jump of the fifth order response. In general each jump of $2n+1$-th order carries a factor of $\hbar^{2n}$. All the factors in Eq.~(\ref{corr2}) are written in the way easily generalizable to the higher order terms. In Sec.~\ref{sec:applications} we will show some examples illustrating how finding these quantum corrections can be implemented in practice. From Eq.~(\ref{corr2}) it is evident that TWA is asymptotically exact at short times. The smaller $\hbar$ the longer the regime of validity of TWA. At long times quantum corrections may or may not become important depending on the details of the problem and the observable $\hat \Omega$. Moreover the moment where TWA breaks down can have very different sensitivity to the actual value of $\hbar$. We will see this considering various examples in Sec.~\ref{sec:applications}. Thus it is very hard to make a general statement of the time of validity of TWA based on Eq.~(\ref{corr2}) without knowing details about the system.

There is another formal representation of Eq.~(\ref{corr2}) through the stochastic quantum jumps:
\beq
&&\langle \hat{\Omega}(\hat{x},\hat{p},t)\rangle\approx \int\int d x_0
d p_0 W_0(x_0,p_0)\nonumber\\
&&\Biggl\{1+{\hbar^2\over 2^2}\sum_{\tau_i}V_{3,i} \int d\xi_i F_3(\xi_i)\biggr|_{\delta p_i=\xi_i\sqrt[3]{\Delta\tau}} -{\hbar^4\over 2^4}\sum_{\tau_i}V_{5,i} \int
d\xi_i F_5(\xi_i)\biggr|_{\delta p_i=\xi_i\sqrt[5]{\Delta\tau}}
\label{diff_6} \\
&&+{\hbar^4\over (2^2)^2}\sum_{\tau_i<\tau_j}V_{3,i}V_{3,j}\int d\xi_i F_3(\xi_i)\biggr|_{\delta p_i=\xi_i\sqrt[3]{\Delta\tau}} \int d\xi_j F_3(\xi_j)\biggr|_{\delta p_j=\xi_j\sqrt[3]{\Delta\tau}}\dots\Biggr\}\Omega_{W}(x(t),p(t), t).\nonumber
\eeq
To simplify notations we introduced $V_{2n+1,i}={\partial^{2n+1}
V(x(\tau_i),\tau_i)\over \partial x(\tau_i)^{2n+1}}$. Similarly to the ordinary Langevin diffusion we discretize the time here in steps of the size $\Delta\tau$: $\tau_i=i\Delta \tau$. The notation $\delta p_i=\xi_i\sqrt[3]{\Delta\tau}$ above implies that at the moment $\tau_i$ the momentum undergoes a quantum jump of the magnitude $\xi_i\sqrt[3]{\Delta\tau}$ where the $\xi_i\in (-\infty,\infty)$. The weight function $F_{3}(\xi_i)$ can be interpreted as a (quasi)-probability distribution of this jump. With this interpretation the jumps become stochastic. We see that in the leading order in quantum fluctuations at some moment in time during the evolution the momentum $p(\tau_i)$ undergoes a single stochastic jump. The distribution function of the magnitude of this jump $F_3(\xi)$ must satisfy the requirements that all its moments are finite and the moments up to the second vanish:
\be
\int_{-\infty}^\infty F_3(\xi)d\xi=0, \; \int_{-\infty}^\infty \xi F_3(\xi)d\xi=0,\;\int_{-\infty}^\infty \xi^2 F_3(\xi)d\xi=0,\; \int_{-\infty}^\infty \xi^3 F_3(\xi)d\xi=1.
\label{moments}
\ee
Similarly in the order $\hbar^4$ we encounter either two third order jumps or a single fifth order jump. In the latter case $\xi$ is distributed according to the function $F_5(\xi)$ where first four moments are equal to zero. In general $F_{2n+1}(\xi)$ must satisfy
\be
\int_{-\infty}^\infty \xi^m F_{2n+1}(\xi)d\xi=0,\quad m<2n,\quad\int_{-\infty}^\infty \xi^{2n+1} F_{2n+1}(\xi)d\xi=1.
\ee
The functions $F_{2n+1}(\xi)$ are not unique. One possible choice is
\be
F_{2n+1}(\xi)={1\over 2^{n+1}\sqrt{\pi}}{1\over (2n+1)!}\,H_{2n+1}\left({\xi\over \sqrt{2}}\right)\mathrm e^{-\xi^2/2},
\label{f_xin}
\ee
where $H_{2n+1}$ is the Hermite polynomial. In particular
\be
F_3(\xi)={1\over 2}\left({\xi^3\over 3}-\xi\right){1\over \sqrt{2\pi}}\mathrm e^{-\xi^2/2}.
\label{f_xi}
\ee

The representation (\ref{diff_6}) bears some similarities with the diffusion but there are important differences too. Thus in ordinary diffusion jumps are always proportional to $\sqrt{\Delta\tau}$, while in Eq.~(\ref{diff_6}) the power of the time interval $\Delta\tau$ is smaller. Likewise in usual diffusion only the first moment of the distribution function should vanish, which allows one to use the Gaussian function for the $F(\xi)$. In Eq.~(\ref{diff_6}) at least the first two moments of $F(\xi)$ vanish so this function can not be positive definite and can not be interpreted as a classical probability of the jump. As in the nonlinear response representation of quantum corrections, each jump explicitly carries a factor of $\hbar^2$ or higher (depending on the order of jump) allowing for a consistent expansion of time evolution in quantum fluctuations. We note that while the representation (\ref{diff_6}) is formally equivalent to the one given by Eq.~(\ref{corr2}) in the limit $\Delta\tau\to 0$, practical convergence of these two expansions can be quite different. In particular, as we will show using several examples, the step $\Delta\tau$ can be chosen to be much bigger than the one determined from the requirement that the two trajectories with and without jump remain close to each other throughout the full evolution. This is also true about usual diffusion where the distribution function is much more robust to the choice of the time step than individual trajectories. Let us mention that for the coherent state phase space the representation of the evolution through the cubic noise was introduced earlier in Refs.~\cite{gevorkyan1, gevorkyan2, plimak1, plimak3} using Ito's stochastic integration~\cite{gardiner-zoller}.

\subsubsection{Generalization to many particles and higher dimensions.}

The formalism above can be straightforwardly generalized to interacting many-particle systems in arbitrary dimensions. Coordinates and momenta become vectors and acquire an additional particle number index. For example Eq.~(\ref{corr2}) generalizes to
\beq
&&\langle \hat{\Omega}(\hat{\bm x},\hat{\bm p},t)\rangle\approx
\int\int d \bm x_0 d \bm p_0 W_0(\bm p_0,\bm x_0)\nonumber\\
&&\left( 1-\int_0^t
d\tau {1\over 3!\,2^2}{\hbar^2\over i^2}{\partial^3 V(\bm x)\over
\partial x_\alpha\partial x_\beta\partial x_\gamma}
{\partial^3\over\partial p_\alpha(\tau)\partial p_\beta(\tau) \partial p_\gamma(\tau)}+\dots\right) \Omega_{W}(\bm
x(t),\bm p(t),t).
\label{corr3}
\eeq
Here indices $\alpha,\beta,\gamma$ run over different components of the D-dimensional phase space, e.g. for a system of $N$ particles in three dimensions $\alpha,\beta,\gamma=x_1,y_1,z_1,\dots x_N, y_N, z_N$. As usually we imply summation over repeated indices. This expression can be again rewritten through the quantum diffusion. Thus instead of Eq.~(\ref{diff_6}) we get
\beq
&&\langle \hat{\Omega}(\hat{\bm x},\hat{\bm p},t)\rangle\approx \int\int d
\bm x_0 d \bm p_0 W_0(\bm x_0,\bm p_0)\nonumber\\
&&\biggl[1+{\hbar^2\over 4}\sum_j \int\int \prod_{m} d\xi_m \sum_{\sigma(\alpha,\beta,\gamma)}{\partial^3 V(\bm x_j)\over \partial x_\alpha \partial x_\beta\partial x_\gamma} F_{\alpha,\beta,\gamma}(\bm \xi)\biggr|_{ \delta p_\alpha(\tau_j)=\xi_\alpha\sqrt[3]{\Delta\tau_j}}\biggr] \Omega_{cl}(\bm x(t),\bm p(t), t),
\label{diff_7}
\eeq
where $\sigma(\alpha,\beta,\gamma)$ stands for all non-equivalent permutations of indices $\alpha,\beta,\gamma$, e.g. $\alpha\leq \beta\leq \gamma$. The functions $F_{\alpha,\beta,\gamma}(\bm\xi)$ should satisfy the conditions that all its moments smaller than two vanish and
\be
\int\int \prod_{m} d\xi_m \xi_\alpha\xi_\beta\xi_\gamma F_{\alpha,\beta,\gamma}(\bm\xi)=1.
\ee
A possible choice for the functions $F_{\alpha,\beta,\gamma}$ is
\beq
&&F_{\alpha,\alpha,\alpha}(\bm\xi)={1\over 2 \sqrt{2\pi}}\left({\xi_\alpha^3\over 3}-\xi_\alpha\right)\mathrm e^{-\xi_\alpha^2/2}\prod_{m\neq\alpha}\delta(\xi_m),\label{f111}\\
&&F_{\alpha,\alpha,\beta}(\bm\xi)={\left(\xi_\alpha^2-1\right)\xi_\beta\over 4\pi}\, \mathrm e^{-(\xi_\alpha^2+\xi_\beta^2)/2}\!\prod_{m\neq\alpha,\beta}\!\delta(\xi_m) ,\\
&&F_{\alpha,\beta,\gamma}(\bm\xi)={\xi_\alpha\xi_\beta\xi_\gamma\over (2\pi)^{3/2}}\,
\mathrm e^{-(\xi_\alpha^2+\xi_\beta^2 +\xi_\gamma^2)/2}\!\! \prod_{m\neq\alpha,\beta,\gamma}\!\!\delta(\xi_m).\phantom{XX}
\label{f123}
\eeq
In Eqs.~(\ref{f111})-(\ref{f123}) $\alpha,\beta$, and $\gamma$ stand for non-equal indices and no summation over identical indices is implied.

\subsection{Coherent state representation.}
\label{sec:main_coherent}

As we pointed in the previous section for bosonic systems, or more generally the systems which support collective bosonic excitations, it can be more convenient to expand around the wave classical limit and to use the coherent state description of the phase space. Examples of classical waves include phonons, photons, Bogoliubov excitations in superfluids, matter waves etc. In this section we are going to address the issue of representing quantum dynamics in this phase space.

Rather than keeping a completely general discussion we will concentrate on bosonic systems with two-body interactions. In particular, such systems appear in the context of cold atoms. The analysis itself, however, does not rely on this assumption and can be applied to arbitrary n-body interactions. As we will see below the structure of the quantum dynamics in the coherent state phase space is identical to that in the coordinate momentum phase space. Let us point that the emerging semiclassical approximation or TWA was first adopted to interacting bosons in Ref.~\cite{steel}. Since then there were many works applying the TWA to particular cold atom experiments. For the review of some of these applications we refer to Ref.~\cite{blakie_08}. The corrections to TWA in the form of nonlinear response were first obtained by the author in Ref.~\cite{ap_twa}.

Let us consider the following Hamiltonian:
\be
\hat{\mathcal H}=\sum_j h^0_{ij}\hat\psi_i^\dagger \hat\psi_j+{1\over 2}\sum_{ijkl}u_{ijkl}\hat\psi_i^\dagger\hat\psi_j^\dagger\hat\psi_k\hat\psi_l,
\label{hamilt_bos}
\ee
where $\psi_i^\dagger$ and $\psi$ are the bosonic creation and annihilation operators. The indices $i,j,k,l$ can refer to any single-particle basis, e.g. they can describe coordinate or momentum states or the single-particle states in an external potential. The Hamiltonian is split into the noninteracting part with the matrix elements $h^0_{ij}$ and the interacting part characterized by the matrix elements of the interaction potential $u_{ijkl}$. Both $h^0$ and $u$ are allowed to depend on time. The operators $\hat \psi^\dagger$ and $\hat \psi$ satisfy the canonical commutation relations:
\be
[\hat\psi_i,\hat\psi_j^\dagger]=\delta_{ij}.
\label{commut_rel}
\ee
 The Hamiltonian~(\ref{hamilt_bos}) is quite general. For example, if indices $i,j$ denote spatial positions on the lattice by the appropriate choice of the matrix elements of $h^0$ and $u$ it reduces to the Hubbard model:
\be
\hat{\mathcal H}_{hub}=-J\sum_{\langle i,j\rangle} (\hat\psi_i^\dagger\hat\psi_j+h.c.)+\sum_j V_j\hat n_j+{1\over 2}\sum_{ij} U_{ij} :\hat n_i\hat n_j:,
\label{hamilt_bh}
\ee
where $\hat n_i=\hat\psi_i^\dagger\hat\psi_i$ is the density operator, $V_j$ is the external potential, $J$ is the tunneling matrix elements between nearest neighbors, and $U_{ij}$ is the matrix element of the density-density interaction. We use capital letters to distinguish parameters of the Hubbard model from those in the general Hamiltonian. The semicolons in the Hubbard Hamiltonian imply the normal-ordered form of the interaction, i.e. the form where the creation operators appear on the left of the annihilation operators. We note that in the continuum limit, when the lattice spacing goes to zero, the Hubbard model reduces to that of interacting bosons without the lattice. The indices $i,j,\dots$ in the Hamiltonian~(\ref{hamilt_bos}) can also refer to momentum states. Then in the absence of external potential $h^0_{k,q}=\epsilon_k\delta_{pq}$, where $\epsilon_k$ is the energy of a single particle with momentum $k$ and $v_{k+q,p-q,p,k}$ are the Fourier components of the interaction potential~\cite{pethick-smith}.

As we discussed in Sec.~\ref{sec:Moyal} the classical wave limit in the coherent state picture corresponds to large occupation numbers of different modes. In this case we can treat the operators $\hat\psi^\dagger$ and $\hat\psi$ as complex $c$-numbers. In the classical limit one substitutes the commutation relations~(\ref{commut_rel}) with the coherent state Poisson brackets~(\ref{coh_pois_brack}): $\{ \psi_i,\psi_j^\star\}_c=\delta_{ij}$.
Under this substitution the Heisenberg equations of motion for the fields $\hat\psi_j$, $\hat\psi_j^\dagger$ become the Gross-Pitaevskii equations:
\be
i\hbar{\partial \psi_i\over\partial t}=\left\{\psi_i,\mathcal H_W\right\}_c={\partial \mathcal H_W\over\partial \psi_i^\star}=\sum_j \tilde h^0_{ij}\psi_j+\sum_{j,k,q}u_{ijkq}\psi_j^\star \psi_k\psi_q,
\ee
where
\be
\tilde h^0_{ij}=h^0_{ij}+{1\over 4}\sum_k u_{ikkj}+u_{ikjk}+u_{kijk}+u_{kikj}.
\label{tilde_h}
\ee
The difference between $\tilde h^0$ and $h^0$ comes from the Weyl ordering of the Hamiltonian.

For example for the Hubbard model (see Eq.~(\ref{hamilt_bh})) the equation above reduces to the familiar discrete Gross-Pitaevskii equation~\cite{tromberttoni01, psg}
\be
i\hbar{\partial\psi_j\over\partial t}=-J\sum_{l\in O_j} \psi_l +\tilde V_j\psi_j +\sum_k U_{jk} |\psi_k|^2\psi_j,
\label{gp_eq}
\ee
where
\be
\tilde V_{ij}=V_{ij}+{1\over 2}\delta_{ij}\left(U_{ij}+\sum_k U_{ik}\right).
\ee
Here $O_j$ denotes the nearest neighbors of the site $j$. We see that for the Hubbard model with translationally invariant interactions ($U_{ij}=U_{i+k,j+k}$ for arbitrary $k$) the Weyl ordering only gives a term proportional to the total number of particles, which can be absorbed into the overall phase or the chemical potential. However, as we will illustrate in Appendix~\ref{appendix_hamilt_ordering}, the extra terms can be very important in other situations with e.g. spatially dependent interactions. Let us note that the Planck's constant in Eq.~(\ref{gp_eq}) can be safely set to unity by either redefining energy or time units. This is in fact customary in cold atom systems where relevant energy scales are often measured in Hz~(see e.g. Ref.~\cite{bloch_review}). The real parameter, which determines the degree of quantum fluctuations is the occupancy of relevant bosonic modes: classical waves correspond to high occupation numbers.

\subsubsection{Truncated Wigner Approximation}

Similarly to the coordinate-momentum picture in the leading semiclassical order (or TWA) quantum corrections appear only through the initial conditions and the form of the observable
\be
\langle \hat{\Omega}(\hat{\bm\psi},\hat{\bm\psi}^\dagger,t)\rangle=\int\int d \bm{\psi}_0 d \bm{\psi}_0^\star\, W_0(\bm{\psi}_0,\bm{\psi}_0^\star)\Omega_{W}(\bm{\psi}(t),\bm{\psi}^\star(t),t),
\label{eq:omega5}
\ee
The integral above is taken over different realizations of all components of initial values of $\bm{\psi}$ and $\bm{\psi}^\star$ so that $d\bm{\psi}_0 d\bm{\psi}_0^\star$ stands for $\prod_j d\psi_{0\,j} d\psi_{0\,j}^\star$. The fields $\bm{\psi}(t)$ and $\bm{\psi}^\star(t)$ are related to $\bm{\psi}_0$ and $\bm{\psi}_0^\star$ via the solution of Eqs.~(\ref{gp_eq}). The  $W_0(\bm{\psi}_0, \bm{\psi}_0^\star)$ is the Wigner transform of the initial density matrix~(\ref{wig_coherent}), and $\Omega_{W}(\bm{\psi}(t),\bm{\psi}^\star(t),t)$ is the Weyl symbol of the operator $\hat{\Omega}$~(\ref{weyl_coherent_def}).

For computing non-equal time correlation functions we can use the recipe similar to the one discussed above in Sec.~\ref{sec:neq_time}. In particular, we can extend the coherent state Bopp operators~(\ref{bopp_psi_dag}) and (\ref{bopp_psi}), to the Heisenberg representation:
\beq
&&\hat{\psi}_j^\dagger(t)\to\psi_j^\star(t)-{1\over 2}{{\partial}\over\partial\psi_j(t)}=\psi_j^\star(t)+{1\over 2}{\overleftarrow{\partial}\over\partial\psi_j(t)},
\label{eq:pa_dag}\\
&&\hat{\psi_j}(t)\to\psi_j(t)+{1\over 2}{{\partial}\over\partial\psi_j^\star(t)},=\psi_j(t)-{1\over 2}{\overleftarrow{\partial}\over\partial\psi_j^\star(t)}\phantom{XX}
\label{eq:pa}
\eeq
interpreting the partial derivatives as response to infinitesimal quantum jumps. Namely, the derivative with respect to $\psi_j(t)$ in the equation for $\hat\psi_j^\dagger (t)$ implies that at moment $t$ the field $\psi_j(t)$ undergoes an infinitesimal jump $\psi_j(t)\to \psi_j(t)+\delta\psi_j(t)$. Then one needs to evaluate the derivative of all operators appearing on the right of $\hat\psi_j^\dagger(t)$ with respect to this jump. The derivatives above can also rewritten in terms of real and imaginary parts of $\psi$ according to the standard rules:
\be
{\partial \over\partial \psi}={1\over 2}{\partial\over \partial \Re
\psi}-{i\over 2}{\partial\over \partial \Im \psi},\;
{\partial\over\partial \psi^\star}={1\over 2}{\partial\over
\partial \Re \psi}+{i\over 2}{\partial\over \partial \Im \psi}.
\label{compl_der}
\ee

\subsubsection{Quantum corrections.}

As in the case of the coordinate-momentum representation, quantum corrections to Eq.~(\ref{eq:omega5}) can be written either in the form of nonlinear response or stochastic quantum jumps. In the first way (cf. Eq.~(\ref{corr3})) we get
\beq
&&\langle \hat{\Omega}(\hat{\psi},\hat{\psi^\dagger},t)\rangle\approx \int\int d \psi_0 d \psi^\star_0 W_0(\psi_0,\psi^\star_0)\nonumber\\
&&\Biggl(1-{i\over 4\hbar}\int_0^t d\tau \sum_{ijkl} u_{ijkl}\biggr[\psi^\star_i(\tau) {\partial^3\over\partial\psi_j(\tau)\partial\psi^\star_k(\tau)\partial\psi^\star_l(\tau)}-c.c.\biggr]\Biggr) \Omega_{W}(\psi(t),\psi^\star(t),t).
\label{corr4}
\eeq
The interpretation of this expression is also similar to what we encountered earlier. In the leading order in quantum fluctuations beyond TWA one has a single quantum jump where the classical field $\psi_j(\tau)$ undergoes the infinitesimal transformation $\psi_j(\tau)\to \psi_j(\tau)+\delta\psi_j(\tau)$. Then this field continues to evolve according to the classical equations of motion. In the end one calculates the nonlinear response of the desired observable to this jump. Further corrections appear as multiple quantum jumps. Note that for the two-body interactions there are no higher order quantum jumps which would correspond to fifth and higher order derivatives with respect to $\psi$ and $\psi^\star$ (cf. Eq.~(\ref{corr3})). As we discussed earlier, near the classical limit fields $\psi$ have large occupation so the derivatives with respect to $\psi$ and $\psi^\star$ give factors of the inverse square root of the occupation number.

One can represent these corrections also using stochastic quantum jumps:
\beq
&&\langle \hat{\Omega}(\hat{\psi},\hat{\psi^\dagger},t)\rangle\approx \int\int d \psi_0 d \psi^\star_0 W_0(\psi_0,\psi^\star_0)\Biggl[1-{i\over 4\hbar}\sum_n\sum_{ijkl} u_{ijkl}\int d\Gamma_{jkl}^\xi\nonumber\\
&&\biggl(\psi_i^\star(\tau_n) F(\xi_j,\xi_k,\xi_l)-\psi_i(\tau_n) F^\star(\xi_j,\xi_k,\xi_l)\biggr)\biggr|_{\delta\psi_{a}(\tau_n)=\xi_{a}\sqrt[3]{\Delta\tau}}
\Omega_{W}(\psi(t),\psi^\star(t),t)\Biggr],\phantom{XX}
\label{corr5}
\eeq
where $a=j,k,l$ and $d\Gamma^{\xi}_{jkl}$ is the phase space volume element with $jkl$ denoting all nonequivalent permutations of indices $i,j,k$. For $j\neq k\neq l$ this phase space volume element is given by $d\Gamma^\xi_{jkl}=d\xi_j d\xi_j^\star d\xi_k d\xi_k^\star d\xi_l d\xi_l^\star$; for $j\neq k=l$ it is
$d\Gamma^\xi_{jkk}=d\xi_j d\xi_j^\star d\xi_k d\xi_k^\star$ and so on. The interpretation of the jump $\delta\psi_a$ in Eq.~(\ref{corr5}) is the same as in the coordinate-momentum representation: at the moment $\tau_n$ the complex field $\psi_a(\tau_n)$ jumps to $\psi_a(\tau_n)+\delta\psi_a(\tau_n)$. The function $F$ can be interpreted as the quasi-probability distribution of the stochastic quantum jump. It should satisfy the requirement that all its moments up to the second vanish and the third moments are equal to:
\be
\int d\Gamma^{\xi}_{jkl}\;\xi_j\xi_k^\star\xi_l^\star F(\xi_j,\xi_k,\xi_l)=(1+\delta_{kl}),
\label{cond_f}
\ee
As in the coordinate-momentum case these requirements do not define $F$ uniquely. A possible choice for this function is
\beq
&&F(\xi_j,\xi_k,\xi_l)=\xi_j^\star\xi_k\xi_l\mathrm e^{-|\xi_j|^2-|\xi_k|^2-|\xi_l|^2},\; j\neq k\neq l,\phantom{XX}\label{f1}\\
&& F(\xi_j,\xi_j,\xi_k)=\xi_k\left(|\xi_j|^2-1\right)\mathrm e^{-|\xi_j|^2-|\xi_k|^2},\; j\neq k\label{f2}\\
&& F(\xi_j,\xi_k,\xi_k)=\xi_j^\star\xi_k^2\mathrm e^{-|\xi_j|^2-|\xi_k|^2},\; j\neq k\label{f3}\\
&& F(\xi_j,\xi_j,\xi_j)=\xi_j^\star\left(|\xi_j|^2-2\right)\mathrm e^{-|\xi_j|^2}.\label{f4}
\eeq
For the Bose-Hubbard model with the Hamiltonian~(\ref{hamilt_bh}) and local interactions $U_{ij}=U\delta_{ij}$ the general formula (\ref{corr5}) reduces to
\beq
&&\langle \hat{\Omega}(\hat{\psi},\hat{\psi^\dagger},t)\rangle\approx \int\int d \psi_0 d \psi^\star_0 W_0(\psi_0,\psi^\star_0)\nonumber\\
&&\Biggl[1-{i U\over 4\hbar}\sum_n\sum_{j}\int\int d\xi_jd\xi_j^\star \biggl(\psi_j^\star(\tau_n) F(\xi_j)-c.c.\biggr)\Omega_{cl}(\psi(t),\psi^\star(t),t)\Biggr],\phantom{XX}
\label{corr_bh}
\eeq
where as in the Eq.~(\ref{corr5}) at the moment $\tau_n$ the classical field $\psi_j$ undergoes a quantum jump $\psi_j\to\psi_j+\xi_j\sqrt[3]{\Delta\tau}$ and $F(\xi_j)$ is given by Eq.~(\ref{f4}).

\subsection{Spin systems.}
\label{sec:spin}

In this section we will discuss expansion of dynamics for interacting spin systems or more generally dynamics of systems with angular momenta. We will work in units where $\hbar=1$. The classical limit corresponds to $S$, size of the spin, being large: $S\gg 1$ so the parameter of the expansion around the classical limit should be $1/S$. The spin operators satisfy the canonical commutation relations:
\be
\left[ \hat{s}_a,\hat{s}_b\right]=i\epsilon_{abc}\hat{s}_c,
\ee
where $\epsilon_{abc}$ is the fully antisymmetric tensor. Let us note that formally spin dynamics can be mapped to the dynamics of bosons using Schwinger representation~\cite{assa}:
\beq
\hat{s}^z={\hat{\alpha}^\dagger\hat{\alpha}-\hat{\beta}^\dagger\hat{\beta}\over 2},\; \hat{s}^{+}=\hat{\alpha}^\dagger\hat{\beta},\; \hat{s}^-=\hat{\beta}^\dagger\hat{\alpha}.
\label{schwing}
\eeq
This representation allows us to apply results from the previous section directly to the spin systems without need to introduce spin-coherent states (see e.g. Ref.~\cite{takahashi}). The bosonic fields $\alpha$ and $\beta$ in Eq.~(\ref{schwing}) should satisfy an additional constraint $\hat{n}=\hat{\alpha}^\dagger\hat{\alpha}+\hat{\beta}^\dagger\hat{\beta}=2S$. In the imaginary time formalism of the equilibrium statistical mechanics this constraint poses an important problem and should be taken care of introducing an auxiliary constraint field at each moment of time (see e.g. Ref.~\cite{assa_arovas}). However, in real-time dynamics the situation is much less complicated because this constraint is preserved in time. Indeed $\hat{n}$ commutes with all spin operators and thus with any Hamiltonian with arbitrary spin-spin interactions. Therefore if the constraint is satisfied at the initial time then the dynamics of spins is equivalent to the unconstrained dynamics of bosons under the mapping~(\ref{schwing}). Thus semiclassical approximation (TWA), quantum-classical correspondence, and quantum corrections for spins can be directly deduced from the results of the previous section.

{\em Quantum classical correspondence} Using Eqs.~(\ref{eq:pa_dag}) and (\ref{eq:pa}) we can find an analogue of the Bopp operators for the spin systems:
\beq
&&\hat{s}_z\to {\alpha^\star\alpha-\beta^\star\beta\over 2}-{1\over 8}\left({\partial^2\over \partial\alpha^\star\partial\alpha}- {\partial^2\over \partial\beta^\star\partial\beta}\right)-{1\over 4}\left(\alpha^\star{\partial\over\partial \alpha^\star} -\alpha{\partial\over\partial \alpha} -\beta^\star{\partial\over\partial \beta^\star}+\beta{\partial\over\partial \beta}\right),\label{S_z}\\
&&\hat{s}_+\to \alpha^\star\beta+{1\over 2}\left(\alpha^\star{\partial\over\partial\beta^\star}-\beta{\partial\over\partial\alpha}\right)-{1\over 4}{\partial^2\over\partial\alpha\partial\beta^\star},\label{S_+}\\
&&\hat{s}_-\to \alpha\beta^\star+{1\over 2}\left(\alpha{\partial\over\partial\beta}-\beta^\star{\partial\over\partial\alpha^\star}\right)-{1\over 4}{\partial^2\over\partial\alpha^\star\partial\beta}.\label{S_-}
\eeq
These equations can be also written using compact notations:
\be
\hat{\bf s}\to {\bf s}-{i\over 2}\left[{\bf s}\times \vec\nabla\right]-{1\over 8}\left[\vec\nabla+ ({\bf s}\cdot\vec\nabla)\vec\nabla-{1\over 2}{\bf s}\nabla^2\right],
\ee
where $\vec\nabla=\partial/\partial {\bf s}$. For example, for $s_z$ this correspondence gives
\be
\hat{s}_z\to s_z-{i\over 2}\left(s_x{\partial\over\partial s_y}-s_y{\partial\over\partial s_x}\right)-{1\over 8}{\partial\over\partial s_z}-{s_z\over 16}\left({\partial^2\over \partial s_z^2}-{\partial^2\over \partial s_x^2}-{\partial^2\over \partial s_y^2}\right)-{s_x\over 8}{\partial^2\over\partial s_x\partial s_z}-{s_y\over 8}{\partial^2\over\partial s_y\partial s_z}.
\ee

These formulae can be used in constructing Weyl symbols for various spin operators. Let us give a few specific examples:
\be
\hat{s}_z\to s_z,\;\hat{s}_z^2\to s_z^2-{1\over 8},\;\hat{s}_z\hat{s}_x\to s_z s_x+{i\over 2}s_y.
\ee
Note that the presence of $i$ in the latter expression is not surprising since the product $\hat{s}_z\hat{s}_x$ is not a Hermitian operator. Like before with coherent states the spin Bopp operators can be also used to find non-equal time correlation functions if we interpret derivatives as a response to infinitesimal jumps.

{\em Wigner transform.} In principle, the mapping (\ref{schwing}) is sufficient to express the Wigner function of any initial state in terms of the bosonic fields $\alpha$ and $\beta$. General expressions can be quite cumbersome, however, one can use a simple trick to find a Wigner transform of any pure single spin state and then generalize it to any given density matrix. Assume that a spin is pointing along the $z$-axis. This can always be achieved by a proper choice of a coordinate system. Then in terms of bosons $\hat{\alpha}$ and $\hat{\beta}$ the initial state is just $|2S+1,0\rangle$, where $S$ is the size of the spin. In other words the wave function is a product of two Fock states one having $2S+1$ particles and one $0$ particles. The corresponding Wigner transform is then~\cite{ap_twa, gardiner-zoller}:
\be
W(\alpha,\alpha^\star,\beta, \beta^\star)=2\mathrm e^{-2|\alpha|^2-2|\beta|^2} L_{2S+1}(4|\alpha|^2),
\label{wig_spin}
\ee
where $L_N(x)$ is the Laguerre's polynomial of order $N$. At large $S$ the Laguerre polynomial is a rapidly oscillating function and the Wigner transform is strongly localized near $|\alpha|^2=2S+1/2$ (see Ref.~\cite{ap_twa}). So in this case to a very good accuracy (up to $1/S^2$) one can use
\be
W(\alpha,\alpha^\star,\beta, \beta^\star)\approx \sqrt{2}\mathrm e^{-2|\beta|^2} \delta(|\alpha|^2-2S-1/2).
\label{wig_spin_1}
\ee
At the same time the fluctuations in $\beta$ are usually much more important since they represent fundamental uncertainty between $\hat{S_z}$ and the transverse components of the spin. Reexpressing the Wigner transform (\ref{wig_spin}) in terms of spin components then we get
\be
W(s_z,s_\perp)={2\over \pi s}\mathrm e^{-4s}L_{2S+1}(4[s_z+s]),
\label{wig_spin_2}
\ee
where $s_\perp=\sqrt{s_x^2+s_y^2}$ and $s=\sqrt{s_z^2+s_\perp^2}$ represent transverse component and the total magnitude of the classical spin respectively. The Wigner function is normalized in such a way that
\be
\int_{-\infty}^\infty ds_z\int_0^\infty s_\perp ds_\perp\int_0^{2\pi}d\phi W(s_z,s_\perp)=1,
\ee
where $\phi$ is the polar angle of the spin in the $xy$ plane. When $S$ is large we have $s_z\approx s$ strongly peaked around $S$ and then Eq.~(\ref{wig_spin_2}) reduces to
\be
W(s_z,s_\perp)\approx {1\over \pi S}\mathrm e^{-s_\perp^2/S}\delta(s_z-S).
\label{wig_spin_approx}
\ee
This Wigner function has a transparent interpretation. If the quantum spin points along the $z$ direction, because of the uncertainty principle, the transverse spin components still fluctuate due to zero point motion so that
\be
\langle s_x^2\rangle=\langle s_y^2\rangle={S\over 2}.
\ee
This is indeed the correct quantum-mechanical result. Clearly from Eq.~(\ref{wig_spin_2}) one can derive the Wigner function for a spin with an arbitrary orientation by the appropriate rotation of the coordinate axes.

For more complicated correlated spin systems the trick with aligning all spins along a particular axis does not work. In this case one can always rely on the Schwinger representation to get the Wigner function and then reexpress this function through the spin components. Alternatively one can directly work with the spin coherent states~\cite{takahashi} and compute the Wigner transform without referring to the bosonic representation of spins. An example of such computation for two spin one half particles can be found in Ref.~\cite{scully_86}.

{\em Quantum corrections.} Let us consider only the situation with two spin interactions of the form:
\be
\hat{\mathcal H}_{\int}=\sum_{i,j} J^{a,b}_{i,j} \hat{s}_{a,i} \hat{s}_{b,j},
\ee
where $a,b=x,y,z$. The corrections to the TWA can be again characterized by quantum jumps. These jumps can be recast in the form of nonlinear response so that within the first order in $1/S$ we have the following expression for the expectation value of an arbitrary observable:
\be
\langle\hat\Omega(t)\rangle\!=\!\int\! d\bm s_0 W(\bm s_0)\Biggl[1+{1\over 8}\int\limits_0^t\! d\tau \sum_{i,j} J^{a,b}_{i,j}\left[{\bf s}_i(\tau)\times \bm\nabla_i\right]_a\left(2\left({\bf s}_j(\tau)\cdot\bm\nabla_j\right)\nabla_{b,j}-
s_{b,j}(\tau)\bm\nabla_j^2\right)\Biggr]\Omega_{W}(\bm s(t),t),
\ee
where $\bm\nabla_{j}=\partial/\partial \bm s_j(\tau)$ and $\nabla_{b,j}=\partial/\partial s_{b,j}(\tau)$. This expression should be understood in the same way as e.g. Eq.~(\ref{corr4}) for the coherent state evolution: at the moment $\tau$ the spins undergo infinitesimal jumps ${\bf s}_j(\tau)\to {\bf s}_j(\tau)+\delta {\bf s}_j$ and at the end one evaluates the nonlinear response of the observable to these jumps. From the equation above it is clear that $1/S$ serves as the expansion parameter determining the strength of quantum fluctuations. Actually the small parameter is $1/S^2$ since the classical limit is achieved when $S\to\infty$, $J\to 0$ such that $JS=$const.

\section{Applications}
\label{sec:applications}
Let us now discuss applications of the formalism described in the previous section to a number of sample problems starting from a single-particle moving in a harmonic potential and then gradually increasing complexity towards interacting many-particle systems. For implementing TWA and quantum corrections one essentially needs two ingredients: calculating the initial Wigner function and solving corresponding classical equations of motion. For complicated initial states finding the Wigner function can be a very challenging task by itself. Since in this paper we are mostly concerned with understanding various aspects of the dynamics we will put the discussion of finding initial Wigner function aside by starting from simple initial states, which correspond to equilibrium in systems with a quadratic Hamiltonian. In many situations complications associated with finding the initial Wigner function can be avoided by either doing various approximations to the initial state like the Bogoliubov's approximation~\cite{blakie_08} or by starting from a simple initial state and then adiabatically evolving the Hamiltonian in time so that the couplings reach the desired values~\cite{pw}. This method is closely related to experimental procedures used in cold atoms, where one usually cools the system first and then adiabatically drives the system towards the desired regime e.g. by turning on optical lattice~\cite{bloch_review}. This method is also closely related to quantum annealing algorithms~\cite{nishimori, santoro, das,farhi}. If the evolution is sufficiently slow then one moves along the constant entropy curve. As long as one remains in the regime of small quantum fluctuations this method of adiabatically evolving the Hamiltonian is very reliable~\cite{pw, rafi}. Though we would like to note that in some low dimensional systems the adiabatic limit can be achieved at anomalously slow rates~\cite{pg_np}. After the desired initial parameters of the Hamiltonian are reached via this procedure one can start real dynamical simulations.

\subsection{Single particle in a harmonic potential.}

 As the first example let us consider a single particle moving in a harmonic potential. The advantage of starting from this simple situation is that all calculations can be done explicitly analytically. The Hamiltonian of a single harmonic oscillator is
\be
\hat{\mathcal H}_0={\hat {p}^2\over 2m}+{m\omega^2\over 2 }\hat{x}^2=\hbar\omega(\hat{\psi}^\dagger \hat{\psi}+1/2),
\ee
where the coordinate and momentum operators $\hat{x}$ and $\hat{p}$ are related to creation and annihilation operators $\hat{\psi}$ and $\hat{\psi}^\dagger$ in a standard way:
\be
\hat \psi=\sqrt{m\omega\over 2\hbar}\left(\hat x+{i\over m\omega}\hat p\right),\quad
\hat \psi^\dagger =\sqrt{m\omega\over 2\hbar}\left(\hat x-{i\over m\omega}\hat p\right).
\ee

Now suppose that the particle is prepared in the ground state and we are suddenly applying a linear potential $V(x)=-\lambda x$. So that the Hamiltonian becomes
\be
\hat{\mathcal H}=\hat{\mathcal H}_0-\lambda \hat x
\ee
Next we compute various observables as a function of time.

{\em Coordinate-momentum representation.} First we will solve this problem using the coordinate-momentum representation. The ground state wave function for a harmonic oscillator is:
\be
\psi_0(x)={1\over (2\pi)^{1/4}\sqrt{a_0}}\,\mathrm e^{-x^2/4a_0^2},
\ee
where $a_0=\sqrt{\hbar/2m\omega}$ is the oscillator length~\cite{LL3}. According to Eq.~(\ref{wig_xp}) the Wigner transform corresponding to this wave function is
\be
W(x_0,p_0)=\int d\xi \psi^\star(x_0+\xi/2)\psi(x_0-\xi/2)\mathrm e^{ip_0\xi/\hbar}=2\exp\left[-{x_0^2\over 2 a_0^2}-{p_0^2\over 2 q_0^2}\right],
\label{wig_osc}
\ee
where $q_0=\hbar/2a_0=\sqrt{m\omega\hbar/2}$. In this case $W(x_0,p_0)$ is positive definite and can be straightforwardly interpreted as the probability distribution of the initial coordinate and momentum. Next we need to solve the classical equations of motion:
\be
{dp\over dt}=-m\omega^2 x+\lambda,\quad {dx\over dt}={p\over m}
\ee
satisfying the initial conditions $x(0)=x_0$, $p(0)=p_0$. Clearly the solution is
\be
x(t)=x_{\rm cl}(t)+x_0 \cos(\omega t)+{p_0\over m\omega}\sin(\omega t),
\ee
where $x_{\rm cl}(t)={\lambda/m\omega^2}(1-\cos(\omega(t)))$ is a classical result describing the motion of the particle which is initially set to rest. Then we need to substitute this solution to the observable corresponding to the quantum operator of interest and find the average over the initial conditions.

For the expectation value of the position we trivially find $\langle \hat{x}(t)\rangle=x_{cl}(t)$, which is just a particular case of the Ehrenfest's principle~\cite{shankar}. Similarly we find
\be
\langle \hat{x}^2\rangle=\overline {x^2(t)}=x_{\rm cl}^2(t)+a_0^2.
\ee
This is of course also the correct result, which can be easily obtained from the solution of the Schr\"odinger equation.

Next let us show how to compute a non-equal time correlation function. In particular, $\langle\hat {x}(t)\hat{x}(t')\rangle$ with $t<t'$. In this case according to Eq.~(\ref{omega_r}) we need to substitute $x(t)\to x(t)+{i\hbar/2}\partial/\partial p_t$ and proceed with the classical calculation. This substitution is equivalent to an infinitesimal quantum jump in the momentum: $p(t)\to p(t)+\delta p$. At the later moment $t'$ we need to evaluate the response of the coordinate to this jump. Then we find
\beq
&&\langle\hat {x}(t)\hat{x}(t')\rangle=\overline{\left(x_{cl}(t)+x_0\cos(\omega t)+{p_0\over m\omega}\sin(\omega t)+{i\hbar\over 2}{\partial\over\partial \delta p}\right)}\nonumber\\
&&\overline{\times\left(x_{cl}(t)+x_0\cos(\omega t')+{p_0\over m\omega}\sin(\omega t')+{\delta p\over m\omega}\sin(\omega(t'-t)\right)}\nonumber\\
&&=x_{\rm cl}(t)x_{\rm cl}(t')+a_0^2\cos(\omega(t-t'))+i a_0^2\sin(\omega(t'-t)).
\eeq
Note that this correlation function is complex because it does not correspond to the expectation value of a Hermitian operator. Similarly for the correlator with the opposite ordering of $t$ and $t'$ we find
\be
\langle\hat {x}(t')\hat{x}(t)\rangle=x_{\rm cl}(t)x_{\rm cl}(t')+a_0^2\cos(\omega(t-t'))-i a_0^2\sin(\omega(t'-t))
\ee
Therefore the symmetric part of the correlator corresponding to the result of a classical measurement~\cite{aash_rmp} is given by
\be
\left<{\hat {x}(t')\hat{x}(t)+\hat {x}(t)\hat{x}(t')\over 2}\right>=x_{\rm cl}(t)x_{\rm cl}(t')+a_0^2\cos(\omega(t-t'))
\ee
and the quantum antisymmetric part is
\be
\langle\hat {x}(t)\hat{x}(t')-\hat {x}(t')\hat{x}(t)\rangle=2i a_0^2\sin(\omega(t'-t)).
\ee
The quantum part vanishes in the limit $\hbar\to 0$ as it should.

{\em Coherent state representation.} Now we will illustrate how the same results can be obtained in the coherent state picture. In the second quantized form the Hamiltonian of the system reads
\be
\mathcal H=\hbar\omega(\hat{\psi}^\dagger \hat{\psi}+1/2)-\lambda a_0 (\hat \psi+\hat \psi^\dagger).
\label{hamilt_osc}
\ee
The classical equation of motion for the complex field $\psi$  can be obtained
from Eqs.~(\ref{coh_pois_brack}) and (\ref{hamilt_osc}):
\be
i\hbar{\partial \psi\over\partial t}=\hbar\omega \psi-\lambda a_0,
\ee
which has the solution
\be
\psi(t)={\lambda a_0\over\hbar\omega}\left(1-\mathrm e^{-i\omega t}\right)+\psi_0\mathrm e^{-i\omega t}.
\ee
Using Eq.~(\ref{wig_coherent}) one can show that the Wigner transform of the initial vacuum state is
\be
W(\psi_0,\psi_0^\star)=2\mathrm e^{-2|\psi_0|^2}.
\ee
Now we can find the desired expectation values
\be
\langle \hat{x}(t)\rangle=a_0\overline{\left( \psi(t)+\psi^\star(t)\right)}={2 a_0^2\lambda \over \hbar\omega}(1-\cos(\omega t))=x_{\rm cl}(t).
\ee
Similarly
\be
\langle \hat{x}^2(t)\rangle =a_0^2\overline{\left(\psi^2(t)+(\psi^\star(t))^2 +2\psi(t)\psi^\star(t)\right)}=x_{\rm cl}^2(t)+a_0^2.
\ee
We obviously got the same answers as before. Similarly one can verify the result for the non-equal time correlation function. Of course it is not surprising that both methods give identical exact results for harmonic systems. However, it is important to realize that once we deal with more complicated interacting models the correct choice of the phase space can significantly simplify the problem. Moreover the expansions around the two possible classical limits are very different. Thus for a system of noninteracting particles moving in some external potential TWA in the coordinate-momentum representation is only approximate unless the potential is harmonic. At the same time TWA in the coherent state representation is exact.

\subsection{Single particle in a Mexican-hat potential.}

\begin{figure}[ht]
\includegraphics[width=12cm]{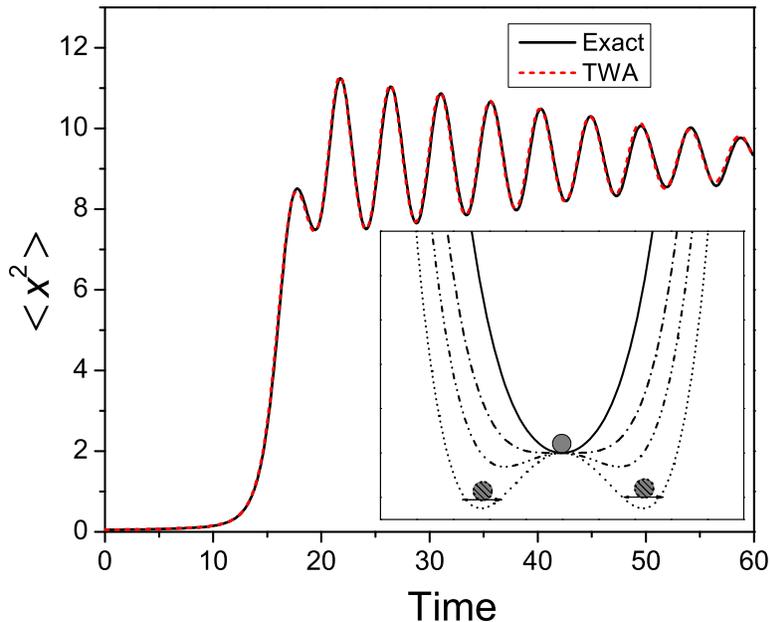}
\caption{Time dependence of the variance of the displacement
of a particle initially prepared in the ground state of
the Hamiltonian~(\ref{hamilt_cat}) with $\kappa=1$ and $\lambda=0$.
Then the potential evolves according to Eq.~(\ref{kappat}) with
$\delta=0.1$. The other parameters are $\lambda=0.1$, $m=10$,
$\omega_0=1$. We work with the units where $\hbar=1$. The solid
black curve represents the exact solution of the Schr\"odinger equation and the dashed red line represents the result of TWA. The inset shows the evolution of the shape of the potential with time.}
\label{fig:cat4}
\end{figure}

Let us now move to a somewhat more complicated problem, where we
will still deal with a single-particle physics but in a quartic
potential so that the Hamiltonian is:
\be
\hat {\mathcal H}(t)={\hat{p}^2\over 2m}+{m\omega^2\over
2}\left(\kappa(t) \hat{x}^2+\lambda {\hat{x}^4\over 2}\right).
\label{hamilt_cat}
\ee
We assume that initially $\kappa(0)=0$ and the particle is in the
ground state. Then $\kappa(t)$ changes in time according to
\be
\kappa(t)=\left\{ \begin{array}{ll} 1-\delta t & 0<t\leq 2/\delta
\\
-1 & t>2/\delta\end{array}\right.
\label{kappat}
\ee
At the moment when $\kappa(t)$ becomes negative classical
equilibrium becomes unstable and the particle should move to one of
the new minima (see the inset in Fig.~\ref{fig:cat4}).
The parameter $\delta$ in Eq.~(\ref{kappat}) controls the rate of the change of the Hamiltonian. Note that classically the particle with the initial conditions $x=0$, $p=0$ will always remain at the
same position. However, zero point fluctuations allow the particle to fall down to one of the
minima and form the ``Schr\"odinger cat'' state. In Fig.~\ref{fig:cat4}
we plot the variance of the coordinate as a function of time for the
parameters $\delta=0.1$, $\lambda=0.1$, $m=10$, and $\omega_0=1$. It is easier to work with the units, where $\hbar=1$. Then it is the inverse mass which plays the role of the quantum parameter: instead of $\hbar=1$ and $m=10$ we could choose $\hbar=0.1$ and $m=1$ and get identical results up to rescaling of units. As it is clear from the figure, the motion near $x=0$ indeed becomes unstable after $\kappa(t)$ becomes negative and $\langle x^2(t)\rangle$ rapidly increases. Then at a certain point $t=2/\delta$ the potential becomes stationary and the
particle oscillates around one of the two new minima. Clearly in this case we find the perfect agreement between the exact and the TWA solutions for considerably long times.

\begin{figure}[ht]
\includegraphics[width=12cm]{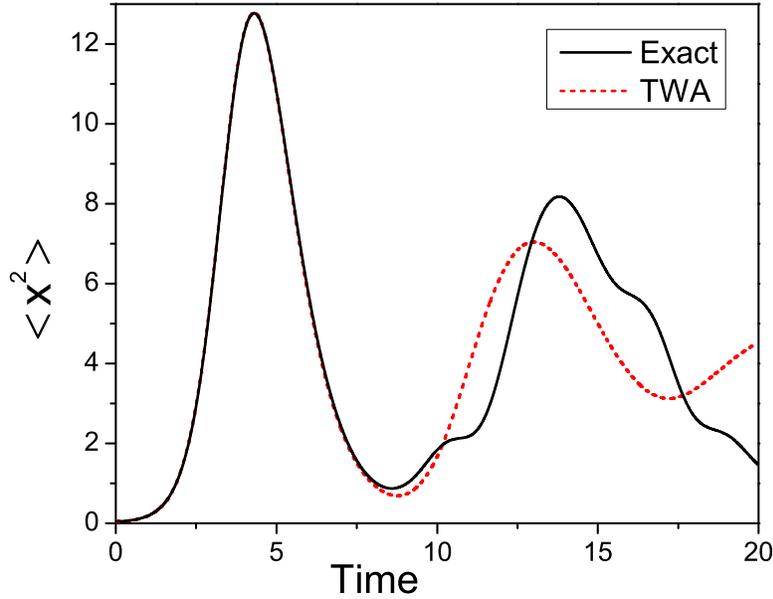}
\caption{Same as in Fig.~\ref{fig:cat4} but for a sudden change of the parameter $\kappa$ from $1$ to $-1$. It is clear that the TWA accurately describes the first oscillation of $\langle {\hat x}^2(t)\rangle$ but then starts to deviate from the exact result.}
\label{fig:cat2}
\end{figure}
\begin{figure}[ht]
\includegraphics[width=12cm]{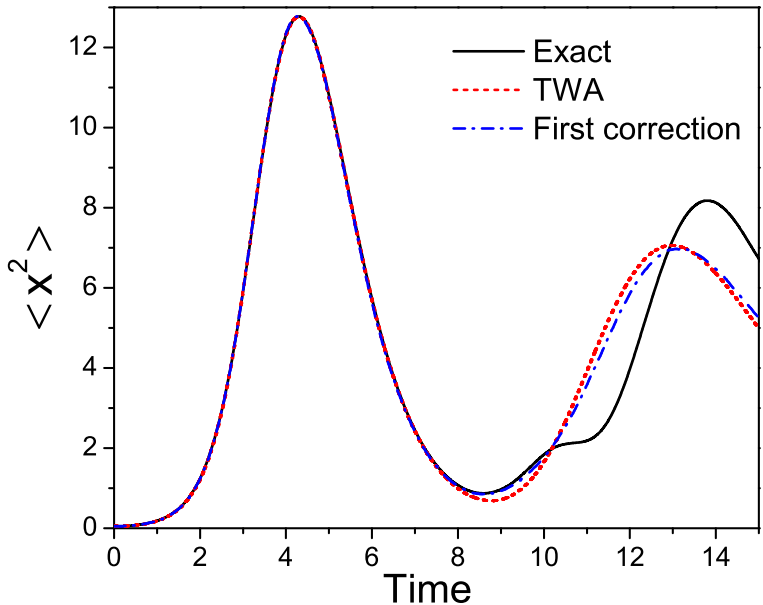}
\caption{Same as in Fig.~\ref{fig:cat2}. The additional blue line represents the first quantum correction evaluated according to Eq.~(\ref{diff_6}) with $\Delta \tau=0.25$. Despite this value of the time step is quite large, we checked that the results are insensitive to this particular choice. It is clear that the quantum correction extends the agreement of TWA with the exact solution to a longer time. }
\label{fig:cat6}
\end{figure}

For a sudden quench $\delta\to\infty$ the classical equations of motion governing the dynamics determined
by the Hamiltonian~(\ref{hamilt_cat}) can be solved explicitly:
\be
x(t)=\sqrt{2\alpha}\, {\rm cs}\left(\sqrt{1+\lambda}(\omega t+\beta)\biggl |1+{2\lambda\alpha\over 1+\sqrt{1+2\lambda\beta}}\right),
\label{sol_cat}
\ee
where ${\rm cs(u|m)}$ is the elliptic function and $\alpha$ and $\beta$ are the free parameters determined by the initial conditions. If the parameter $\lambda$ in Eq.~(\ref{hamilt_cat}) is small then at short times the quartic term in the Hamiltonian is irrelevant and Eq.~(\ref{sol_cat}) simplifies to
\be
x(t)\approx \sqrt{2\alpha}\sinh(\beta+\omega t)
\label{sol_cat1}
\ee
so that
\be
x_0\approx \sqrt{2\alpha}\sinh\beta,\;p_0\approx \sqrt{2\alpha}\omega/m\,\cosh(\beta).
\ee
These relations allow one to express the unknown constants $\alpha$ and $\beta$ through initial conditions and then average an observable of interest over these initial conditions with the Wigner function. In particular, at short times when the approximation (\ref{sol_cat1}) still holds (or when $x^2\ll 1/\lambda$) we find that
\be
\langle x^2\rangle\approx {\hbar\over 2m\omega}\cosh2\omega t
\label{sol_cat2}
\ee
From this expression we can estimate the time it takes the particle to reach one of the minima of the mexican hat potential
\be
t^\star\approx {1\over 2\omega}\ln\left[2m\omega\over \hbar\lambda\right].
\ee
This result can be of course obtained by more elementary methods. Using properties of the elliptic functions one can analyze corrections to Eq.~(\ref{sol_cat2}). Our goal here is, however, to understand the limits of applicability of the semiclassical approach and analyze the lowest quantum corrections to the dynamics. For this purpose we will proceed with numerical results.

In Fig.~\ref{fig:cat2} we plot the results of the semiclassical (TWA) and exact calculations for the variance of the displacement of the particle after a sudden quench. It is clear that TWA describes very well the initial spread of the wave packet representing the particle and its following collapse. However, at later times the exact and semiclassical curves start to depart from each other and the quantitative agreement is lost. In Fig.~\ref{fig:cat6} we show the same plot on a shorter time scale with an additional line representing the first quantum correction evaluated according to Eq.~(\ref{diff_6}). This correction extends the validity of TWA, however, always remaining relatively small and thus failing to indicate where TWA breaks down. As we will see in the  next example the situation can be quite different and the quantum correction can accurately predict where TWA stops being valid. Note that for evaluating the correction formally one needs to take the limit $\Delta \tau\to 0$. However, in practice we used a relatively big time step $\Delta\tau=0.25$ and yet got accurate results, which do not change if $\Delta \tau$ becomes smaller. From this simple example it is evident that the quantum diffusion gives very robust results similarly to the classical Langevin type diffusion and thus can be used for practical calculations. We note that the formal evaluation of the response according to Eq.~(\ref{corr2}) gives very poorly converging integrals because of the unstable classical dynamics near $x=0$ and thus it is not suitable for numerical purposes.

\subsection{Sine-Gordon model.}
\label{sec:sg}

Further increasing the complexity of the system in this section we will show how the formalism can be applied to the sine-Gordon (SG) model, which is described by the following Hamiltonian~\cite{giamarchi}:
\be
\hat{\mathcal H}=\int {dx\over 2}\left[ \hat n^2(x)+(\partial_x\hat \phi(x))^2-2V\cos(\beta\hat \phi(x))\right],
\label{h_sg}
\ee
where $\hat n(x)$ and $\hat \phi(x)$ are conjugate variables with canonical commutation relations (like coordinate and momentum) and $\beta$ is the parameter, which characterizes the strength of quantum fluctuations and plays the role of $\hbar$. This model has numerous applications in different areas of physics. Both classical and quantum SG models are integrable~\cite{Lukyanov97}. The discrete version of the SG model, where  $\partial_x\phi\to \phi_{j+1}-\phi_{j}$, is known under the name of the Frenkel-Kontorova model~\cite{chaikin-lubensky}. Its Hamiltonian is
\be
\hat{\mathcal H}=\sum_j \left[ {\hat n_j^2\over 2}+{1\over 2a^2}(\hat\phi_j-\hat\phi_{j+1})^2-V\cos(\beta\hat\phi_j)\right].
\label{FK}
\ee
Here $a$ is the lattice constant with $a\to 0$ corresponding to the continuous SG model. The model above has identical low-energy properties as the SG model. We will use the Frenkel-Kontorova model in actual numerical simulations and will set $a=1$.

If the SG model represents commensurate bosons in an optical lattice potential then $\beta=2\sqrt{\pi /K}$, where $K$ is the Luttinger-liquid parameter~\cite{buchler}\footnote{There is some controversy in the literature in defining either the Luttinger-Liquid parameter or its inverse enters the expression for $\beta$. We use the convention discussed in e.g. Refs.~\cite{cazalilla, claudia}, where large $K$ corresponds to weak interactions}. For $\beta<2\sqrt{2\pi}$ the cosine potential is relevant in equilibrium and for $\beta>2\sqrt{2\pi}$ it is irrelevant~\cite{giamarchi}. This means that in the former case even for the infinitesimal value of $V$ the system opens a gap in the spectrum~\cite{donohue, ludwig}. In this regime the structure of the spectrum of the SG model is quite rich. Thus at $\beta\leq 2\sqrt{\pi}$ the elementary excitations, solitons, can bind together to form new quasiparticles $B_1$ breathers. Then at $\beta\leq \sqrt{2\pi}$ $B_1$ breathers can bind to form $B_2$ breathers and so on~\cite{vova_quench, vova_response}. Details of the spectrum of the excitations are important for understanding non-equilibrium dynamics. In particular, it was shown that for $\beta<2\sqrt{\pi}$ quench in the SG model can lead to damped or undamped oscillations at frequencies corresponding to the breather's masses~\cite{vova_quench}. For large $\beta$, where the cosine term is irrelevant and the spectrum is gapless, one can expect strong damping of the oscillations or no oscillatory behavior at all.

For the purpose of this paper it is important to realize that $\beta$ plays the role of the Planck's constant. This can be seeing, for example, from the rescaling $\phi\to\phi/\beta$ and noticing that $1/\beta$ becomes the saddle point parameter of the quantum evolution operator. Thus for $\beta\ll 1$ we expect that the semiclassical description is valid and for $\beta\gtrsim 1$ it fails. To simulate the time evolution in the SG model we chose a particular dependence of the coupling $V$ on time: $V(t)=0.1\tanh (0.2 t)$. For a system of commensurate bosons this corresponds to turning on an optical lattice~\cite{claudia}. We will be interested in the expectation value of $\cos(\beta\hat\phi)$ as a function of time. On general grounds we anticipate a large response for small $\beta$, where the potential is strongly relevant, and conversely a small response at large $\beta$.

First let us assume that the system is initially prepared in the ground state. The latter corresponds to the product of ground states of harmonic oscillators corresponding to different momenta $q$. Thus the Wigner function of this state is a product of Gaussians. In particular, for the Frenkel-Kontorova model with $a=1$ we have
\be
W(\phi_q^0,n_q^0)=\prod_q \exp\left[-{|\phi_q^0|^2\over 2\sigma_q^2}-2\sigma_q^2 |n_q^0|^2\right],
\label{wig_sg}
\ee
where $\sigma_q=1/\sqrt{2\omega_q}$, $\omega_q=2\sin(q/2)$, and $q=2\pi n/L$ with $n=0,1,.. L-1$ ($L$ is the system size). We note that the zero momentum mode is characterized by $n_q=0$ and random $\phi_q$. From this distribution we can generate the initial values for $\phi_j^0$ and $n_j^0$ in real space:
\be
\phi_j^0={1\over \sqrt{L}}\sum_q Re\left(\phi_q^0\mathrm e^{iqj}\right),\;
n_j^0={1\over \sqrt{L}}\sum_q Re\left(n_q^0\mathrm e^{iqj}\right)
\label{jq}
\ee
Within TWA we need to solve the classical equations motion:
\be
{d^2\phi_j\over dt^2}=(\phi_{j+1}+\phi_{j-1}-2\phi_j)-\beta V(t)\sin(\beta\phi_j)
\label{eq:sg}
\ee
subject to the random initial conditions $\phi_j(t=0)=\phi_j^0$, $\dot\phi_j(t=0)=n_j^0$ distributed according to Eqs.~(\ref{wig_sg}) and (\ref{jq}). The first quantum correction to the semiclassical approximation can be evaluated according to Eq.~(\ref{diff_6}), where the third derivative of the potential $V_{3,j}(\tau_i)=-V(\tau_i)\beta^3\sin(\beta\phi_j(\tau_i))$.

Before showing numerical results we would like to emphasize the difference between expanding evolution in quantum fluctuations ($\beta$) and in the strength of interaction $V$. Note that since for noninteracting problems TWA is exact the expansion in $\beta$ is also the expansion in $V$. However these two expansions are very different. The expansion in $\beta$ requires solving nonlinear classical equations of motion taking into account $V$ at the classical level in all orders of perturbation theory. Conversely linear response (and higher order corrections in $V$) assumes that the potential is a weak perturbation but treats all orders in $\beta$ exactly. It is clear that the perturbative expansion in $V$ should work better at large values of $\beta$ where the cosine potential is less relevant. Conversely the expansion in $\beta$ should work better when $\beta$ is small and the cosine potential is strongly relevant. The expectation value of $\cos(\beta\hat\phi)$ within the linear response theory is given by~\cite{abrikosov}:
\be
\langle\cos \beta\hat{\phi}(t)\rangle\approx i\int\limits_0^t dt\int\limits_0^L\! dx\, V(t')\left< \left[\cos(\beta\hat\phi(x,t')), \cos(\beta\hat\phi(0,t))\right]\right>_0,
\ee
where the index ``0'' indicates that the averaging is taken over the initial state with the evolution given by the unperturbed Hamiltonian (without the cosine term). The correlation function above is related to the imaginary part of the retarded correlation function~\cite{abrikosov}, which can in turn be evaluated by analytic continuation of the corresponding imaginary time correlation function~\cite{giamarchi} to the real time axis. Then we find:
\be
\langle\cos \beta\hat{\phi}(t)\rangle\approx \int\limits_0^t d\tau V(t-\tau)\sum_j\exp\left[-
{\beta^2\over 2L}\sum_{q=0}^{L-1} {1-\cos(q j)\cos(\omega_q\tau)\over \omega_q}\right]\sin\left[
{\beta^2\over 2L}\sum_{q=0}^{L-1} {\cos(q j)\sin(\omega_q\tau)\over \omega_q}\right].
\label{sg_lr_q}
\ee
We note that the $q=0$ mode here has to be included in the expression above with the $0/0$ uncertainty evaluated according to the L'Hospital's rule.

We can also evaluate the analytical expression for the linear response within the semiclassical truncated Wigner approach:
\be
\langle\cos\beta\hat{\phi}(t)\rangle_{sc}\approx \int\limits_0^t d\tau V(t-\tau)\sum_j\exp\left[-
{\beta^2\over 2L}\sum_{q=0}^{L-1} {1-\cos(q j)\cos(\omega_q\tau)\over \omega_q}\right]\,\left[
{\beta^2\over 2L}\sum_{q=0}^{L-1} {\cos(q j)\sin(\omega_q\tau)\over \omega_q}\right].
\label{sg_lr_cl}
\ee
The expressions (\ref{sg_lr_q}) and (\ref{sg_lr_cl}) almost coincide except for the last multiplier. Clearly in the limit of small $\beta$ the argument of the sine function in the exact expression is small and the difference between the two results vanishes. Therefore within the linear response the TWA result is accurate up to the terms $O(\beta^6)$. Comparison between Eqs.~(\ref{sg_lr_q}) and (\ref{sg_lr_cl}) also shows that TWA gives accurate results at short times even if $\beta$ is not small. This is in accord with our general expectations that TWA is asymptotically exact at small $t$. Whether TWA always remains accurate or diverges at long times is on the other hand highly non-universal. For example if the sum over momenta $q$ in the argument of the sine function in Eq.~(\ref{sg_lr_q}) is bounded at all times, which would be the case in higher dimensions, then the difference between Eqs.~(\ref{sg_lr_q}) and (\ref{sg_lr_cl}) remains small at all times provided that $\beta\ll 1$. One can expect that in this situation TWA remains accurate for infinitely long times even beyond the linear response approximation.

Next we compare the results for the expectation value of $\cos(\beta\hat\phi)$ within TWA and the additional quantum correction with the full linear response calculations for different values of $\beta$. As we mentioned above we take $V(t)=0.1\tanh (0.2 t)$ so that the potential always remains small and the linear response is justified at least for relatively short times. Thus we can use the linear response analysis to determine the accuracy of the semiclassical approach and the quantum corrections at short times.

\begin{figure}[ht]
\includegraphics[width=12cm]{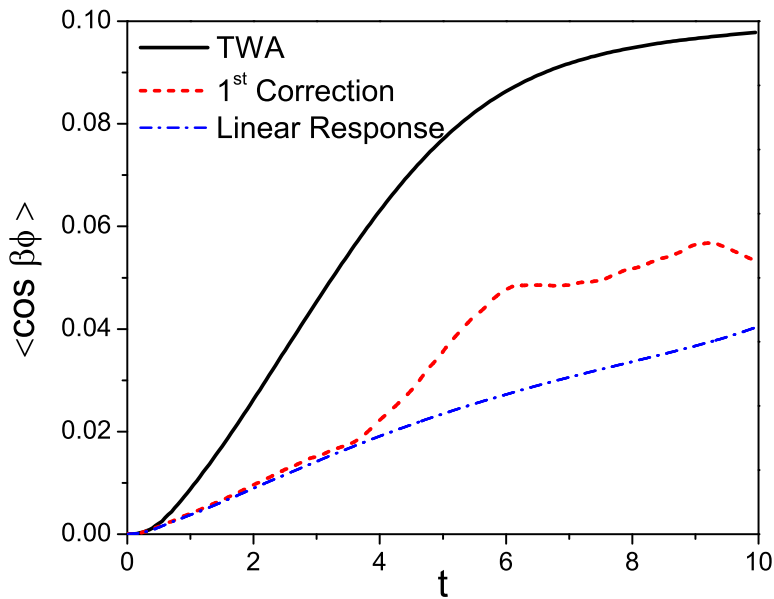}
\caption{Time dependence of $\langle \cos(\beta\hat\phi)\rangle$ for the discretized SG model described by the Hamiltonian~(\ref{h_sg}) with the system size $L=20$, $\beta=2\sqrt{\pi}$, and the cosine potential turning on as $V(t)=0.1\tanh(0.2)$. The three plots correspond to the TWA, TWA with the first quantum correction, and the linear response. }
\label{fig:sg4}
\end{figure}

In Fig.~\ref{fig:sg4} we show time dependence of $\langle\cos(\beta\hat\phi)\rangle$ for $\beta=2\sqrt{\pi}$ and $L=20$. The three plots correspond to the semiclassical approximation, the first quantum correction evaluated according to Eq.~(\ref{diff_6}) with $\Delta \tau=0.2$ (see Appendix~\ref{sec:app_implement} for additional details), and the linear response. While the cosine potential is relevant in this regime the linear response is remains valid for sufficiently long times and thus gives an accurate prediction for the exact result. It is clear from the plot that TWA is valid only at very short times and then rapidly breaks down. The first quantum correction significantly extends the TWA approach and unambiguously shows where TWA breaks down.

\begin{figure}[ht]
\includegraphics[width=12cm]{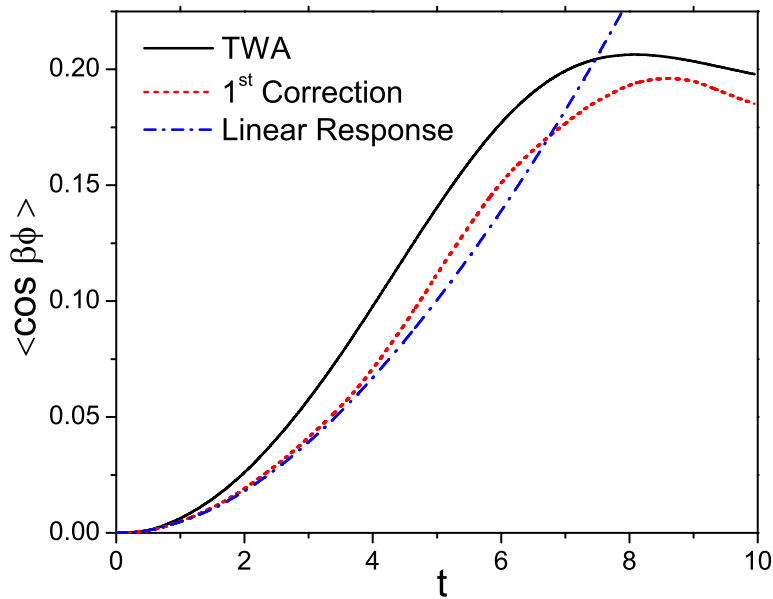}
\caption{Same as in Fig.~(\ref{fig:sg4}) but for $\beta=\sqrt{2\pi}$. }
\label{fig:sg24}
\end{figure}

\begin{figure}[ht]
\includegraphics[width=12cm]{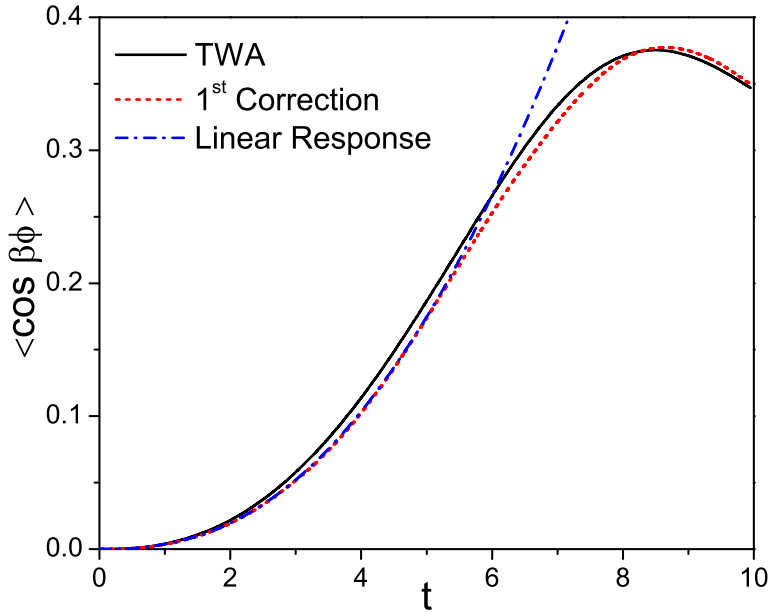}
\caption{Same as in Fig.~(\ref{fig:sg4}) but for $\beta=\sqrt{\pi}$. }
\label{fig:sg25}
\end{figure}

The situation changes dramatically if we consider smaller values of $\beta$, i.e. if we go closer to the classical limit. Thus in Fig.~\ref{fig:sg24} we plot $\langle\cos\beta\hat\phi\rangle$ for $\beta=\sqrt{2\pi}$, i.e. a factor of $\sqrt{2}$ smaller than in Fig.~\ref{fig:sg4}. Despite this is a relatively small change in the parameter $\beta$ the results are quite different. First the linear response is clearly valid for much shorter times $t\lesssim 5$. After that the linear response calculation is not reliable leading to a spurious divergence of $\langle\cos\beta\hat\phi\rangle$. Second the accuracy and the time domain of validity of the semiclassical approximation is significantly improved compared to the previous case. The first quantum correction again clearly extends the domain of applicability of the semiclassical approach. And finally in Fig.~\ref{fig:sg25} we plot the same dependences for even smaller $\beta=\sqrt{\pi}$. Now the linear response result shows a very good agreement with the TWA calculation for relatively short times $t\lesssim 5$. The first quantum correction makes this agreement almost ideal. At longer times the linear response clearly breaks down, however the first quantum correction to the semiclassical result continues to be small. Thus we can expect that the semiclassical approximation is very accurate here. It is interesting that by changing the parameter $\beta$ (which plays the role of $\hbar$) by only a factor of two we expand the regime of validity of TWA by at least two orders of magnitude.

Instead of starting from the ground state one can consider a situation, where the system is initially prepared at finite temperature $T$. The Wigner function for the thermal density matrix can be straightforwardly evaluated using Eq.~(\ref{wig_bos_coh}) (see also the supplementary information of Ref.~\cite{pg_np}):
\be
W(\phi_q^0,n_q^0)=\prod_q \exp\left[-{|\phi_q^0|^2\over 2\sigma_q^2 r_q}-{2\sigma_q^2\over r_q} |n_q^0|^2\right],
\label{wig_sg_temp}
\ee
where $r_q=\coth(\omega_q/(2T))$. In the limit $T\ll \omega_q$ we have $r_q\to 1$ and we recover the zero temperature result (\ref{wig_sg}). In the opposite limit $T\gg\omega_q$ we get
\be
W(\phi_q^0,n_q^0)\to \prod_q \exp\left[-{|\phi_q^0|^2 \omega_q^2+|n_q^0|^2\over 2T}\right],
\label{wig_sg_temp1}
\ee
which is nothing but the Boltzmann's distribution. We note that for any nonzero temperature the distribution of the zero momentum mode ($q=0$) is always Boltzmann's because there is no zero point energy associated with this mode, which simply describes translation of the whole chain. It is interesting to observe that the uncertainty between conjugate variables $\phi_q^0$ and $n_q^0$ at finite temperatures satisfies:
\be
\delta|\phi_q^0|\delta|n_q^0|={1\over 2}r_q^2.
\ee
Thus we recover the minimum uncertainty relation at $T\to 0$ and $\delta|\phi_q^0|\delta|n_q^0|\sim 2T^2/|\omega_q|^2\gg 1$ at large $T$.

It is also straightforward to generalize Eq.~(\ref{sg_lr_q}) to the case of nonzero temperatures:
\beq
&&\langle\cos \beta\hat{\phi}(t)\rangle\approx \int_0^t d\tau V(t-\tau)\nonumber\\
&&\times\sum_j\exp\left[-
{\beta^2\over 2L}\left(\sum_{q=1}^{L-1} {1-\cos(q j)\cos(\omega_q\tau)\over \omega_q \tanh(\omega_q/2T)}+T\tau^2\right)\right]\sin\left[
{\beta^2\over 2L}\sum_{q=0}^{L-1} {\cos(q j)\sin(\omega_q\tau)\over \omega_q}\right].
\label{sg_lr_t}
\eeq

\begin{figure}[ht]
\includegraphics[width=12cm]{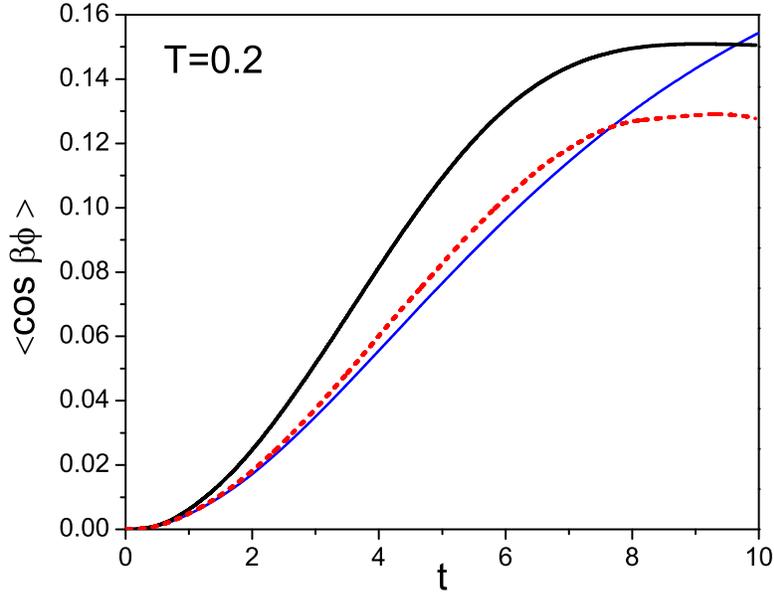}
\caption{Same as in Fig.~(\ref{fig:sg4}) but for $\beta=\sqrt{2\pi}$ and initial temperature $T=0.2$. }
\label{fig:sg27}
\end{figure}

The results of the TWA simulations, the first quantum correction as well as the linear response calculations for $T=0.2$ and $\beta=\sqrt{2\pi}$ are plotted in Fig.~\ref{fig:sg27}. Perhaps a somewhat unexpected feature of the simulations is that even at finite initial temperatures the semiclassical approximation still shows significant deviations from the exact result. As in the zero temperature case the first quantum correction considerably extends the time domain of the applicability of the semiclassical approximation. We want to stress again that the small parameter of the theory is $\beta$. In simulations we used $\beta=\sqrt{2\pi}$ corresponding to the regime of relatively strong quantum fluctuations. We intentionally used this regime to emphasize the significance of the quantum correction. For smaller value of $\beta$ the accuracy of TWA improves dramatically like in Fig.~\ref{fig:sg25}.

\subsection{Dynamics in the coherent state representation.}

In this section we will address various problems in the coherent state phase space. We will review both examples discussed in some earlier works and present a couple of new ones. Most of the examples here have direct relevance to the cold atom experiments. We want to emphasize that the main goal of this section is to verify accuracy of the formalism discussed here. So some of the problems discussed here will be somewhat artificial. A detailed overview of applying TWA to experimentally relevant situations in cold atom systems can be found in the review~\cite{blakie_08} and in more recent works~\cite{cohen, ludwig1, daw-wei_wig, ruostekoski_09, hipolito}.

\subsubsection{Collapse of a coherent state}

The first example we will briefly review here deals with the evolution of the initial coherent state with on average $N$ particles per site subject to the Hamiltonian containing only the interaction term (see Ref.~\cite{ap_twa} for more details):
\be
\hat{\mathcal H}={U\over 2} \hat{\psi}^\dagger \hat{\psi}(\hat{\psi}^\dagger \hat{\psi}-1).
\label{ham_collapse}
\ee
This problem is closely related the collapse-revival experiment by M.~Greiner et. al.~\cite{bloch_collapse}. Because the problem does not have a kinetic term it can be easily solved analytically. In particular, the expectation value of the annihilation operator is given by~\cite{ap_twa}:
\be
\langle \hat\psi(t)\rangle= \sqrt{N} \exp\left[N (\mathrm e^{-i U t}-1)\right].
\label{examples5}
\ee
This result can be obtained by expanding the coherent state in the Fock basis:
\be
|\psi_0\rangle=\mathrm e^{-N/2}\sum_n {{\sqrt N}^n\over\sqrt{n!}}|n\rangle
\ee
and then trivially propagating this state in time. The classical limit is achieved here taking the limit $N\to\infty$, $U\to 0$ while keeping the product $UN=\lambda$ fixed.
\begin{figure}[ht]
\includegraphics[width=12cm]{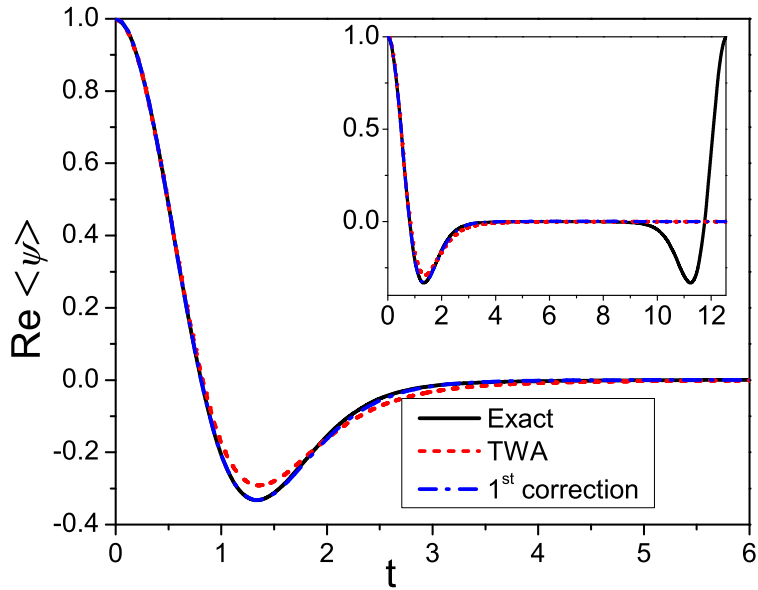}
\caption{Dependence of the real part of $\langle \hat{\psi}\rangle$ on time for the initial coherent state evolving according to the Hamiltonian~(\ref{ham_collapse}). The interaction is taken to be $U=0.5$ and the average number of particles in the initial state is $N=4$. }
\label{fig:collapse}
\end{figure}
The exact evolution shows a rapid collapse of the coherent state at time $t_c\sim 1/UN=1/\lambda$ and then the consequent revival at time $t_r\sim 1/U=N/\lambda$ (see Fig.~\ref{fig:collapse}). In the classical limit $t_r\to\infty$ thus the revival is a purely quantum phenomenon.

Next we solve the problem using TWA and quantum corrections. For doing this we need the Weyl symbol for the Hamiltonian~(\ref{ham_collapse}):
\be
\mathcal H_W(\psi^\star,\psi)={U\over 2}|\psi|^2(|\psi|^2-2)+{U\over 4},
\ee
which yields the following classical equation of motion
\be
i{\partial \psi(t)\over\partial t}=U (|\psi(t)|^2-1)\psi(t).
\label{eq:single_site}
\ee
This equation can be trivially solved using that $|\psi(t)|^2=|\psi_0|^2$ is the integral of motion. The solution should be supplemented by random initial conditions distributed according to the Wigner function~\cite{gardiner-zoller, ap_twa}:
\be
W(\psi_0,\psi_0^\star)=2\exp[-2|\psi_0-\sqrt{N}|^2].
\ee
Using the explicit analytic solution of Eq.~(\ref{eq:single_site}) and the Wigner function above one can analytically calculate both the TWA result ($\psi_0(t)$) and the first (and higher)  quantum correction ($\psi_1(t)$)~\cite{ap_twa}:
\be
\psi_{0}(t)=\sqrt{N} \exp\left[-\frac{i UN t}{1 + iUt/2}\right]\exp[i U t]\frac{1}{ (1 + i U t/2)^2},
\label{examples6}
\ee
\be
\label{examples7}
\psi_1(t)=\psi_{0}(t){U^2t^2\over 4}\left(1 -
    \frac{i UN t}{3\left( 1 + i Ut/2 \right)^2 } -
  \frac{i U t}{3\left(1 + i U t/2\right)}\right).
\ee
It can be explicitly checked that the Taylor expansion of the exact result for $\langle \hat{\psi}(t)\rangle$ in powers of $1/N$ (at a fixed $\lambda=UN$) coincides with the similar expansion of the TWA result $\psi_0(t)$ up to the terms of the order of $1/N^2$ and with the expansion of $\psi_0(t)+\psi_1(t)$ up to the terms of the order of $1/N^4$. To illustrate these dependences we plot the real part of the expectation value of $\langle \hat{\psi}(t)\rangle$ versus time in Fig.~\ref{fig:collapse} for the parameters $U=0.5$ and $N=4$ (corresponding to $\lambda=2$). It is clear that at short times TWA gives a very good approximation to the exact result and the next quantum correction improves it even further. However, both TWA and the correction fail to describe the revival phenomena which have a purely quantum origin. From Eq.~(\ref{examples5}) one can see that in order to correctly describe the revival one needs to include exponentially large number of terms. Mathematically this happens because the functions of the type $\exp[\exp[ix]]$ have very bad convergence of the Taylor expansion at large $x$.

\subsubsection{Evolution towards the Schr\"odinger cat state.}
\label{sec:cat}

Next we will consider an example largely following Ref.~\cite{ap_cat} dealing with dynamics of the system of bosons in a double well potential. We assume that the only two lowest single-particle (vibrational) states are relevant so the system can be described by the Bose-Hubbard Hamiltonian (see Ref.~\cite{trotzky2008} for discussion of applicability of the Bose-Hubbard model to cold atoms in optical lattices):
\be
\hat{\mathcal H}=-J\sum_j \left(\hat{\psi}_j^\dagger \hat{\psi}_{j+1}+\hat{\psi}_{j+1}^\dagger \hat{\psi}_j\right)+{U(t)\over 2}\,\hat{\psi}_j^\dagger\hat{\psi}_j(\hat{\psi}_j^\dagger\hat{\psi}_j-1),
\label{ham_bhm}
\ee
where $j=1,2$, $J$ is the tunneling matrix element and $U$ is the interaction constant. We assume that initially the system of $2N$ particles is prepared in the noninteracting ground state and then slowly attractive interactions are ramped up. We note that alternatively one can initially prepare the system in the most excited state and then turn on repulsive interactions~\cite{ap_cat}. Beyond some critical strength the classical symmetric equilibrium, where both wells are equally populated, becomes unstable and the system evolves towards the ``Schr\"odinger cat'' state. As in the previous example the classical limit is achieved  when the average number of particles per site $N$ becomes large while the product $UN$ is kept fixed. The corresponding classical (Gross-Pitaevskii) equations read
\beq
&& i {\partial \psi_1(t)\over\partial t}=-J\psi_2(t)+U(t) \left(|\psi_1(t)|^2-1\right)\psi_1(t)\label{eq:gp_a} \\
&& i {\partial \psi_2(t)\over\partial t}=-J\psi_1(t)+U(t) \left(|\psi_2(t)|^2-1\right)\psi_2(t).
\label{eq:gp}
\eeq
As in Eq.~(\ref{eq:single_site}) extra ``$-1$'' in Eqs.~(\ref{eq:gp_a}) and (\ref{eq:gp}) appear because of the Weyl ordering of the Hamiltonian. Since these terms only result in the overall phase of the wave function they can be dropped. We assume that initially the noninteracting system of $2N$ particles is prepared in the ground state and then one slowly introduces an attractive interaction:
\be
U(t)=-{1\over N}{\tanh(\delta t)\over 1-\delta t},
\label{u(t)}
\ee
where $\delta$ is the parameter controlling the ramp speed. We have chosen such dependence of $U(t)$ so that at time $t=1/\delta$ the ratio of the interaction to the tunneling becomes infinite. Physically this corresponds to linearly shutting down the tunneling between the two sites since only the ratio $J/U$ is physically important. The parameter $\delta$ in Eq.~(\ref{u(t)}) controls the speed of the ramp. We note that this problem is very closely related to the motion of the particle in the Mexican hat potential if we use number imbalance $\hat{n}=\hat{\psi}_1^\dagger \hat{\psi}_1 - \hat{\psi}_2^\dagger \hat{\psi}_2$ as an effective coordinate (see Refs.~\cite{psg, ap_cat, hipolito} for details). At sufficiently weak interactions the particles evenly distributed between the two wells and the effective potential has a minimum at $n=0$. As interactions exceed some critical value $U_c\sim J/N$ the equilibrium $n=0$ becomes unstable and the particles prefer to concentrate in one of the wells.
\begin{figure}[ht]
\includegraphics[width=12cm]{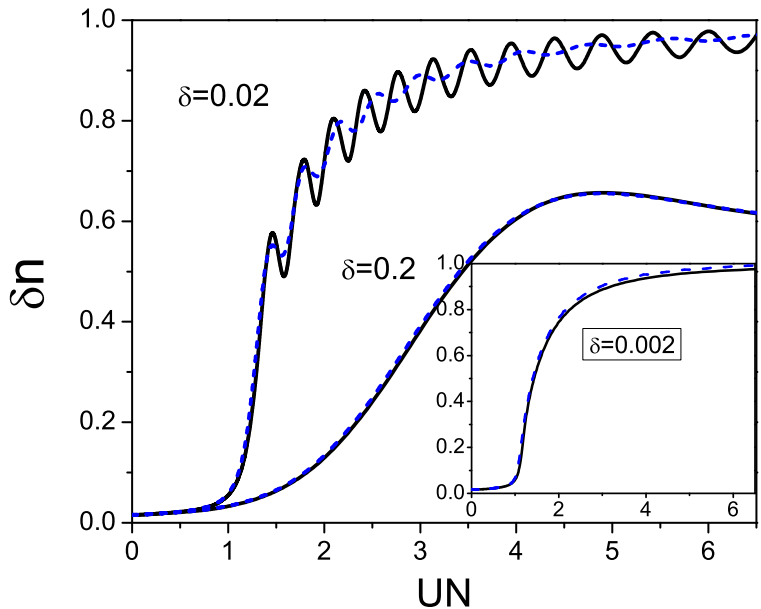}
\caption{Dependence of the normalized number variance $\delta n=\sqrt{\langle\hat{ n}^2\rangle}/2N$ on the interaction strength. The number of particles per cite is $N=32$ and the interaction depends on time according to Eq.~(\ref{u(t)}). The solid black line represents the exact solution and the dashed blue line is the result of the semiclassical approximation (TWA). This  plot is taken from Ref.~\cite{ap_cat}.}
\label{cat_bosons}
\end{figure}
For large number of particles the total number constraint is not important and the initial state is almost equivalent to the product of coherent states, for which the Wigner function reads~\cite{ap_cat}:
\be
W(\psi_j,\psi_j^\star)=\prod_j 2\exp\left[-2|\psi_j-\sqrt{N}\exp(i\phi)|^2\right].
\label{wig_bos_coh}
\ee
The overall site independent phase $\phi$ here is arbitrary and we choose $\phi=0$. For the initial state where the number of bosons is fixed the Wigner function is also readily obtained (see Eq.~(43) in Ref.~\cite{ap_cat}), however it contains highly oscillating component and is thus much harder to deal with. In large systems the difference between coherent state and fixed number symmetric state is not important. In Fig.~\ref{cat_bosons} we show the dependence of the number variance on the interaction strength for three different ramp rates $\delta$. It is clear that the semiclassical approach gives very accurate predictions for slowest and fastest rates with small deviations occurring only at the intermediate rate.

\subsubsection{Turning on interactions in a system of lattice bosons.}
\label{sec:ramp}

Let us now consider a process of heating bosonic atoms due to non-adiabatic changes of the coupling in time. As in Sec.~\ref{sec:cat} we  assume that the system is described by the Bose-Hubbard model but with more than two sites: $j=1,\dots, M$. For simplicity we chose periodic boundary conditions. We analyze a process where one starts from the ground state in the noninteracting regime and then ramps on now repulsive interactions according to
\be
U(t)=U_0\tanh(\delta t).
\label{u_ramp}
\ee
The classical Gross-Pitaevskii equations of motion are straightforward multi-site generalization of Eqs.~(\ref{eq:gp}). To avoid dealing with highly oscillating Wigner function, describing the initial conditions, it is convenient to start from the coherent state rather than the state with fixed number of particles. Then the Wigner function is given by Eq.~(\ref{wig_bos_coh}). To see the effects of quantum corrections we intentionally choose parameters corresponding to strong quantum fluctuations: $J=1$, $U_0=1$ and on average one particle per site. As in Ref.~\cite{pg_np} we will analyze the energy density (or the energy per site) in the system.

First we consider a small periodic one-dimensional chain consisting from eight sites so that we can do comparisons with exact results obtained by directly solving the Schr\"odinger equation.
\begin{figure}[ht]
\includegraphics[width=12cm]{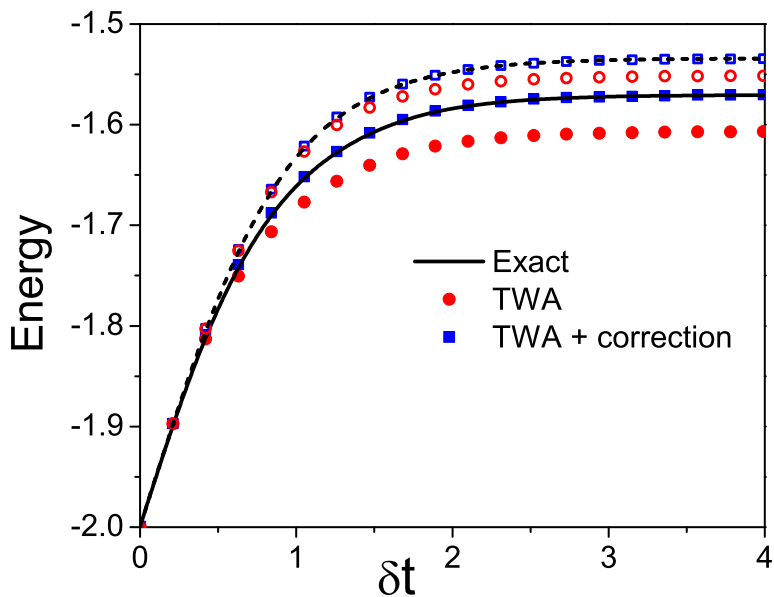}
\caption{Dependence of the energy density on the product $\delta t$ (see Eq.~(\ref{u_ramp})) for a periodic Bose-Hubbard chain consisting of eight sites with on average one particle per site. The hopping amplitude $J=1$ and interaction changes in time according to Eq.~(\ref{u_ramp}) with $U_0=1$. Initially the system is assumed to be in the coherent zero-momentum state. The two sets of curves correspond to two different ramping rates $\delta=0.5$ (solid symbols and a solid line) and $\delta=2.5$ (open symbols and a dashed line). Red circles correspond to TWA and blue squares include the first quantum correction evaluated according to Eq.~(\ref{corr4}). Black lines represent exact results obtained by direct solution of the Schr\"odinger equation.}
\label{fig:hubbard_ramp}
\end{figure}
In Fig.~\ref{fig:hubbard_ramp} we show comparison between exact results as well as TWA and TWA + first correction (evaluated according to Eq.~(\ref{corr4})) for the two values of ramping rate: $\delta=0.5$ and $\delta=2.5$. We discuss details of numerical implementation of the quantum correction in the Appendix~\ref{sec:app_implement}. It is evident that TWA gives rather accurate results even though the system ends up in the strongly interacting regime. The first quantum correction significantly improves the accuracy of TWA making the results practically indistinguishable from the exact ones. This example highlights again that the quantum correction can be used not only to significantly improve the accuracy of TWA but also as a method to find where TWA starts to deviate from the exact result.

Next we will consider the same setup in a two-dimensional system of $32\times 32$ sites. This system is far beyond the limits of exact quantum simulations. Moreover this parameter regime corresponds to relatively strong interactions and significant heating
\begin{figure}[ht]
\includegraphics[width=12cm]{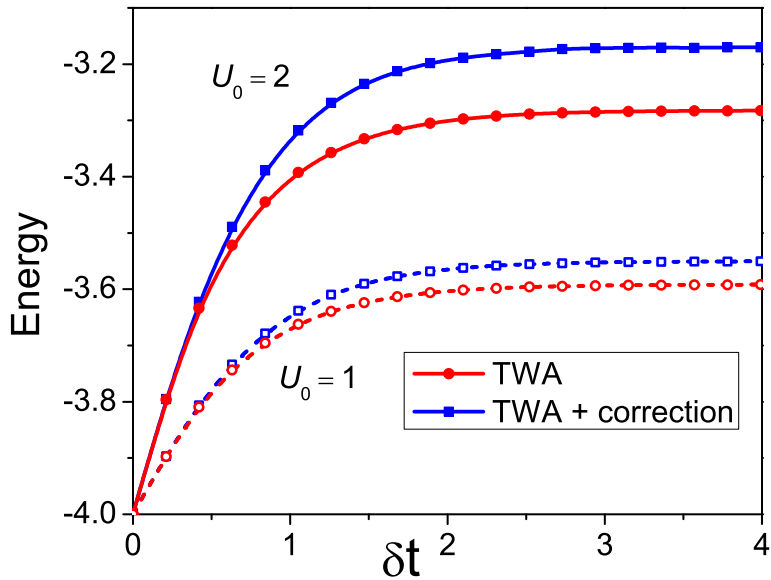}
\caption{Same as in Fig.~\ref{fig:hubbard_ramp} but for a two-dimensional $32\times 32$ lattice. We use $\delta=1$, $J=1$, on average one particle per site. The solid lines correspond to $U_0=2$ while the dashed lines do to $U_0=1$. As before red circles represent TWA and blue squares describe TWA + first quantum correction.}
\label{fig:hubbard_ramp_2d}
\end{figure}
However, it can be readily simulated using the phase space methods discussed here. In Fig.~\ref{fig:hubbard_ramp_2d} we plot energy per site for the ramp with $\delta=1$ and two values of the final interaction $U_0=2$ (solid lines) and $U_0=1$ (dashed lines). We also use $J=1$ and the unit filling. As expected TWA is more accurate for smaller interactions, which is reflected in a smaller difference between the blue and the red graphs.

\subsection{Thermalization within the Bose-Hubbard model.}

\begin{figure}[h]
\includegraphics[width=12cm]{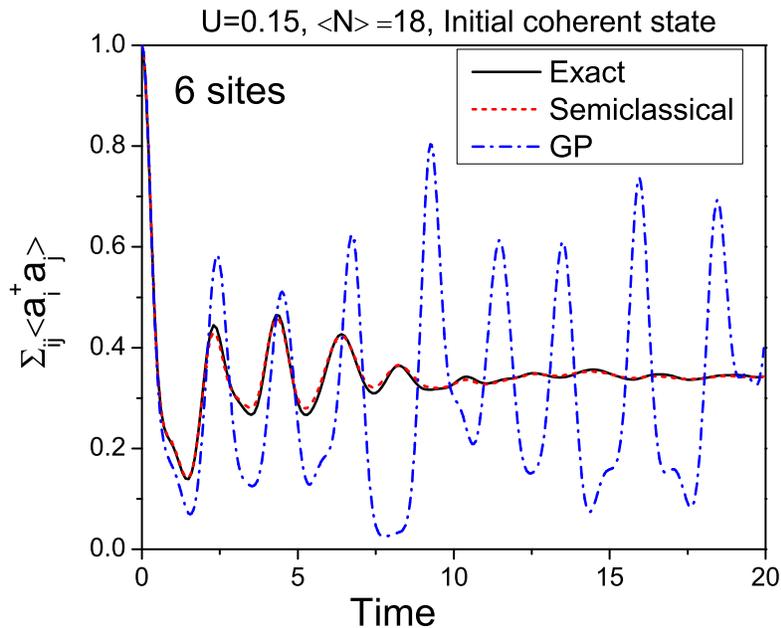}
\caption{Dependence of the occupancy of the zero momentum state on time for the initial state where all particles populate a coherent state in a single site. The total number of sites is $6$, the average number of particles is $N=18$, the interaction strength is $U=0.15$, and the tunneling is $J=1$. The blue curve describes the classical (Gross-Pitaevskii) simulations, which show no sign of thermalization. The black curve is the result of exact simulations and the red curve corresponds to TWA. }
\label{fig:thermal_coh}
\end{figure}

Let us now analyze the applicability of TWA to a very important issue of thermalization. We will again focus on the Bose-Hubbard model discussed above. As a specific situation we assume that initially all atoms are placed into one of the sites and then they are allowed to spread over. Since the system is nonintegrable one can expect that it will thermalize at long times. However, there is a famous example of the Fermi-Pasta-Ulam problem~\cite{fpu}, which shows that the thermalization is not necessarily guaranteed even on nonintegrable systems. Overall the issue of thermalization in closed Hamiltonian systems attracted recently a lot of theoretical attention~\cite{olshanii_nature, rigol_therm, kollath_altman_quench, reimann}. Because of lack of available tools to address dynamics in nonintegrable systems one either has to rely on exact simulations like in Ref.~\cite{olshanii_nature} or on DMRG and related methods like in Ref.~\cite{kollath_altman_quench}. The former can deal only with small system sizes, while the latter have long time limitations and generally applicable only to systems close to equilibrium. From this prospect phase space methods, in particular TWA, provide a very natural framework to study thermalization dynamics. In particular, TWA and further corrections give asymptotically accurate results at short times. At long times, if the system thermalizes, quantum fluctuations become subdominant if the temperature characterizing the steady state is not small. Then quantum fluctuations should remain subdominant and TWA results are expected to continue to be valid. While this discussion is somewhat speculative and requires further thorough analysis we will show here that these simple ideas are valid
for our simple toy example. Thermalization within the phase space approach was also analyzed in Refs.~\cite{rafi, pg_np, hipolito, pw, ludwig1, wright_08} and others. We assume that the system is initially prepared in the coherent state in a single site with an average number of atoms $N$. In Fig.~\ref{fig:thermal_coh} we show the dependence of the occupancy of the zero momentum state for such initial conditions for a periodic chain of six sites.

\begin{figure}[h]
\includegraphics[width=12cm]{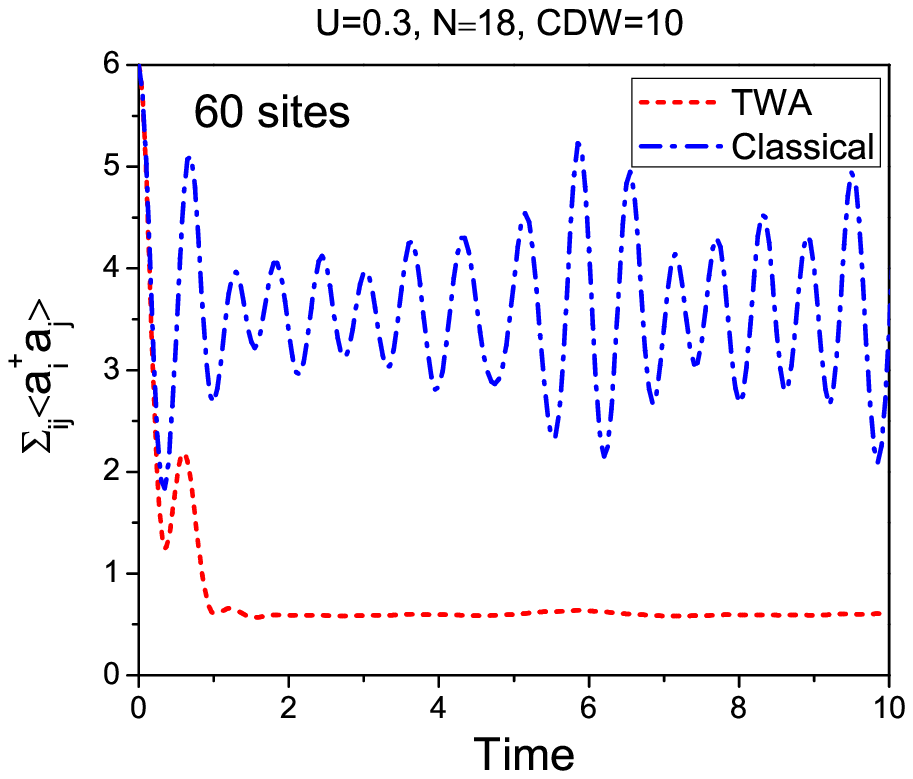}
\caption{Same as in Fig.~\ref{fig:thermal_coh} but for an array of 60 sites. The initial state is such that each tenth site is in a coherent state with 18 particles on average and the same phase. The two curves show the results of the semiclassical (TWA) and the classical GP simulations. The TWA apparently predicts rapid thermalization, while the classical evolution is clearly non-ergodic.}
\label{fig:thermal_coh1}
\end{figure}

The classical simulations show very little sign of thermalization. In this sense this problem is very analogous to the Fermi-Pasta-Ulam problem with the main difference that the initial state is localized in the coordinate rather than momentum state. Once we include quantum fluctuations already at the level of TWA we clearly see that the system approaches a steady state. Comparing these results with those of the exact diagonalization we see that TWA gives extremely accurate predictions. Obviously for the situation shown the quantum corrections are negligible (and this indeed can be checked numerically). It can be checked that at longer times the exact quantum simulations show revival of the initial state due to a small system size. As in earlier examples TWA misses that quantum revival. We expect that these revivals disappear as we go to larger system sizes. Since the goal of this paper is to review quantum dynamics using the phase space methods, we will set aside the question whether the steady state seen in Fig.~\ref{fig:thermal_coh} corresponds to the thermal equilibrium and address it in a separate publication.

We can easily extend TWA and classical GP-simulations to larger arrays, where the exact integration of the Schr\"odinger equation is no longer possible. In Fig.~\ref{fig:thermal_coh1} we show the evolution of the zero momentum occupancy for the initial array of coherent states having the same phase and occupying each tenth site. (We start from coherent states because they correspond to the well defined classical initial conditions). TWA again clearly indicates very fast thermalization while the classical curve shows non ergodic behavior.

\subsection{Spin dynamics in a linearly changing magnetic field: multi-level Landau-Zener problem.}

In this section we will illustrate how phase space methods can be applied to another exactly solvable problem. Namely we will consider a motion of an arbitrary spin $S$ in a linearly changing magnetic field:
\be
\mathcal H=2 h_z(t) s^z+2 g s^x,
\label{ham_spin}
\ee
where $h_z(t)=\delta t$. We assume that the system is initially prepared in some state or superposition of states at $t=-t_0$ and we will be interested in finding expectation values of various observables at $t=t_0$, where $t_0$ is large so that $h_z(t_0)\gg g$.

As we discussed in Sec.~\ref{sec:spin} one can map the time evolution of noninteracting spins to the evolution of noninteracting bosons using the Schwinger representation. Therefore TWA is exact in this case. Using Eqs.~(\ref{schwing}) the Hamiltonian~(\ref{ham_spin}) becomes:
\be
\mathcal H=h_z(t)(\alpha^\dagger \alpha-\beta^\dagger \beta)+g(\alpha^\dagger \beta+\beta^\dagger \alpha).
\ee
The classical equations of motion corresponding to this Hamiltonian are
\beq
&&i{d\alpha\over dt}=\delta t \alpha + g\beta,\label{LZ1}\\
&&i{d\beta\over dt}=g\alpha-\delta t\beta.\label{LZ2}
\eeq
These equations should be supplemented by the initial conditions distributed according to the Wigner transform of the initial wave function (or the density matrix).

Note that Eqs.~(\ref{LZ1}) and (\ref{LZ2}) map exactly to the equations describing the conventional Landau-Zener problem. Then the evolution can be described by a unitary $2\times 2$ matrix:
\be
\alpha_{\infty}=T\alpha_0+R\mathrm e^{i\phi}\beta_0,\quad\beta_\infty=-R\mathrm e^{-i\phi} \alpha_0+T\beta_0,
\ee
where (see Refs.~\cite{kg, zener})
\be
T=\mathrm e^{-\pi\gamma},\; R=\sqrt{1-T^2},\;\phi=\gamma\left[\ln(\gamma)-1\right]-2\gamma\ln(\sqrt{2\delta}T),
\ee
and  $\gamma=g^2/(2\delta)$ is the Landau-Zener parameter.

Finally we need to reexpress different spin components at $t=\infty$ through the initial values of the spins. Thus we have
\begin{widetext}
\beq
&&s^z_\infty=(T^2-R^2){\alpha_0^\star\alpha_0-\beta_0^\star\beta_0\over 2}+\alpha_0^\star\beta_0  RT\mathrm e^{i\phi}+\alpha_0\beta_0^\star  RT\mathrm e^{-i\phi}\nonumber\\
&&~~~~=(T^2-R^2) s^z_0+2 RT\cos(\phi)s^x_0  -2  RT\sin(\phi)s_0^y,\\
&& s^x_\infty=-2RT\cos(\phi) s^z_0+(T^2-R^2\cos(2\phi))s^x_0+R^2\sin(2\phi)s^y_0,\nonumber\\
&& s^y_\infty=2RT\sin(\phi) s^z_0+R^2\sin(2\phi)s^x_0+(T^2+R^2\cos(2\phi))s^y_0.\nonumber
\eeq
\end{widetext}

Now using these expressions and the results of Sec.~\ref{sec:spin} we can compute expectation values of various quantities.  For simplicity we assume that initially the spin is prepared in the eigen state of $\hat s_z$. On the language of bosons this corresponds to the Fock state $|S-n,n\rangle$, where a particular value of $n$ corresponds to the state with $s_z=S-n$. Let us give a few specific examples:
\beq
&&\langle \hat{s}^z_\infty\rangle = (T^2-R^2) s^z_0\nonumber\\
&& \langle \left(\hat{s}^z_\infty\right)^2\rangle = \left[T^4+R^4-4T^2R^2\right] (s^z_0)^2+2T^2R^2s(s+1),\nonumber\\
&& \langle \hat{s}^z_\infty \hat{s}^x_\infty+\hat{s}^z_\infty \hat{s}^x_\infty\rangle=2RT(T^2-R^2)\cos(\phi)\left[s(s+1)-3\langle(\hat{s}^z_0)^2\rangle\right].
\eeq
Note that for the conventional Landau-Zener problem corresponding to the spin $s=1/2$ the last two equations become trivial: $\langle (\hat{s}^z_\infty)^2\rangle=1$ and $\langle \hat{s}^z_\infty \hat{s}^x_\infty+\hat{s}^z_\infty \hat{s}^x_\infty\rangle=0$. But for larger values of the spin these correlation functions are nontrivial with e.g. $\langle \hat{s}^z_\infty \hat{s}^x_\infty+\hat{s}^z_\infty \hat{s}^x_\infty\rangle$ being an oscillating function of the total time $t_0$ and the Landau-Zener parameter $\gamma$.

\subsection{Dicke model.}
\label{sec:dicke}

As a final illustration of the phase space approach to quantum dynamics we will consider the Dicke model (or more accurately the Jaynes-Cummings model), which maps to the bosonic mode interacting with a large spin:
\be
\hat{\mathcal H}=-\lambda \hat{b}^\dagger \hat{b}+{g\over\sqrt{2S}}\left(\hat{b}^\dagger\hat{S}^-+\hat{b}\hat{S}^+\right),
\label{hamilt_dicke}
\ee
where $\hat{b}$ is the bosonic annihilation operator and $\hat{S}^+,\hat{S}^-$ are the raising and lowering spin operators (see Ref.~\cite{agkp} for additional discussions). Here $S$ is the value of the spin and the classical limit is achieved at $S\to\infty$. We will specifically consider a setup where one starts at $\lambda$ large and negative in the state where bosonic mode is unoccupied and the spin is polarized along the z-axis. This problem was analyzed in Ref.~\cite{agkp} using various methods including TWA. Here we will repeat the details of TWA analysis to this problem because it illustrates the power of phase space methods to describe all aspects of quantum dynamics. This problem also
illustrates the possibility of mixing different phase space representations together, namely the coherent state phase space describing the bosonic mode and the $SO(3)$ vector describing the spin degree of freedom. This problem is particularly important also in view that using the Anderson mapping~\cite{anderson_spin} the Hamiltonian~(\ref{hamilt_dicke}) describes the system of degenerate fermions interacting with a bosonic mode~\cite{agkp}. Thus the dynamics in this case represents a simplified version of the dynamics through the BEC-BCS crossover in cold atoms (see E.g. Ref.~\cite{regal_thesis}). Under this mapping in the classical limit (which is equivalent to the mean-field approximation in the BCS model) the dynamics of the system can be fully mapped to the system of classical equations of motion analyzed in Refs.~\cite{bls, agr} for a more general case with the dispersion. TWA and corrections show the way to go beyond the purely classical limit and indicate that the whole approach of this paper can be extended to interacting fermionic systems at least in certain situations. We note that the model (\ref{hamilt_dicke}) also appears in various other contexts in optics, condensed matter and atomic physics.

We choose a specific process where $\lambda$ changes linearly in time: $\lambda=2\delta t$. Initially at $t\to-\infty$ there are no bosons and the spin points along the $z$ direction: $\langle \hat{S}^z\rangle=S$. As the time increases the bosonic mode becomes resonant with the spin forcing it to rotate. Eventually when time becomes large and positive the bosonic mode becomes off resonant again. For $S=1/2$ this problem is identical to the Landau-Zener problem. For larger values of the spin it becomes a particular generalization of the Landau-Zener problem to multiple energy levels. However, unlike in the previous example, this generalization results in an interacting quantum problem which has very nontrivial adiabatic ($\delta\to 0$) limit~\cite{ag, agkp}. Using Eqs.~(\ref{wig_spin_approx}) and (\ref{wig_bos_coh}) with $N=0$ we find that the Wigner distribution of the initial conditions for large $S$ can be well approximated by
\be
W(b^\star,b, S_x, S_y, S_z)\approx 2\mathrm e^{-2|b|^2} {1\over \pi S}\mathrm e^{-(S_x^2+S_y^2)/ S}\delta(S_z-S).
\ee
The classical equations of motion corresponding to the Hamiltonian (\ref{hamilt_dicke}) read:
\beq
&&i{\partial b\over\partial t}=-\lambda(t)b+{g\over\sqrt{2S}}S^-,\nonumber\\
&&{\partial {\bf S}\over\partial t}={2g\over \sqrt{2S}}{\bf B}\times {\bf S},
\label{eq_motion_dicke}
\eeq
where ${\bf B}=(\Re b, \Im b, 0)$ is the effective magnetic field acting on the spin.

\begin{figure}[ht]
\includegraphics[width=12cm]{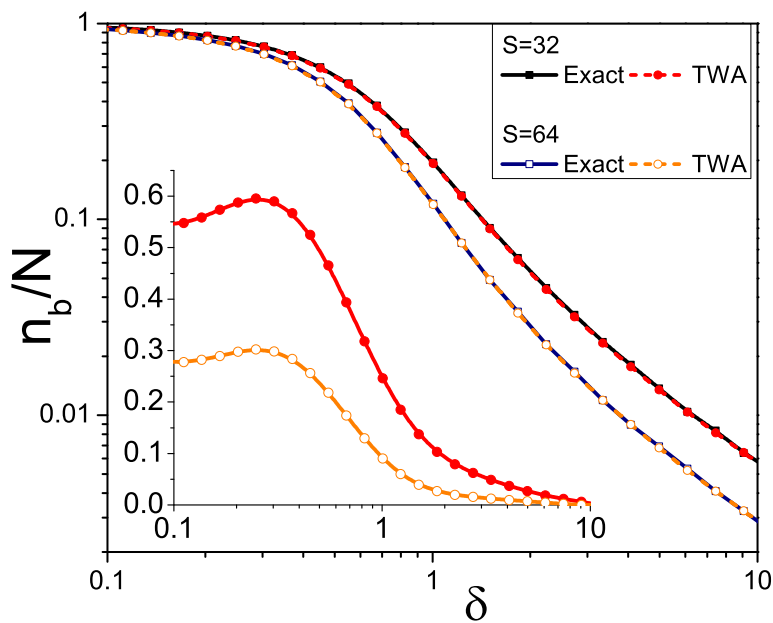}
\caption{Average relative number of bosons $n_b=\langle \hat{b}^\dagger \hat{b}\rangle/2S$ created during the process as a function of the rate of change of the chemical potential $\delta$. Initially the bosonic level is empty and the spin points in the positive $z$ direction. Shown are the semiclassical and exact results for two values of spin $S=32$ and $S=64$. The inset shows the difference between exact and semiclassical results multiplied by a factor of 100. The parameters of the model for the simulations are identical to those in Ref.~\cite{agkp}}
\label{fig:vg1}
\end{figure}

In the purely classical limit the initial conditions are $b=0, S_z=S, S_x=S_y=0$. It is easy to see that the solution $b(t)\equiv 0$, $S_z(t)\equiv S$ satisfies both the classical equations of motion and the initial conditions. It is thus necessary to include quantum fluctuations to see the nontrivial dynamics. In Ref.~\cite{agkp} it was shown that using conservation laws Eqs.~(\ref{eq_motion_dicke}) can be simplified further and be reduced to a single second order differential equation. The latter in turn can be approximately solved using either methods of adiabatic invariants~\cite{agkp} or more accurately using the exact solution of the Painlev\'e equation~\cite{itin}. This solution indicates a very nontrivial adiabatic limit for large values $S$. Since our goal here is to understand the validity of the TWA we will stop short of discussing details of this solution and focus on the comparison between the results of exact simulations and TWA. In Fig.~\ref{fig:vg1} we show the average relative number of created bosons $n_b=\langle \hat{b}^\dagger \hat{b}\rangle/2S$ during the process described above as a function of the rate $\delta$. Slow rates correspond to the adiabatic limit and thus to nearly $100\%$ conversion efficiency (nearly maximum possible number of bosons in the final state). We plot the exact and TWA results for two values of spin $S=32$ and $S=64$. The accuracy of the semiclassical calculation in the whole range of $\delta$ does not exceed $1\%$ for $S=32$ and is about a factor of two less for $S=64$ (see the inset in Fig.~\ref{fig:vg1}).

\begin{figure}[ht]
\includegraphics[width=12cm]{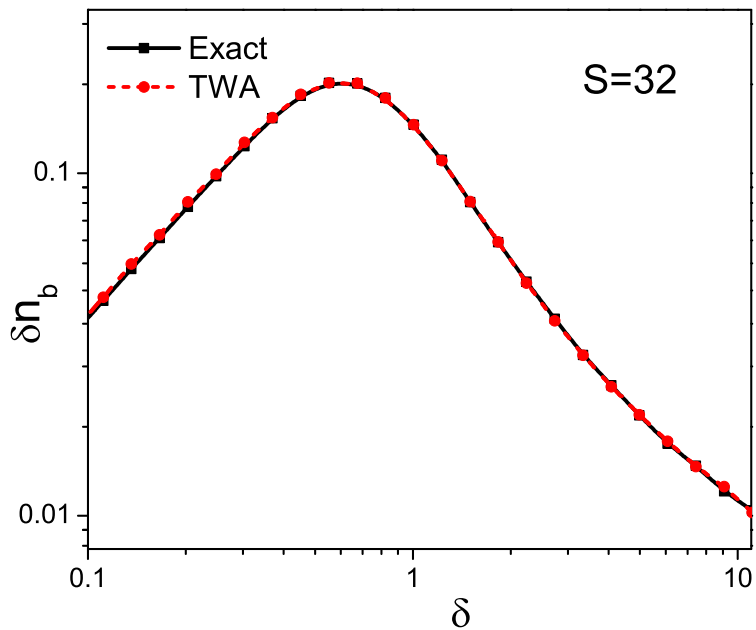}
\caption{Fluctuations of the relative number of bosons $\delta n_b=\sqrt{\langle (\hat{b}^\dagger \hat{b})^2\rangle-\langle \hat{b}^\dagger \hat{b}\rangle^2}/2S$ as a function of $\delta$. The parameters are the same as in Fig.~\ref{fig:vg1} with total spin $S=32$. This plot is taken from Ref.~\cite{agkp}.}
\label{fig:vg3}
\end{figure}

We can also check that the semiclassical approximation is accurate for other observables. For example in Fig.~\ref{fig:vg3} we plot fluctuations of the relative number of bosons at $t\to\infty$ as a function of $\delta$. The number fluctuations are small both in the limit of large $\delta$ where almost no bosons are created and in the limit of small $\delta$, where there is almost $100\%$ conversion. However, at intermediate rates the fluctuations are large, exceeding $50\%$ of the mean number of bosons. We see from the graph that again the agreement between the exact and semiclassical curves is excellent and we expect it to be better for larger values of $S$. Moreover, one can show that the whole distribution function of the boson occupation number, which interpolates between the exponential distribution for fast quenches and the Gumbel distribution at slow quenches, is accurately reproduced by TWA~\cite{agkp, kiegel}.

\section{Derivation of the results within the path integral formalism.}
\label{sec:der}

A convenient framework connecting quantum and classical dynamics is given by the Feynman's path integral representation of the time evolution. Then classical trajectories appear in the form of the saddle point of the action in the path integral~\cite{shankar}. As it is well known from the standard quantum mechanics the classical limit is naturally recovered in the Heisenberg representation, where operators corresponding to various observables depend on time, while the wave function (or the density matrix) only describes the initial state of the system~\cite{LL3}.  In this section we will show how combining path integrals and the Heisenberg representation reproduces all the results of the previous sections. The formalism here will be somewhat resembling the one used in the derivation of the Keldysh diagrammatic technique~\cite{kamenev, kamenev1, kamenev2}. However, there are important differences. Namely the Keldysh technique is usually derived in the Schr\"odinger representation and the elementary objects there are the single-particle Green's functions. Also the majority of applications of the Keldysh technique deal with steady state situations, where the initial conditions are unimportant~\cite{kamenev2}. While here we will work in the Heisenberg representation directly with the phase space variables and the initial conditions are of primary importance to us. The Heisenberg representation has another advantage over the Schr\"odinger representation that one does not have to calculate the density matrix, which is a tremendously complicated object in many-particle interacting systems containing exponentially large amount of information. Within the Heisenberg picture we aim to calculate expectation values of particular observables.

Partially the results of Sec.~\ref{sec:main} were obtained in literature earlier analyzing the von Neumann's equation for the density matrix in the Wigner form. We will briefly review this approach in the next section. The path integral formalism, in our opinion, gives a more natural and complete way to obtain these results. Besides these earlier approaches failed to reproduce the structure of quantum corrections and non-equal time correlation functions. Another important goal of this section is to show that the Wigner function, Weyl ordering, Bopp operators and other related concepts naturally emerge in the path integral representation of time evolution without the need to make any extra assumptions or justify usage of these concepts {\em a-posteriori}. To keep the discussion relatively short we will outline only the key steps in the derivations skipping unessential intermediate details.

\subsection{Coordinate-momentum representation.}

\subsubsection{Truncated Wigner Approximation.}

Let us start from a simple situation, where we are interested in the time evolution of the expectation value of some single-particle observable $\hat \Omega(\hat p,\hat x,t)$. According to general rules of quantum mechanics the latter is given by
\be
\langle \hat\Omega(\hat x,\hat p,t)\rangle={\rm Tr}\left[\hat\rho\, T_\tau\,\mathrm e^{i/\hbar\int_0^t \hat{\mathcal H}(\tau) d\tau} \hat\Omega(\hat x,\hat p,t)
e^{-i/\hbar\int_0^t \hat{\mathcal H}(\tau) d\tau} \right],
\label{eq:omega}
\ee
where $\rho$ is the (time independent) density matrix, $T_\tau$ stands for the Keldysh time ordering~\cite{kamenev}, where later moments in time always appear closer to the observable $\Omega$. Next we split the evolution operator into the product:
\be
T_\tau \mathrm e^{i/\hbar\int_0^t \hat{\mathcal H}(\tau) d\tau}\approx\prod_{i=1}^N \mathrm e^{i/\hbar  \hat{\mathcal H}(\tau_i) \Delta\tau},
\label{prod}
\ee
where $\Delta\tau=t/N$ and $\tau_i=i\Delta\tau$. Finally we insert the resolution of identity $|p_i\rangle \langle p_i|x_i\rangle \langle x_i|$ between all the terms in the product. At the end we will take the limit $N \to\infty$. We use the standard convention for the overlaps between the coordinate and momentum basis states:
\begin{displaymath}
\langle x_i|p_i\rangle=\exp[i p_i x_i/\hbar],\quad \langle p_i|x_i\rangle={1\over 2\pi\hbar}\exp[-i p_i x_i/\hbar].
\end{displaymath}
In this way we map the quantum evolution to the path integral over propagating classical phase space variables: $x_i\equiv x(\tau_i)$ and $p_i\equiv p(\tau_i)$. Because we have two exponents in the path integral we have to introduce two sets of phase space variables: ($x_{fi}$, $p_{fi}$) and ($x_{bi}$, $p_{bi}$), where indices $f$ and $b$ imply ``forward'' and ``backward''. It is intuitively clear (see also Refs.~\cite{kamenev, kamenev2, ap_twa}) that the symmetric combination of the phase space variables: $x_i=[x_{fi}+x_{bi}]/2$ and $p_i=[p_{fi}+ p_{bi}]/2$ correspond to the classical coordinate and momentum, while their differences $\xi_i=x_{fi}-x_{bi}$ and $ \eta_i=p_{fi}-p_{bi}$ describe quantum fluctuations. Indeed in the classical limit there is a unique trajectory corresponding to the fixed initial conditions, while quantum mechanically one has to sum over all possible trajectories. Our goal is thus to develop a systematic expansion of $\langle \hat\Omega(\hat x,\hat p,t)\rangle$ in $\xi_i$ and $\eta_i$, while treating classical variables $x_i$ and $p_i$ exactly. After inserting the complete basis and changing the variables to $x_i, \xi_i, p_i, \eta_i$ Eq.~(\ref{eq:omega}) assumes the form:
\beq
&&\langle \hat\Omega(\hat x,\hat p,t)\rangle=\int\int d x_0 d p_0 W_0(p_0,x_0)\int\int D x(\tau) D p(\tau) D  \xi(\tau) D \eta(\tau)\nonumber\\
 &&\exp\Biggl\{{i\over \hbar}\int_0^t d\tau \Biggl[\xi(\tau){\partial p(\tau)\over \partial\tau}-\eta(\tau)
{\partial x(\tau)\over\partial \tau}+\mathcal H_W\left(p(\tau)+{\eta(\tau)\over 2},x(\tau)+{\xi(\tau)\over 2},\tau\right)\nonumber\\
&&~~~~~~~~~~~~~~~~~~~~- \mathcal H_W\left(p(\tau)-{\eta(\tau)\over 2}, x(\tau)-{\xi(\tau)\over 2},\tau\right)\Biggr]\Biggr\}\;\Omega_{W}(x(t),p(t),t),
\label{eq:omega1}
\eeq
where $\mathcal H_W(p(\tau),x(\tau),\tau)$ is the Weyl symbol of the Hamiltonian $\hat{\mathcal H}(\hat x,\hat p,\tau)$. The derivation of this expression is straightforward and very similar to the one described in Ref.~\cite{ap_twa} in the coherent state basis. The expression in the brackets is nothing but the difference between forward ($S_f$) and backward ($S_b$) classical actions appearing in the path integral of the evolution operator~\cite{shankar}:
\be
S_{f,b}=i\int_0^t d\tau\; x_{f,b}(\tau){d p_{f,b}(\tau)\over d\tau}+\mathcal H(x_{f,b}(\tau), p_{f,b}(\tau),\tau).
\ee
There are two subtle points in this derivation. First, one needs to carefully follow the boundary terms at $\tau=0$ and $\tau=t$. This can be straightforwardly done by working with the discrete time steps in the functional integral and taking the continuum limit at the very end (see Ref.~\cite{ap_twa} for the analogous discussion in the coherent state basis). One can check that integrating over $\xi(0)$ and $\eta(0)$ yields the Wigner transform of the density matrix (given by Eq.~(\ref{wig_xp})), which is interpreted as the distribution function of the classical initial conditions $x_0$ and $p_0$. Integrating over $\xi(t)$ and $\eta(t)$ results in the emergence of the Weyl symbol of the operator $\hat{\Omega}({\hat x},{\hat p},t)$: $\Omega_{W}(x(t), p(t),t)$, which is given by Eq.~(\ref{weyl}). Remarkably both the Wigner function and the Weyl symbol appear automatically in the path integral. The second subtle point refers to the Weyl ordering of coordinates and momenta in the Hamiltonian in the exponent of Eq.~(\ref{eq:omega1}). For the Hamiltonians of the type of (\ref{hamilt}) which split into the sum of kinetic and potential energies, this issue is not important because the Weyl ordering does not produce any additional terms. However, in a more general situation, where the Hamiltonian can contain cross-products like $\hat x^2\hat p^2$ the issue of the correct ordering becomes relevant. Naively in this case one should use the ``normal'' ordering where $\hat x$ operators appear on the left of $\hat p$ operators because in this case $\langle x_i| \exp[i\hat{\mathcal H}(\hat x,\hat p,\tau_i)\Delta \tau]|p_i\rangle\approx \exp[i\mathcal H_n(x_i,p_{i+1},\tau_i)\Delta\tau]$, where $n$ emphasizes that quantum-classical correspondence is taken in the Hamiltonian written in the normal form. However, this turns out to be not accurate. The fact that the time indices in $x$ and $p$ in the Hamiltonian are not the same (or more precisely that the momentum $p$ is evaluated at a slightly later time than the coordinate $x$) introduces the correction to the classical Hamiltonian $\mathcal H_n$ modifying it to $\mathcal H_W$. We will discuss this issue in detail in the Appendix~\ref{appendix_hamilt_ordering}

Expression (\ref{eq:omega1}) has in fact very transparent interpretation. The first integral is taken over all possible initial values of the classical coordinate and momentum $x_0$ and $p_0$ weighted with the Wigner function. The latter thus plays the role of the probability distribution of the initial conditions. Next in Eq.~(\ref{eq:omega1}) there is a functional integral over classical and quantum fields with classical fields satisfying the initial conditions $x(t=0)=x_0$ and $p(t=0)=p_0$. And finally $\Omega_{W}(x(t), p(t),t)$ is the Weyl symbol of the operator $\hat{\Omega}({\hat x},{\hat p},t)$, which expectation value is computed. As written, Eq.~(\ref{eq:omega1}) is formally exact. However, to make use of this representation one needs to find an efficient way of computing the functional integral over all trajectories. In this work we are focusing in expanding this integral in quantum fluctuations near the classical limit. In principle one can consider other expansions, e.g. in the interaction strength. The action in Eq.~(\ref{eq:omega1}) can also have other non-classical saddle points, which potentially describe such phenomena as tunneling~\cite{kamenev2}. The latter can not be understood through the analytic expansion of the dynamics in powers of $\hbar$. We already argued that $\eta(\tau)$ and $\xi(\tau)$ are responsible for the quantum fluctuations, so it is useful to expand the integrand in Eq.~(\ref{eq:omega1}) in powers of $\eta$ and $\xi$. In the leading order we have
\beq
&&\mathcal H_W\left[p(\tau)+{\eta(\tau)\over 2},x(\tau)+{\xi(\tau)\over 2},\tau\right]- \mathcal H_W\left[p(\tau)-{\eta(\tau)\over 2},x(\tau)-{\xi(\tau)\over 2},\tau\right]\approx\nonumber\\
&&\approx \eta(\tau) {\partial \mathcal H_W(p,x,\tau)\over\partial p(\tau)}+\xi(\tau) {\partial \mathcal H_W(p,x,\tau)\over\partial x(\tau)}.
\label{taylor_1}
\eeq
The derivatives of the Hamiltonian with respect to canonical coordinate and momentum are just the classical Poisson brackets~\cite{LLI}:
\beq
&&{\partial \mathcal H_W(p,x,\tau)\over\partial x(\tau)}=\left\{\mathcal H_W(p,x,\tau), p\right\},\nonumber\\
&&{\partial \mathcal H_W(p,x,\tau)\over\partial p(\tau)}=-\left\{\mathcal H_W(p,x,\tau), x\right\}.
\label{classical_eqs}
\eeq

If we stop at this order of the expansion in $\eta(\tau)$ and $\xi(\tau)$ then the functional integral in Eq.~(\ref{eq:omega1}) is trivial. Indeed, integrating out $\eta(\tau)$ and $\xi(\tau)$ gives the $\delta$-function constraint on $p(\tau)$ and $x(\tau)$:
\be
{dx\over d\tau}=\left\{x,\mathcal H_W\right\},\; {d p\over d\tau}=\left\{p, \mathcal H_W\right\}.
\ee
Thus taking the functional integral over the classical fields $x(\tau)$ and $p(\tau)$ becomes equivalent to simply solving the Hamiltonian equations of motion. Only the integral over the initial conditions remains. So in the leading order in quantum fluctuations we recover the semiclassical (truncated Wigner) approximation given by Eq.~(\ref{eq:omega2}). Straightforward generalization of this derivation to many-particle system immediately gives the same result with the only difference that now integration is taken over coordinates and momenta of all particles in the system.

\subsubsection{Nonequal time correlation functions.}
\label{sec:neq_time_1}

Above we described how one can determine the expectation value of a single-time observable $\hat\Omega(\hat x, \hat p, t)$. We saw that one needs to use the Weyl symbol for the observable, which corresponds to first symmetrically ordering operators $\hat x$ and $\hat p$ and then substituting quantum operators by the corresponding classical variables. Now we will be interested in a more general case of finding the expectation value of the product of two (or more) observables evaluated at different times:
\be
\langle \hat{\Omega}_1(\hat{x},\hat{p},t_1) \hat{\Omega}_2(\hat{x},\hat{p},t_2)\rangle.
\label{non_eq_time_1}
\ee
 Let us first assume that $t_1\leq t_2$. Later we will generalize our results to multiple time correlation functions and other types of ordering of the operators. In the coherent state picture the non-equal time correlation functions were analyzed in Refs.~\cite{plimak1, plimak_new} using related but somewhat different approach.

 Before proceeding with a formal derivation let us discuss where the potential complications come from. Classically the expectation value~(\ref{non_eq_time_1}) can be understood as average of the result of a joint measurement of first the quantity $\Omega_1$ and then $\Omega_2$. At the classical level (the ideal) measurement itself does not have any effect on the consequent dynamics thus one can write
\beq
\overline{ \Omega_1(x,p,t_1) \Omega_2(x,p,t_2)}&=&\int\int dx_0 dp_0\,W_{cl}(x_0,p_0) \Omega_1(x(t_1),p(t_1),t_1)\Omega_2(x(t_2),p(t_2),t_2)\nonumber\\
&=&\int\int d\omega_1 d\omega_2\, \omega_1 \omega_2 P(\omega_1,t_1) P(\omega_2, t_2|\omega_1,t_1),
\label{neq_time_corr_cl}
\eeq
where we use overline for the classical averaging to distinguish it from the quantum expectation value,
\be
P(\omega_1,t_1)=\int \int dx_0 dp_0\,W_{cl}(x_0, p_0)\Omega_1(x(t_1),p(t_1),t_1)\delta(\omega_1- \Omega_1(x(t_1),p(t_1),t_1))
\ee
is the probability that at the time $t_1$ the measurement of $\Omega_1$ will give the value $\omega_1$. Similarly $P(\omega_2, t_2|\omega_1, t_1)$ is the conditional probability that the measurement of $\Omega_2$ at the moment $t_2$
will give the value $\omega_2$ given the measurement of $\Omega_1$ at $t_1$ gave the value $\omega_1$. Equation~(\ref{neq_time_corr_cl}) can be interpreted that the result of the measurement of $\Omega_1$ simply restricts the possible initial conditions $x_0$ and $p_0$ to those which result in $\Omega_1(x(t_1),p(t_1),t_1)=\omega_1$ but does not affect the consequent dynamics. In quantum mechanics the situation is more complicated. Indeed the very fact of measurement of $\hat\Omega_1$ has an effect on the consequent evolution and thus on the measurement of $\hat \Omega_2$. Mathematically it should be manifested in some form of back action at moment $t_1$, which affects the consequent time evolution. The other important issue, which differentiates between quantum and classical multi-time measurements is the correct ordering of operators in the corresponding correlation functions. This discussion goes beyond the scope of this paper; it is available in literature for various specific situations (see e.g. Refs.~\cite{nazarov, aash_rmp}). We will just discuss how one can compute the correlation functions of the type~(\ref{non_eq_time_1}) within the phase space representation without connecting them to actual measurement procedures.

According to Eq.~(\ref{weyl}) the Weyl symbol corresponding to the product of two operators is formally defined as:
\beq
&&\left(\Omega_1(t_1)\Omega_2(t_2)\right)_W= \int {d\xi d\eta\over 4\pi} \biggl<x(t_1)+\xi/2\biggl|\hat\Omega_1(\hat{x},\hat{p},t_1)T_\tau \mathrm e^{i/\hbar\int_{t_1}^{t_2} H(\tau) d\tau}\nonumber\\
&&\times\hat\Omega_2(\hat{x},\hat{p},t_2) \mathrm e^{-i/\hbar\int_{t_1}^{t_2} H(\tau) d\tau} \biggr|p(t_1)-\eta/2\biggr>\exp\left[-{i\over \hbar}\left(p(t_1)x(t_1)+{\xi p(t_1)-\eta x(t_1)\over 2}\right)\right].
\label{non_eq_time_2}
\eeq
It is easy to check that if $\Omega_2\equiv 1$ this expression indeed becomes equivalent to Eq.~(\ref{weyl}). The equation (\ref{non_eq_time_2}) can be further analyzed similarly to Eq.~(\ref{eq:omega}). The result of this analysis is that at the moment $t_1$ the phase space variables $x(t)$ and $p(t)$ become discontinuous acquiring jumps $\delta x_1$ and $\delta p_1$ with the probability distribution, which depends on details of the operator $\hat\Omega_1$. If the latter is written in the ordered form where all $\hat {x}$ operators appear on the left of $\hat{p}$ operators the probability distribution of these jumps assumes the following form:
\be
W_{\Omega_1}(\delta x_1,\delta p_1)=\int\int {d\xi d\eta\over (2\pi)^2}\, \mathrm e^{2i\delta p_1\delta x_1/\hbar}\mathrm e^{i\xi\delta p_1/\hbar+i\eta\delta x_1/\hbar}\Omega_1(x(t_1)+\xi/2,  p(t_1)-\eta/2),
\label{prob_omega_1}
\ee
where we put the subindex $\Omega_1$ to highlight that these jumps depend on the form of the operator $\hat\Omega_1$. If $\hat\Omega_1$ is the identity operator then the integral becomes a product of two $\delta$-functions enforcing $\delta x_1$ and $\delta p_1$ to be equal to zero. Then there is no jump in either $x(t_1)$ or $p(t_1)$ as it should be. If the operator $\hat\Omega$ is written as a product of powers of $\hat x$ and $\hat p$: $\hat\Omega_{nm}=\hat{x}^n \hat{p}^m$, then the probability of the corresponding jump can be found using the generating function:
\be
R(\alpha,\beta,\delta x,\delta p)=\int\int {d\xi d\eta\over 2\pi}\, \mathrm e^{-2i\alpha (x+\xi/2)+2i\beta(p-\eta/2)}\mathrm e^{2i\delta p\delta x/\hbar}\mathrm e^{i\xi\delta p/\hbar+i\eta\delta x/\hbar}.
\label{gen_funct1}
\ee
To shorten notations we dropped explicit time indices for phase space variables $x$ and $p$. Then
\be
W_{n,m}(\delta x,\delta p)={i^{n-m}\over 2^{n+m}}\partial^n_\alpha\partial^m_\beta R(\alpha,\beta,\delta x,\delta p)\bigg|_{\alpha,\beta=0} .
\ee

The integral in Eq.~(\ref{gen_funct1}) can be easily evaluated and we find
\be
R(\alpha,\beta,\delta x,\delta p)=\mathrm e^{-2i\alpha x+2i\beta p+2i\delta x\delta p/\hbar}\delta(\delta p-\alpha\hbar)\delta(\delta x-\beta\hbar).
\ee
There are two particularly important cases $n=1, m=0$ corresponding to the coordinate operator ($\hat\Omega_{1,0}=\hat x$) and $n=0, m=1$ corresponding to the momentum operator ($\hat\Omega_{0,1}=\hat p$). In the former case we find
\be
W_{1,0}(\delta x,\delta p)=\mathrm e^{2i\delta x\delta p/\hbar}\left(x-{i\hbar\over 2}{\partial\over\partial\delta p} \right)\delta(\delta p)\delta(\delta x).
\label{W10}
\ee
Using standard rules of manipulating with $\delta$-functions and noting that the derivative acting on the first exponent does not give any contribution we find
\be
W_{1,0}(\delta x,\delta p)=\delta(\delta p)\delta(\delta x)\left(x+{i\hbar\over 2}{\partial\over \partial \delta p}\right).
\ee
This expression is in turn equivalent to the representation (\ref{omega_r}):
\be
\hat x(t_1)\to x(t_1)+{i\hbar \over 2}{\partial \over\partial \delta p(t_1)}.
\label{omega_r1}
\ee
From Eq.~(\ref{W10}) it is clear that the derivative here is understood as the response of the operator $\hat \Omega_2$ to the infinitesimal jump $p(t_1)\to p(t_1)+\delta p(t_1)$. In Eq.~(\ref{omega_r}) we used $\partial/\partial p$ instead $\partial/\partial \delta p$ to shorten notations. In the same way we find:
\be
W_{0,1}(\delta x,\delta p)=\delta(\delta p)\delta(\delta x)\left(p-{i\hbar\over 2}{\partial\over \partial \delta x}\right)
\ee
leading to the substitution (\ref{omega_p}):
\be
\hat p(t_1)\to p(t_1)-{i\hbar \over 2}{\partial \over\partial \delta x(t_1)},
\label{omega_p1}
\ee
In the same spirit one can obtain the expression for the product $\hat x\hat p$:
\be
W_{1,1}(\delta x,\delta p)=\mathrm e^{2i\delta x\delta p/\hbar}\left(x-{i\hbar\over 2}{\partial\over\partial\delta p} \right)\left(p-{i\hbar\over 2}{\partial\over\partial\delta x} \right)\delta(\delta p)\delta(\delta x)
\ee
Now there is an additional subtlety that integrating $\delta$-functions by parts we will get a nonzero contribution from the exponent coming from taking the double derivative with respect to $\delta x$ and $\delta p$ so the result is~(\ref{omega_rp}):
\be
\hat{x}\hat{p}\to x p+{i\hbar\over 2}-{i\hbar\over 2} x{\partial\over\partial\delta x}+{i\hbar\over 2} p{\partial\over\partial\delta p}+{\hbar^2\over 4}{\partial^2\over\partial\delta x\partial\delta p}.
\label{omega_rp1}
\ee

In a similar fashion one can obtain representation of higher powers of $\hat x$ and $\hat p$. However, there is a much simpler way to do this using just Eqs.~(\ref{omega_r1}) and (\ref{omega_p1}). Note that in the derivation of the representation of $\hat\Omega_1$ we never used any information about the second operator $\hat\Omega_2$. In particular, the same representation will be true if we have product of arbitrary number of operators ordered such that later time appears on the right e.g. $\hat \Omega_1(t_1) \hat \Omega_2(t_2)\hat \Omega_3(t_3)$ with $t_1\leq t_2\leq t_3$ and so on.One simply needs to understand derivatives appearing in Eqs.~(\ref{omega_r1}), (\ref{omega_p1}), and (\ref{omega_rp1}) as acting on all operators appearing later in time (or on the right) of $\hat\Omega_1$. Then we can recover expression (\ref{omega_rp1}) from Eqs.~(\ref{omega_r1}) and (\ref{omega_p1}) noting that
\be
\hat x(t_1)\hat p(t_1) =\hat x(t_1)\hat p(t_{1+})\to \left(x(t_1)+{i\hbar \over 2}{\partial \over\partial \delta p(t_1)}\right)\left(p(t_{1+})-{i\hbar \over 2}{\partial \over\partial \delta x(t_{1+})}\right),
\ee
where $t_{1+}$ stands for time approaching to $t_1$ from above. Moving the derivative with respect to $\delta p$ to the right and noting that we will pick up additional constant term $i\hbar/2$ we recover Eq.~(\ref{omega_rp1}). Similarly one can get the representation of other products of operators. E.g. $\hat p\,\hat x=\hat x \hat p -i\hbar$. So the representation of $\hat p\,\hat x$ is given by Eq.~(\ref{omega_rp1}) with the negative sign in front of $i\hbar/2$. The same result is also recovered by multiplying Eqs.~(\ref{omega_p1}) and (\ref{omega_r1}) in this order. We want to emphasize again that the same rule applies even in the case when $\hat \Omega_2$ is the identity operator. Then all derivative terms in Eqs.~(\ref{omega_r1}), (\ref{omega_p1}) and (\ref{omega_rp1}) drop and as a result we recover the Weyl symbol of the operator $\hat \Omega_1$. This can be indeed checked by comparing the results of integration in Eq.~(\ref{weyl}) and applying Eqs.~(\ref{omega_r1}) and (\ref{omega_rp1}) to specific operators. Thus for equal time correlation functions we just recover Bopp representation of the coordinate and the momentum which automatically reproduce Weyl symbols of arbitrary operators as we discussed in Sec.~\ref{sec:Moyal} (see also Refs.~\cite{hillery, bopp, kubo}). It is notable that the Bopp representation automatically emerges from the path integral with the interpretation of the derivative terms as a response of future observables to the quantum jumps.

 In analyzing the correlation function~(\ref{non_eq_time_1}) we actually never used the fact that $t_1<t_2$: the time index in the path integral can increase or decrease since it is only the label for different phase space variables. So the correspondence (\ref{omega_r}) and (\ref{omega_p}) works if we are interested in correlation functions with the opposite ordering of times to (\ref{non_eq_time_1}), namely
\be
\langle \hat{\Omega}_2(\hat{x},\hat{p},t_2) \hat{\Omega}_1(\hat{x},\hat{p},t_1)\rangle.
\label{non_eq_time_3}
 \ee
It is convenient to assume again that $t_1<t_2$ but change the actual order of the operators so that the earlier one appears on the right. In this case using previous results one can reach the time $t_2$ and then simply propagate the equations of motion backwards in time until the moment $t_1$. However, this representation is clearly not casual: quantum jumps at the later time $t_2$ affect the observable $\Omega_1$ evaluated at an earlier time $t_1$. It turns out that the casuality can be restored by rewriting Eq.~(\ref{non_eq_time_2}) in the following form
\beq
&&\left(\Omega_2(t_2)\Omega_1(t_1)\right)_W= \int {d\xi d\eta\over 4\pi} \biggl<x(t_1)+\xi/2\biggl|T_\tau \mathrm e^{i/\hbar\int_{t_1}^{t_2} H(\tau) d\tau} \hat\Omega_2(\hat{x},\hat{p},t_2)\mathrm e^{-i/\hbar\int_{t_1}^{t_2} H(\tau) d\tau}\nonumber\\
&&\times\hat\Omega_1(\hat{x},\hat{p},t_1)  \biggr|p(t_1)-\eta/2\biggr>\mathrm e^{-ip(t_1)x(t_1)/\hbar-i\xi p(t_1)/(2\hbar)+i\eta x(t_1)/(2\hbar)}.
\eeq
This equation can be analyzed similarly to Eq.~(\ref{non_eq_time_2}) and the result is identical to (\ref{prob_omega_1}) with the only change that $\delta x_1, \delta p_1\to -\delta x_1,-\delta p_1$. In turn this leads to the change in sign in the derivatives in the representations (\ref{omega_r1}) and (\ref{omega_p1}). This change in sign is most easily interpreted as a result of simple calculus. Indeed the derivative with respect to $\delta x_1$ and $\delta p_1$ now acts on the operator $\Omega_2$, which is on the left of $\Omega_1$. So Eqs.~(\ref{omega_r1}) and (\ref{omega_p1}) have to be integrated by parts. We can combine the result for both types of orderings using the notations of left and right derivatives as in Eqs.~(\ref{omega_r}) and (\ref{omega_p}). The choice of the left or right derivative is arbitrary. However, it is unique if we want to preserve the casuality of our description. If earlier time appears on the left like in Eq.~(\ref{non_eq_time_1}) then we should use right derivatives in the representation of $\hat\Omega_1$ and conversely in the ordering like in (\ref{non_eq_time_3}) we have to use left derivatives.

 These results can be further generalized to multi-time correlation functions. At each moment of time we use one of the representations (\ref{omega_r}) or (\ref{omega_p}) depending on the convenience. As we saw in the case of the two-time correlation functions the choice of the representation is not unique. However, it becomes unique if we insist on the casuality. In general there are two special types of correlation functions (i) those which do not require any jumps so that one can use a simple quantum classical correspondence $\hat x\to x$, $\hat p\to p$ and (ii) correlations which have casual representation. The first type of correlations can be directly calculated in TWA. The second type of correlations is very important, because only such correlations should appear in physical observables if we insist on casuality of the phase space representation of the quantum evolution, i.e. on that this representation is physical.

The first type of correlations correspond to special, time symmetric, ordering of coordinate and momentum operators, which was first introduced in Refs.~\cite{plimak1, plimak2} under a somewhat different name and later used in Ref.~\cite{plimak_new}. Let us first point that for two operators the time-symmetric ordering simply coincides with the symmetric ordering. Indeed if we are interested in the correlation functions of the type
\be
\langle \hat x(t_1)\hat x(t_2)+ \hat x(t_2)\hat x(t_1)\rangle,
\ee
where we again assume $t_1<t_2$, then according to Eq.~(\ref{omega_r}) we should use the following substitution
\beq
&&\langle \hat x(t_1)\hat x(t_2)+ \hat x(t_2)\hat x(t_1)\rangle\to \left(x(t_1)+{i\hbar \over 2}{\overrightarrow\partial\over \partial p(t_1)}\right) x(t_2)+x(t_2)\left(x(t_1)-{i\hbar \over 2}{\overleftarrow\partial\over \partial p(t_1)}\right)=\nonumber\\
&&= \left(x(t_1)+{i\hbar \over 2}{\partial\over \partial p(t_1)}\right) x(t_2)+\left(x(t_1)-{i\hbar \over 2}{\partial\over \partial p(t_1)}\right)x(t_2)=2x(t_1) x(t_2).
\eeq
Similarly one can check that the symmetric combination of $\hat x$ and $\hat p$ operators does not acquire any jumps:
\be
\hat x(t_1)\hat p(t_2)+\hat p(t_2)\hat x(t_1)\to 2 x(t_1) p(t_1).
\ee
This result can be straightforwardly generalized to larger number of operators noting that
\be
\hat x(t_1) \hat A +\hat A^T \hat x(t_1) \to x(t_1) (\hat A+\hat A^T),
\label{time_symm_1}
\ee
where $\hat A$ is an arbitrary combination of $\hat x(t_j)$ and $\hat p(t_j)$, $\hat A^T$ is the transposed operator where the same combination of operators $\hat x(t_j)$ and $\hat p(t_j)$ appears in the reverse order. E.g. if $\hat A=\hat x(t_2)\hat p(t_3)$ then $\hat A^T=\hat p(t_3)\hat x(t_2)$. Equation (\ref{time_symm_1}) is still true if instead of $\hat x(t_1)$ we will use $\hat p(t_1)$. To avoid jumps appearing in the representation of $\hat A$, the latter should have similar structure i.e. $\hat A=\hat x(t_2) \hat B +\hat B^T\hat x(t_2)$ or $\hat A=\hat p(t_2) \hat B +\hat B^T\hat p(t_2)$. The operator $\hat B$ should again have this form and so on. Following Refs.~{\cite{plimak1, plimak_new}} we can define time-symmetric ordering symbol according to the following recursion relation:
\be
\mathcal T [\hat \xi(t_1)\hat \xi(t_2),\dots \hat\xi(t_n)]=
{1\over 2}\left[\hat \xi(t_1) \mathcal T[\hat \xi(t_2),\dots, \hat \xi(t_n)]+\mathcal T[\hat \xi(t_2),\dots, \hat \xi(t_n)]\hat\xi (t_1)\right],
\label{time_symm_2}
\ee
where $\hat \xi$ stands for either $\hat x$ or $\hat p$. We would like to emphasize that in Ref.~\cite{plimak_new} time-symmetric ordering was defined with the additional requirement that $t_1\leq t_2\leq \dots, t_n$. As we will show next this enforces the casuality of the time evolution. Note that even though the operators $\hat x(t)$ and $\hat p(t)$ map to numbers $x(t)$ and $p(t)$ in the time-symmetrically ordered operators for arbitrary order of times $t_1,\dots, t_n$, one can not still reorder them i.e. the average of $x(t_1) x(t_2) x(t_3)$ is not equivalent to the average of $x(t_2) x(t_1) x(t_3)$. The reason is that the full representation of quantum evolution (beyond TWA) contains quantum jumps, which are sensitive to the ordering of $x(t_j)$. Only within the semiclassical (TWA) approximation are such permutations allowed because for each initial condition the classical trajectory is unique and one can arbitrarily reorder classical numbers.

Now suppose that we are interested in a non-symmetrized correlation function of the type
\be
\langle\hat \xi(t_1)\hat \xi(t_2)\dots\hat \xi(t_n)\rangle
\label{corr_func}
\ee
We want to understand for which orderings of time one can represent this correlator through quantum jumps in a casual way, i.e. in a way that jumps in $x$ or $p$ at moments $t_\alpha$ only affect observables appearing at later moments in time. We saw above that if there are only two times involved then such representation is always possible. However, if we have a multi-time correlation function this is no longer true. Indeed let us consider a particular situation of a three time correlation function, e.g.
\be
\langle\hat x(t_1)\hat x(t_2)\hat x(t_3)\rangle
\ee
Then for $t_1\leq t_2\leq t_3$ there is an obvious casual representation where we use right derivatives in representation of both $\hat x(t_1)$ and $\hat x(t_2)$:
\be
\hat x(t_1)\hat x(t_2)\hat x(t_3)\to \left(x(t_1)+{i\hbar \over 2}{\partial\over \partial p(t_1)}\right)\left(x(t_2)+{i\hbar \over 2}{\partial\over \partial p(t_2)}\right)x(t_3).
\ee
We note that in the leading order in $\hbar$ the jumps in the momentum at $t_1$ and $t_2$ are additive. In the next order in $\hbar^2$ there is a double jump appearing in the form of the second order response of $x(t_3)$ to $p(t_1)\to p(t_1)+\delta p(t_1)$ and $p(t_2)\to p(t_2)+\delta p(t_2)$. Similarly for the opposite ordering of times $t_3\leq t_2\leq t_1$ one should use left derivatives resulting in
\be
\hat x(t_1)\hat x(t_2)\hat x(t_3)\to \left(x(t_3)-{i\hbar \over 2}{\partial\over \partial p(t_3)}\right)\left(x(t_2)-{i\hbar \over 2}{\partial\over \partial p(t_2)}\right)x(t_1),
\ee
where we reordered times in the casual way and removed arrows over the derivatives.

There are two additional orderings for which there is a casual representation where one has to use combination of left and right derivatives. Thus if $t_1\leq t_3\leq t_2$ we should use
\be
\hat x(t_1)\hat x(t_2)\hat x(t_3)\to \left(x(t_1)+{i\hbar \over 2}{\partial\over \partial p(t_1)}\right)\left(x(t_3)-{i\hbar \over 2}{\partial\over \partial p(t_3)}\right)x(t_2),
\ee
and finally for $t_3\leq t_1\leq t_2$ the causal representation gives
\be
\hat x(t_1)\hat x(t_2)\hat x(t_3)\to \left(x(t_3)-{i\hbar \over 2}{\partial\over \partial p(t_3)}\right)\left(x(t_1)+{i\hbar \over 2}{\partial\over \partial p(t_1)}\right)x(t_2),
\ee

Note that the two remaining orderings where $t_2$ is the earliest time do not have any casual representation. Indeed whether we use the right derivative for $\hat x(t_1)$ or left derivative for $\hat x(t_3)$ we will necessarily have to calculate the response of $\hat x(t_2)$ to the jump occurring at a later time.

It is now straightforward to generalize this discussion to arbitrary multi-time correlation functions of type (\ref{corr_func}). Casual representation exists only if the earlier time is never sandwiched between two later times. It is clear that there are $2^n$ orderings of times $t_1, t_2,\dots t_n$ satisfying this condition of casuality (compared to $n!$ of total number of orderings). Indeed to avoid sandwiching the earliest time should appear either first or last in the sequence so there are only two distinct possibilities: either $t_1$ or $t_n$ is the earliest time. Suppose that it is $t_1$. Then there are only two possibilities for either $t_2$ or $t_{n}$ being the next earliest time (if $t_n$ is the earliest time then either $t_1$ or $t_{n-1}$ should be the next earliest time) and so on. Interestingly only the orderings of correlation functions allowed by the casuality (so called Schwinger-Keldysh orderings) appear in the quantum theory of measurements~\cite{nazarov, aash_rmp} (see e.g. Eqs. (22) and (23) in Ref.~\cite{nazarov}). This suggests that there are potentially deep connections between casuality of the representation of the multi-time correlation function in the phase space and the possibility of measuring them. I.e. it is likely that only casual correlation functions are physical.

\subsubsection{Beyond truncated Wigner approximation.}

So far treating the exact expression~(\ref{eq:omega1}) for the evolution we considered only the leading linear terms in quantum fields $\xi(\tau)$ and $\eta(\tau)$ (see Eq.~(\ref{taylor_1})). These terms resulted in the semiclassical (truncated Wigner) approximation. To get further quantum corrections one needs to include higher powers in the Taylor expansion of the Hamiltonian in quantum fields. For the quadratic (harmonic) Hamiltonians the higher order terms are absent and thus Eq.~(\ref{taylor_1}) and TWA are exact. For conventional non-relativistic Hamiltonians (\ref{hamilt}) the kinetic energy is quadratic in $p$. Therefore higher order powers of quantum fields come only from the expansion in $\xi$. If one is interested in e.g. quantum corrections to the time evolution of a relativistic system, where the kinetic energy is of the form $\sqrt{m^2+p^2}$, then it is necessary to consider higher order terms in the field $\eta$ in a similar fashion. Expanding Eq.~(\ref{taylor_1}) to the third order in $\bm\xi$ and substituting this expansion into Eq.~(\ref{eq:omega1}) we find
\beq
&&\langle \hat{\Omega}(\hat{x},\hat{p},t)\rangle\approx \int\int d x_0 d p_0 W_0(x_0,p_0)\int\int D x(\tau) D  \xi(\tau)\nonumber\\
 &&\exp\Biggl\{{i\over \hbar}\int_0^t d\tau \Biggl[\xi(\tau){\partial p(\tau)\over \partial\tau}+\xi(\tau) {\partial H\over \partial x(\tau)}+{1\over 24}\xi^3(\tau) {\partial^3 V(x)\over \partial^3 x(\tau)}\Biggr]\Biggr\}\;\Omega_{W}(x(t),p(t),t).
\label{taylor_2}
\eeq
For the Hamiltonian (\ref{hamilt}) the momentum $p(\tau)$ still satisfies the classical Hamiltonian equation of motion $p(\tau)=m \,\dot{x}(\tau)$ (because there are no cubic terms in $\eta$) and thus the functional integral in Eq.~(\ref{taylor_2}) is taken only over coordinates. This integral can not be evaluated exactly. However, one can find it perturbatively by expanding the exponent containing the third power of $\xi$ into the Taylor series:
\be
\exp\left({i\over 24\hbar}\int_0^t d\tau \xi^3(\tau){\partial^3 V(x)\over \partial x(\tau)^3}\right)\approx 1+{i\over 24\hbar}\int_0^t d\tau\xi^3(\tau){\partial^3 V(x)\over \partial x(\tau)^3}+\dots
\label{taylor_3}
\ee
Combining Eqs.~(\ref{taylor_2}) and (\ref{taylor_3}) and noting that
\begin{displaymath}
\int {dx\over 2\pi} x^3 \exp(i\alpha x)={1\over i^3}{\partial^3\over
\partial \alpha^3}\left[\delta(\alpha)\right]=-{1\over i^3}\delta(\alpha){\partial^3\over
\partial \alpha^3}
\end{displaymath}
we find
\be
\langle \hat{\Omega}(\hat(x),\hat{p},t)\rangle\approx \int\int d x_0
d p_0 W_0(p_0,x_0)\left( 1-\int_0^t d\tau {i\over
24\hbar}{\hbar^3\over i^3}{\partial^3 V(x)\over \partial x(\tau)^3}
{\partial^3\over\partial \delta p(\tau)^3}\right)
\Omega_{W}(x(t),p(t),t).
\label{corr1}
\ee

This expression is understood in the following sense. At some moment $\tau$ the momentum $p$ undergoes an infinitesimal quantum jump $p(\tau)\to p(\tau)+\delta p(\tau)$. Then the evolution proceeds according to the classical equations of motion. At the end one evaluates the nonlinear response of the observable to this quantum jump. Note that this jump indeed represents the quantum correction since it is accompanied by a factor of $\hbar^2$. It is straightforward to generalize Eq.~(\ref{corr1}) to include higher order terms in $\hbar$. In particular, one can either have more quantum jumps occurring at different moments of time or one can have higher order quantum jumps corresponding to higher order terms in the Taylor expansion of Eq.~(\ref{taylor_1}). For example up to the fourth order in $\hbar$ one recovers Eq.~(\ref{corr2})

We can observe that formally Eq.~(\ref{corr1}) can be rewritten through the quantum diffusion (see Eq.~(\ref{diff_6}), which we repeat here for completeness):
\beq
&&\langle \hat{\Omega}(\hat{x},\hat{p},t)\rangle\approx \int\int d x_0
d p_0 W_0(x_0,p_0)\nonumber\\
&&\Biggl\{\Omega_{W}(x(t),p(t),t)+{\hbar^2\over 4}\sum_{\tau_i}{\partial^3 V(x_i)\over \partial x_i^3} \int
d\xi F_3(\xi)\Omega_{W}\left[x(t),p(t), t, \delta p_{i}=\xi
\sqrt[3]{\Delta\tau}\right]\Biggr\},
\label{diff_6a}
\eeq
where $\tau_i=i t/N$, $\Delta\tau=t/N$, $x_i, p_i=x(\tau_i), p(\tau_i)$. Formally we need to take the limit $N\to\infty$ or $\Delta\tau\to 0$. Equations~(\ref{corr1}) and (\ref{diff_6a}) are equivalent if the function $F_3(\xi)$, which can be interpreted as the probability distribution for the jumps, satisfies the following conditions:
\be
\int_{-\infty}^\infty F_3(\xi)d\xi=0, \; \int_{-\infty}^\infty \xi F_3(\xi)d\xi=0,\;\int_{-\infty}^\infty \xi^2 F_3(\xi)d\xi=0,\; \int_{-\infty}^\infty \xi^3 F_3(\xi)d\xi=1.
\label{moments1}
\ee
The equivalence between the two representations of the correction (\ref{corr1}) and (\ref{diff_6a}) can be seen by expanding the observable $\Omega_{W}$ in Eq.~(\ref{diff_6a}) into Taylor series in $\delta p_i$:
\beq
&&\Omega_{W}\left[x(t),p(t), t, \delta p_{i}=\xi
\sqrt[3]{\Delta\tau_i}\right]\approx \Omega_{W}\left[x(t),p(t), t\right]+ {\partial \Omega_{W}\left[x(t),p(t), t,\delta p_i\right]\over\partial \delta p_i}\xi\sqrt[3]{\Delta \tau}\nonumber\\
&&+ {1\over 2}{\partial^2 \Omega_{W}\left[x(t),p(t), t,\delta p_i\right]\over\partial \delta p_i^2}\xi^2\sqrt[3]{\Delta \tau^2}+ {1\over 6}{\partial^3 \Omega_{W}\left[x(t),p(t), t,\delta p_i\right]\over\partial \delta p_i^3}\xi^3\Delta \tau+O(\Delta\tau^{4/3}).
\label{taylor_4}
\eeq
Here all the derivatives are evaluated at $\delta p_i=0$. Now let us observe that the first three terms in this equation vanish upon integrating over $\xi$ weighted by $F_3(\xi)$ because the moments of $\xi$ up to the second are equal to zero (see Eq.~(\ref{moments1})). The terms containing powers of $\Delta\tau$ higher than one vanish in the limit $\Delta\tau\to 0$. So the only term contributing to Eq.~(\ref{diff_6a}) from the Taylor expansion (\ref{taylor_4}) is the one containing the third order derivative with respect to $\delta p_i$. Combining this observation with the last of Eqs.~(\ref{moments1}) we see that Eq.~(\ref{diff_6a}) is indeed equivalent to Eq.~(\ref{corr1}). Clearly the requirements~(\ref{moments1}) do not define the distribution function $F_3(\xi)$ uniquely. One possible choice would be a Gaussian multiplied by a cubic polynomial (see Eq.~(\ref{f_xi})). Another possible choice for $F_3(\xi)$ satisfying Eqs.~(\ref{moments1}) is the combination of $\delta$-functions:
\be
F_3(\xi)={\delta(\xi-2)-\delta(\xi+2)-2\delta(\xi+1)+2\delta(\xi+1)\over 12}
\ee
This representation of $F_3(\xi)$ explicitly reduces Eq.~(\ref{diff_6a}) to the finite difference representation of Eq.~(\ref{corr1}).

Similarly one finds that for the higher order jumps the distribution function must satisfy
\beq
&&\int_{-\infty}^\infty \xi^m F_{2n+1}(\xi)d\xi=0,\;m<2n,\nonumber\\
&&\int_{-\infty}^\infty \xi^{2n+1} F_{2n+1}(\xi)d\xi=1.
\eeq
and the expansion takes the form of Eq.~(\ref{corr2}) where higher order corrections in $\hbar$ either come from higher order jumps or from having several low order jumps.

{\em Multidimensional many-particle generalization.} The expression of quantum evolution through response to infinitesimal jumps can be straightforwardly generalized to the situation where we deal with many degrees of freedom. For example instead of Eq.~(\ref{corr1}) we should use Eq.~(\ref{corr3}), which we repeat here for completeness:
\beq
&&\langle \hat{\Omega}(\hat{\bm x},\hat{\bm p},t)\rangle\approx
\int\int d \bm x_0 d \bm p_0 W_0(\bm p_0,\bm x_0)\nonumber\\
&&\left( 1-\int_0^t
d\tau {1\over 3!\,2^2}{\hbar^2\over i^2}{\partial^3 V(\bm x)\over
\partial x_\alpha\partial x_\beta\partial x_\gamma}
{\partial^3\over\partial p_\alpha(\tau)\partial p_\beta(\tau) \partial p_\gamma(\tau)}+\dots\right) \Omega_{W}(\bm
x(t),\bm p(t),t).
\label{corr3a}
\eeq
Here indices $\alpha,\beta,\gamma$ can run over different spatial and particle indices. For example for $N$ particles in three dimensions they take the values $x_1, y_1, z_1, x_2, y_2, z_2,\dots x_N, y_N, z_N$. This expression can be again rewritten through stochastic quantum jumps yielding Eq.~(\ref{diff_7}):
\beq
&&\langle \hat{\Omega}(\hat{\bm x},\hat{\bm p},t)\rangle\approx \int\int d
\bm x_0 d \bm p_0 W_0(\bm x_0,\bm p_0)\nonumber\\
&&\biggl[1+{\hbar^2\over 4}\sum_j \int\int \prod_{m} d\xi_m \sum_{\sigma(\alpha,\beta,\gamma)}{\partial^3 V(\bm x_j)\over \partial x_\alpha \partial x_\beta\partial x_\gamma} F_{\alpha,\beta,\gamma}(\bm \xi)\biggr|_{ \delta p_\alpha(\tau_j)=\xi_\alpha\sqrt[3]{\Delta\tau_j}}\biggr] \Omega_{W}(\bm x(t),\bm p(t), t),\phantom{XX}
\label{diff_7a}
\eeq
where $\sigma(\alpha,\beta,\gamma)$ stands for all inequivalent combinations of indices $\alpha,\beta,\gamma$ (e.g. all combinations with $\alpha\leq\beta\leq\gamma$) and $F_{\alpha,\beta,\gamma}(\bm\xi)$ is a function which should satisfy the conditions that all its moments smaller than two vanish and
\be
\int\int \prod_{m} d\xi_m \xi_\alpha\xi_\beta\xi_\gamma F_{\alpha,\beta,\gamma}(\bm\xi)=1.
\label{int_f}
\ee
A possible choice for the function $F_{\alpha\beta\gamma}$ is given by Eqs.~(\ref{f111})-(\ref{f123}):
\beq
&&F_{\alpha,\alpha,\alpha}(\bm\xi)={1\over 2 \sqrt{2\pi}}\left({\xi_\alpha^3\over 3}-\xi_\alpha\right)\mathrm e^{-\xi_\alpha^2/2}\prod_{m\neq\alpha}\delta(\xi_m),\label{f111a}\\
&&F_{\alpha,\alpha,\beta}(\bm\xi)={\left(\xi_\alpha^2-1\right)\xi_\beta\over 4\pi}\, \mathrm e^{-(\xi_\alpha^2+\xi_\beta^2)/2}\!\prod_{m\neq\alpha,\beta}\!\delta(\xi_m) ,\\
&&F_{\alpha,\beta,\gamma}(\bm\xi)={\xi_\alpha\xi_\beta\xi_\gamma\over (2\pi)^{3/2}}\,
\mathrm e^{-(\xi_\alpha^2+\xi_\beta^2 +\xi_\gamma^2)/2}\!\! \prod_{m\neq\alpha,\beta,\gamma}\!\!\delta(\xi_m).\phantom{XX}
\label{f123a}
\eeq
Here the different labels $\alpha,\beta$, and $\gamma$ stand for different indices.

The equivalence between representations of quantum corrections (\ref{diff_7}) and (\ref{corr3}) (or Eqs.~(\ref{diff_7a}) and (\ref{corr3a})) can be established by Taylor expanding Eq.~(\ref{diff_7a}) in powers of $\delta p_\alpha(\tau_j)$ like in the single-particle case. A simple observation shows that only third order derivatives survive both the integration of $\xi_\alpha$ and the limit $\Delta\tau\to 0$. Using that the functions $F_{\alpha,\beta,\gamma}$ satisfy the property (\ref{int_f}) we recover that indeed Eq.~(\ref{diff_7a}) reduces to Eq.~(\ref{corr3a}) in the limit $\Delta\tau\to 0$. The sum over non-equivalent permutations of indices $\alpha,\beta,\gamma$ in Eq.~(\ref{diff_7a}) ensures that we avoid double counting of identical terms.

\subsection{Coherent state representation}
\label{sec:der_coherent}

In this section we will show details of derivation of the results of Sec.~(\ref{sec:main_coherent}). The derivation closely mimics the one shown above in the number-phase representation. However, there are some distinct features as well.

\subsubsection{Truncated Wigner approximation}

First let us consider a single time expectation value of some operator $\hat\Omega(\hat\psi,\hat\psi^\dagger,t)$. To shorten notations we will skip the single-particle indices in the bosonic fields $\hat\psi$ and $\hat\psi^\dagger$ and reinsert them only when needed. The derivation essentially repeats that of the previous section. Partially the results were published previously in Ref.~[\onlinecite{ap_twa}]. In the Heisenberg representation this expectation value reads
\be
\langle \hat\Omega(\hat\psi,\hat\psi^\star,t)\rangle={\rm Tr}\left[\rho\, T_\tau\,\mathrm e^{i\int_0^t \hat{\mathcal H}(\tau) d\tau} \hat\Omega(\hat\psi,\hat\psi^\dagger,t) e^{-i \int_0^t \hat{\mathcal H}(\tau) d\tau} \right],
\label{eq:omega_coh}
\ee
Because in the coherent state picture the Planck's constant plays the mere role of conversion between time and energy units we set $\hbar=1$ throughout this section to simplify notations.Next we split the integral over time into a product as in Eq.~(\ref{prod}) and insert the resolution of identity $|\psi(\tau_i)\rangle\langle\psi(\tau_i)|$ between different terms in this product. Then we get (cf. Eq.~(\ref{eq:omega1})):
\beq
&&\langle \hat{\Omega}(\hat{\psi},{\hat\psi^\dagger},t)\rangle=\int\int d \psi_0 d \psi_0^\star W_0(\psi_0,\psi_0^\star)\int\int D \psi(\tau) D \psi^\star(\tau) D  \eta(\tau) D \eta^\star(\tau)\nonumber\\
&& \exp\Biggl\{\int_0^t d\tau\Biggl[\eta^\star(\tau){\partial \psi(\tau)\over \partial\tau} -\eta(\tau) {\partial \psi^\star(\tau)\over\partial \tau}+i \mathcal H_W\left(\psi(\tau)+{\eta(\tau)\over 2},\psi^\star(\tau) +{\eta^\star(\tau)\over 2},\tau\right)\nonumber\\
&&- i \mathcal H_W\left(\psi(\tau) -{\eta(\tau)\over 2},\psi^\star(\tau) -{\eta^\star(\tau)\over 2},\tau\right)\Biggr]\Biggr\}\;\Omega_{W}(\psi(t),\psi^\star(t),t),
\label{eq:omega4}
\eeq
where $\int\int d\psi d\psi^\star\dots = \int\int d\Re\psi\, d\Im\psi/\pi\dots$ Here $W_0(\psi_0, \psi_0^\star)$ is the Wigner transform of the initial density matrix (see Eq.~(\ref{wig_coherent})):
\be
W_0(\psi_0,\psi_0^\star)={1\over 2}\int d\eta_0^\star d\eta_0 \left<
\psi_0-{\eta_0\over 2}\right|\rho \left| \psi_0+{\eta_0\over
2}\right>\mathrm e^{-|\psi_0|^2-{1\over
4}|\eta_0|^2}\,\mathrm e^{{1\over
2}(\eta_0^\star\psi_0-\eta_0\psi_0^\star)}.
\label{wig_coherent_a}
\ee
The quantities $\Omega_{W}(\psi(t),\psi^\star(t),t)$ and $\mathcal H_W(\psi(\tau),\psi^\star(\tau),\tau)$ are the Weyl symbol of the operator $\hat{\Omega}$ and the Hamiltonian $\hat{\mathcal H}$ (see Eqs.~(\ref{weyl_coherent_def}) and (\ref{Weyl_coherent})). The emergence of the Weyl ordering of the Hamiltonian in Eq.~(\ref{wig_coherent_a}) is somewhat subtle (see related discussion in the coordinate-momentum representation after Eq.~(\ref{eq:omega1})). We will discuss this issue in Appendix~\ref{appendix_hamilt_ordering}~\footnote{We note that in Ref.~\cite{ap_twa} the issue of the Weyl rather than normal ordering of the Hamiltonian was overlooked. As we discussed in Sec.~\ref{sec:main_coherent}, see Eq.~(\ref{gp_eq}), for the Hamiltonians with two body uniform interactions this difference only affects the overall phase.}. In the leading order in quantum fluctuations we expand the integrand in Eq.~(\ref{eq:omega4}) up to the linear terms in $\eta$. Then the functional integral over $\eta(t)$ enforces the $\delta$-function Gross-Pitaevskii constraint on the classical field $\psi(t)$:
\be
i\partial_t \psi={\partial H_W(\psi(t),\psi^\star(t),t)\over\partial\psi^\star(t)}\equiv \left\{\psi(t),H_W(\psi(t),\psi^\star(t),t)\right\}_c
\ee
and we recover the semiclassical (truncated Wigner) approximation (\ref{eq:omega5}).

\subsubsection{Non-equal time correlation functions.}

The discussion for finding non-equal time correlation functions in the coherent state representation closely mimics that of Sec.~\ref{sec:neq_time_1} in the coordinate-momentum representation. So we will only give a brief review of the derivation. We note that the results presented in this section were discussed in Ref.~\cite{plimak_new} using somewhat different derivation.

As in Sec.~\ref{sec:neq_time_1} we start from considering a two-time correlation function:
\be
\langle \hat{\Omega}_1(\hat{\psi},\hat{\psi}^\dagger,t_1) \hat{\Omega}_2(\hat{\psi},\hat{\psi}^\dagger,t_2)\rangle.
\label{non_eq_time_4}
\ee
 Assuming that $t_1<t_2$ and repeating the discussion of Sec.~\ref{sec:neq_time_1} we find that at the moment $t_1$ the classical fields $\psi(t_1)$ and $\psi^\star(t_1)$ undergo quantum jumps: $\psi(t_1)\to \psi(t_1)+\delta\psi_1$,  $\psi^\star(t_1)\to \psi^\star(t_1)+\delta\psi_1^\star$ described by the probability distribution
which can be deduced from the generating function
\be
R(\delta\psi,\delta\psi^\star,\epsilon,\epsilon^\star) =\pi\exp[\epsilon\psi^\star+\epsilon^\star\psi-|\epsilon|^2/2] \exp\biggl[{\epsilon\over 2}{\partial\over\partial\delta\psi} -{\epsilon^\star\over 2}{\partial\over\partial\delta\psi^\star}\biggr] \delta(\Re\delta\psi)\delta(\Im\delta\psi).
\ee
In this way $\hat\Omega_{nm}=(\hat{\psi^\dagger})^n\hat{\psi}^m $ produces the following probability distribution for the jumps
\be
W_{nm}(\delta\psi,\delta\psi^\star)=\partial_\epsilon^n\partial_{\epsilon^\star}^m R(\delta\psi,\delta\psi^\star,\epsilon,\epsilon^\star)\biggr|_{\epsilon=0}.
\ee
As before the two most important cases are $\hat\Omega_{1,0}=\hat\psi^\dagger$ corresponding to
\be
W_{1,0}(\delta\psi,\delta\psi^\star)=\pi\left(\psi^\star+{1\over 2}{\partial\over\partial\delta\psi} \right)\delta(\Re\delta\psi)\delta(\Im\delta\psi),
\ee
and $\hat\Omega_{0,1}=\hat\psi$ corresponding to
\be
W_{0,1}(\delta\psi,\delta\psi^\star)=\pi\left(\psi-{1\over 2}{\partial\over\partial\delta\psi^\star} \right)\delta(\Re\delta\psi)\delta(\Im\delta\psi).
\ee
For the number density $\hat\Omega_{1,1}=\hat\psi^\dagger\hat\psi$ we find
\be
W_{1,1}(\delta\psi,\delta\psi^\star)=\pi\biggl[|\psi|^2-{1\over 2}\left(
\psi^\star{\partial\over\partial\delta\psi^\star}-
\psi{\partial\over\partial\delta\psi}\right)-{1\over 4} {\partial^2\over\partial\delta\psi\partial\delta\psi^\star} -{1\over 2}\biggr]\delta(\Re\delta\psi)\delta(\Im\delta\psi).
\ee
As in the coordinate momentum case one can note that it is sufficient to establish the correspondence between operators and classical fields based on expressions for $W_{1,0}$ and $W_{0,1}$ and using the property of the $\delta$-function: $\delta^\prime(x)=-\delta(x) d/dx$:
\be
\hat{\psi^\dagger}\to\psi^\star-{1\over 2}{\partial\over\partial\delta\psi},\quad\hat{\psi}\to\psi+{1\over 2}{\partial\over\partial\delta\psi^\star},
\label{eq:pa1}
\ee
These are nothing but the coherent state Bopp operators~(\ref{bopp_psi_dag}) and (\ref{bopp_psi}) where the derivatives are understood as a response to the infinitesimal quantum jumps. If we reinsert the time label $t_1$ in the fields $\psi$ and $\psi^\star$ we will recover Eqs.~(\ref{eq:pa_dag}) and (\ref{eq:pa}). As we discussed earlier for finding a single time expectation value Bopp representation of $\hat\psi$ and $\hat\psi^\dagger$ can be used to reproduce the Weyl symbol of an arbitrary operator.

The analysis of the correlation function with opposite time ordering, where the earlier time appears on the right, goes along the same lines as in Sec.~\ref{sec:neq_time_1}. We thus will not repeat the derivation and only quote the final result that in this case one can use right derivatives with opposite signs as it is indicated in Eqs.~(\ref{eq:pa_dag}) and (\ref{eq:pa}). This correspondence gives the casual phase space representation of the expectation value of $\hat\Omega_2(t_2)\hat\Omega_1(t_1)$ for $t_1<t_2$.

Generalization of two-time correlation functions to multi-time correlation functions also goes completely along the lines of the discussion given in the previous section for the coordinate-momentum representation. In particular, the correspondence is very simple, $\hat\psi\to \psi$, $\hat\psi^\dagger\to\psi^\star$, for time-symmetrically ordered correlation functions given by Eq.~(\ref{time_symm_2}), where now $\hat\xi$ stands for either $\hat\psi$ or $\hat\psi^\dagger$. For example:
\beq
{1\over 4}\left[\hat\psi^\dagger(t_1),\left[\hat\psi(t_2),\hat\psi(t_3)\right]_+\right]_+\to \psi^\star(t_1)\psi(t_2)\psi(t_3),
\eeq
where $[\hat A,\hat B]_+$ stands for the anti-commutator of the operators $\hat A$ and $\hat B$. Formally this substitution works for any ordering of times $t_1,t_2,t_3$. However, only when $t_1\leq t_2\leq t_3$ the casuality of the description is preserved and one does not have to propagate equations of motion backwards in time. Generally (like in the coordinate momentum case) the casuality is preserved for operators appearing according to the Schwinger-Keldysh ordering, where the operator corresponding to an earlier time can never appear between the two operators evaluated at a later time. The casuality then corresponds to a particular choice of left or right derivatives at different times. For example for $t_1\leq t_2\leq t_3$ we have
\be
\hat \psi^\dagger (t_1)\hat\psi(t_3)\hat\psi(t_2)\to\left(
\psi^\star(t_1)-{1\over 2}{\partial\over\partial\psi(t_1)}\right)
\left(
\psi(t_2)-{1\over 2}{\partial\over\partial\psi^\star(t_2)}\right)\psi(t_3),
\ee
while
\be
\hat \psi^\dagger (t_1)\hat\psi(t_2)\hat\psi(t_3)\to\left(
\psi^\star(t_1)-{1\over 2}{\partial\over\partial\psi(t_1)}\right)
\left(\psi(t_2)+{1\over 2}{\partial\over\partial\psi^\star(t_2)}\right)\psi(t_3).
\ee

\subsection{Beyond the truncated Wigner approximation.}

Calculating quantum corrections to Eq.~(\ref{eq:omega5}) also follows along the lines of the previous section. We need to expand the Hamiltonian in Eq.~(\ref{eq:omega4}) to higher orders in the quantum field $\eta$. Note that for the nointeracting systems TWA is exact. For the Hamiltonians with two-body interaction~(\ref{hamilt_bos}) only the third power of $\eta$ appears in the expansion. To simplify the discussion we will focus on the derivation of quantum corrections in the system described by the Hubbard model with the Hamiltonian~(\ref{hamilt_bh}):
\be
\hat{\mathcal H}_{hub}=-J\sum_{\langle i,j\rangle} (\hat\psi_i^\dagger\hat\psi_j+\psi_j^\dagger\psi_i)+\sum_j V_j\hat \psi_j^\dagger\psi_j+{1\over 2}\sum_{\langle i,j\rangle} U_{ij} \psi_i^\dagger\psi_j^\dagger\psi_j\psi_i.
\label{hamilt_bh_1}
\ee
Then Eq.~(\ref{eq:omega4}) assumes the following form:
\beq
&&\langle \hat{\Omega}(\hat{\bm\psi},{\hat{\bm \psi}^\dagger},t)\rangle=\int\int d {\bm\psi}_0 d {\bm \psi}_0^\star W_0({\bm\psi}_0,{\bm \psi}_0^\star)\int\int D {\bm\psi}(\tau) D {\bm\psi}^\star(\tau) D  {\bm\eta}(\tau) D {\bm\eta}^\star(\tau) \nonumber\\
&&\exp\Biggl\{i\int_0^t d\tau\Biggl[\sum_j \eta_j^\star(\tau)\left(-i{\partial \psi_j(\tau)\over \partial\tau}+{\partial \mathcal H_W\over\partial\psi_j^\star(\tau)}\right)
+\eta_j(\tau) \left(i{\partial \psi_j^\star(\tau)\over\partial \tau}+{\partial \mathcal H_W\over\partial\psi_j(\tau)}\right)\nonumber\\
&&+{1\over 4}\sum_{\langle i,j\rangle} U_{ij}\left[\eta_i^\star(\tau)\psi_i(\tau) +\eta_i(\tau)\psi_i^\star(\tau)\right]|\eta_j(\tau)|^2\Biggr] \Omega_{W}(\psi(t),\psi^\star(t),t).
\label{eq:omega6}
\eeq
Next we expand the exponent containing cubic terms in $\eta_j$ and to the leading order we find
\begin{widetext}
\beq
&&\langle \hat{\Omega}(\hat{\bm\psi},{\hat{\bm \psi}^\dagger},t)\rangle\approx\int\int d {\bm\psi}_0 d {\bm \psi}_0^\star W_0({\bm\psi}_0,{\bm \psi}_0^\star)\int\int D {\bm\psi}(\tau) D {\bm\psi}^\star(\tau) D  {\bm\eta}(\tau) D {\bm\eta}^\star(\tau) \nonumber\\
&&~~~~~\left(1+{i\over 4}\int_0^t d\tau\sum_{\langle i,j\rangle} U_{ij}\left[\eta_i^\star(\tau)\psi_i(\tau) +\eta_i(\tau)\psi_i^\star(\tau)\right]|\eta_j(\tau)|^2\right)
\label{eq:omega7}\\
&&\times\exp\Biggl\{i\int\limits_0^t d\tau\Biggl[\sum_j \eta_j^\star(\tau)\left(-i{\partial \psi_j(\tau)\over \partial\tau}+{\partial \mathcal H_W\over\partial\psi_j^\star(\tau)}\right)+\eta_j(\tau) \left(i{\partial \psi_j^\star(\tau)\over\partial \tau}+{\partial \mathcal H_W\over\partial\psi_j(\tau)}\right)\Biggr]\Omega_{W}(\psi(t),\psi^\star(t),t).\nonumber
\eeq
\end{widetext}
The functional integral over $\eta$ and $\eta^\star$ can now be taken formally by introducing source terms into the path integral~(\ref{eq:omega7}):
\be
S=\exp\left[\sum_j \sum_{\tau_\alpha=0}^t \left( s_j^\star(\tau_\alpha)\eta_j(\tau_\alpha)-s_j(\tau_\alpha)\eta_j^\star(\tau_\alpha)\right)\right],
\ee
where to avoid dealing with functional derivatives we discretized the time: $\tau_\alpha=\alpha\Delta\tau$, $\alpha=0,1,\dots t/\Delta\tau$. As usual in the end we will take the limit $\Delta\tau\to 0$. Then we have
\be
\eta_j^\star(\tau_\alpha)=-{\partial\over\partial s_j(\tau_\alpha)}S,\quad
\eta_j(\tau_\alpha)={\partial\over\partial s_j^\star(\tau_\alpha)}S,
\ee
where all the derivatives are evaluated at $s_j(\tau)= 0$. Then we can rewrite Eq.~(\ref{eq:omega7}) in the following way:
\beq
&&\langle \hat{\Omega}(\hat{\bm\psi},{\hat{\bm \psi}^\dagger},t)\rangle=\int\int d {\bm\psi}_0 d {\bm \psi}_0^\star W_0({\bm\psi}_0,{\bm \psi}_0^\star)\int\int D {\bm\psi}(\tau) D {\bm\psi}^\star(\tau) D  {\bm\eta}(\tau) D {\bm\eta}^\star(\tau)\nonumber\\
&&~~~~~\left(1-{i\over 4}\sum_{\tau_\alpha=0}^t \Delta\tau\sum_{\langle i,j\rangle} U_{ij}\left[\psi_i(\tau_\alpha){\partial\over\partial s_i(\tau_\alpha)}- \psi_i^\star(\tau_\alpha){\partial\over\partial s_i^\star(\tau_\alpha)}\right] {\partial^2\over\partial s_j(\tau_\alpha)s_j^\star(\tau_\alpha)}\right)\\
&&\times\exp\Biggl\{i\sum_{\tau_\alpha=0}^t \Delta\tau\Biggl[\sum_j \eta_j^\star(\tau_\alpha)\left(-i{\psi_j(\tau_\alpha+\Delta\tau)-\psi_j(\tau_\alpha)-s_j(\tau_\alpha)\over \Delta\tau}+{\partial \mathcal H_W\over\partial\psi_j^\star(\tau_\alpha)}\right)\nonumber\\
&&+\eta_j(\tau_\alpha) \left(i{\psi_j^\star(\tau_\alpha+\Delta\tau) - \psi_j^\star(\tau_\alpha)-s_j^\star(\tau_\alpha) \over\Delta \tau}+{\partial \mathcal H_W\over\partial\psi_j(\tau_\alpha)}\right)\Biggr] \Omega_{W}(\psi(t),\psi^\star(t),t).\nonumber
\label{eq:omega8}
\eeq
The source terms $s_j(\tau)$ should be set to zero in the end of the calculation. Now one can integrate over ${\bm \eta}(\tau)$ and ${\bm \eta}^\star(\tau)$. As a result we get that the classical fields $\psi_j(\tau)$ satisfy the Gross-Pitaevskii equations of motion with additional source terms, which play the role of infinitesimal shift of $\psi_j(\tau)$: $\psi_j(\tau)\to\psi_j(\tau)+s_j(\tau)$. Changing notations $s_j(\tau_\alpha)=\delta\psi_j(\tau_\alpha)$, taking the continuum limit $\Delta\tau\to 0$, and using the property of $\delta$-function $\partial_s\delta(s)=-\delta(s)\partial_s$ we find that Eq.~(\ref{eq:omega8}) reduces to
\beq
&&\langle \hat{\Omega}(\hat{\bm\psi},{\hat{\bm\psi}^\dagger},t)\rangle\approx \int\int d {\bm\psi}_0 d {\bm\psi}_0^\star W_0({\bm\psi}_0,{\bm\psi}_0^\star)
\label{eq:omega9}\\
&&\times\left(1-{i\over 4}\int_0^t d\tau\sum_{\langle i,j\rangle}U_{ij}\left[\psi_i^\star(\tau){\partial\over\partial \delta\psi_i^\star(\tau)}- \psi_i(\tau){\partial\over\partial \delta\psi_i(\tau)}\right] {\partial^2\over\partial \delta\psi_j(\tau)\partial\delta\psi_j^\star(\tau)}\right)\Omega_{W}(\psi(t),\psi^\star(t),t),\nonumber
\eeq
where at the time $\tau$ the classical fields $\psi_i(\tau)$ and $\psi_j(\tau)$ undergo infinitesimal jumps $\delta\psi_i(\tau)$ and $\delta\psi_j(\tau)$ and later at the time $t$ the nonlinear response of the observable $\Omega$ is evaluated with respect to these jumps. Note that Eq.~(\ref{eq:omega9}) coincides with Eq.~(\ref{corr4}) for a particular choice of the interaction $u_{ijkl}=U_{ij}\delta_{kj}\delta_{il}$. Similarly higher quantum corrections to TWA appear in the form of multiple quantum jumps.

As in the coordinate momentum case one can represent nonlinear response through stochastic quantum jumps. Then one arrives to Eq.~(\ref{corr5}). It is straightforward to check that indeed Eq.~(\ref{corr5}) is equivalent to Eq.~(\ref{corr4}) by Taylor expanding Eq.~(\ref{corr5}) in the powers of $\xi_j$ and noting that the terms up to the second order in this expansion give zero contribution to the integral because of the requirement (\ref{cond_f}). The fourth and higher order derivative terms disappear in the limit $\Delta\tau\to 0$. So the only terms in the Taylor expansion contributing to Eq.~(\ref{corr5}) are the ones containing third order derivatives. These terms precisely reproduce Eq.~(\ref{corr4}) because of the condition (\ref{cond_f}). As it is straightforward to check the functions (\ref{f1})-(\ref{f4}) give a possible choice for the distribution of jumps satisfying the conditions~(\ref{cond_f}).

\section{Derivation of the truncated Wigner approximation from the von Neumann's equation.}
\label{sec:liouv_wig}

The results of the previous section can be partially recovered directly analyzing the von Neumann equation for the density matrix (see Ref.~\cite{hillery} for details):
\be
i\hbar\dot{\hat\rho} =[\hat{\mathcal
 H},\hat\rho].
 \label{liouville_quant}
\ee
Taking the Weyl transform of both sides of the equation and using Eq.~(\ref{moyal_bracket}) for the Moyal bracket we find that in the coordinate-momentum representation this equation becomes:
\be
\dot W=\left\{\mathcal H_W,W\right\}_{MB}=-{2\over \hbar}\mathcal H_W\sin\left[{2\over\hbar}\Lambda\right] W.
\label{liouville}
\ee
If we expand the Moyal bracket in the powers of $\hbar$ and stop in the leading order then the Moyal bracket reduces to the Poisson bracket and the von Neumann's equation (\ref{liouville}) reduces to the classical Liouville equation for the Wigner function:
\be
\dot W\approx \left\{\mathcal H_W,W\right\},
\label{liouville_cl}
\ee
where the Wigner function plays the role of the classical probability distribution. The semiclassical (truncated Wigner) approximation is recovered by noting that the classical trajectories are the characteristics of Eq.~(\ref{liouville_cl}) along which the Wigner function is conserved. Noting that the expectation value of any operator is equal to the average of the corresponding Weyl symbol weighted with the Wigner function~\cite{hillery} we indeed recover that Eq.~(\ref{liouville_cl}) is equivalent to Eq.~(\ref{eq:omega2}). The interpretation of each trajectory in TWA, unlike in the purely classical limit, is not very straightforward for the reason that the Wigner function is not positive definite and can not be interpreted as a probability of a particular realization of the initial conditions. So the individual trajectory does not immediately represent a possible outcome of a single experiment corresponding to a particular realization of the initial conditions.

The quantum corrections to Eq.~(\ref{liouville_cl}) appear because of the higher order terms in expansion of the Moyal bracket in powers of $\hbar$. Thus to the order of $\hbar^2$ we find (see also Ref.~\cite{zurek_rmp})
\be
\dot W=\left\{\mathcal H,W\right\}-{\hbar^2\over 2^2 3!}\sum_{j,k,l} {\partial^3 U\over \partial x_j \partial x_k\partial x_l}{\partial W^3\over\partial p_j\partial p_k\partial p_l}+\dots
\label{liouville_cl1}
\ee
This partial differential equation is much harder to solve than the classical Liouville equation since the method of characteristics does not apply. One can attempt to solve it in the case of say a single degree of freedom, but then there is no advantage in this method over the direct solution of the Schr\"odinger equation. Another possibility is to write the evolution operator for this equation in the path integral form (see Ref.~\cite{plimak_new}) and try to treat this term perturbatively. In this case one should recover the results obtained in this paper. There are no other methods of efficiently simulating Eq.~(\ref{liouville_cl1}) known in the literature.

The truncated Wigner approximation also was independently developed in quantum optics as an approximate solution of the quantum Liouville equation for the Wigner function written in the coherent state basis~\cite{walls-milburn, gardiner-zoller}. This approximation later was extended to the systems of interacting bosons~\cite{steel} and recently found many applications to cold atom systems (see e.g. Ref.~\cite{blakie_08}). TWA immediately follows from the application of the coherent state Moyal brackets~(\ref{coh_moyal_bracket}) to the von Neumann's equation for the density matrix~(\ref{liouville_quant}):
\be
i\hbar \dot W=\left\{\mathcal H_W, W\right\}_{MBC}=2\mathcal H_W\sinh\left[{1\over 2}\Lambda_c\right]W.
\label{liouville_coh_ex}
\ee
Expanding this Moyal bracket in the powers of $\Lambda_c$ (which is equivalent in the expansion in $1/N$) to the leading order we find
\be
i\hbar \dot W\approx \left\{\mathcal H_W, W\right\}_{C}
\label{liouville_coh}
\ee
recovering the classical Liouville equation for the density matrix. This equation can be solved using the method of characteristics which exactly coincide with the classical Gross-Pitaevskii equations:
\be
i\hbar\dot \psi_j={\partial \mathcal H_W\over\partial\psi_j^\star(t)}
\ee
These characteristics again represent trajectories along which the Wigner function is approximately conserved.

If we go beyond TWA then we need to expand the coherent state Moyal bracket up to the third order in $\Lambda_c$. Then we recover a third order partial differential equation analogous to Eq.~(\ref{liouville_cl1}):
\be
i\hbar \dot W=\left\{\mathcal H_W,W\right\}_c+{1\over 8}\sum_{\alpha,\beta,\gamma}{\partial^3 \mathcal H_W\over \partial\psi_\alpha\partial\psi_\beta^\star\partial\psi_\gamma^\star} {\partial^3 W\over \partial\psi_\alpha^\star\partial\psi_\beta \partial\psi_\gamma}- {\partial^3 \mathcal H_W\over \partial\psi_\alpha^\star\partial\psi_\beta\partial\psi_\gamma} {\partial^3 W\over \partial\psi_\alpha\partial\psi_\beta^\star\partial\psi_\gamma^\star}.
\label{liouville_coh_1}
\ee
For the Hamiltonians with two-body interactions (or generally with not necessarily number conserving interactions involving no more than quartic terms in the fields $\hat\psi$ and $\hat\psi^\dagger$) there are no higher than third order terms in the expansion of Eq.~(\ref{liouville_coh_ex}) in powers of $\Lambda_c$ and thus Eq.~(\ref{liouville_coh_1}) is exact (see also Refs.~\cite{steel, blakie_08}). As in the coordinate-momentum representation this equation is hard to solve because the method of characteristics does not apply.

We note that the (probably unfortunate) name truncated Wigner approximation appeared in the literature for Eq.~(\ref{liouville_coh}) (which is also equivalent to Eq.~(\ref{eq:omega5})) because at this level of approximation one truncates the Liouville equation~(\ref{liouville_coh_1}) to the linear order in derivatives with respect to $\psi$ and $\psi^\star$. Thus this terminology is akin terming the Poisson bracket as the truncated Moyal bracket.

\acknowledgements
 The author is very grateful to L.~Plimak for numerous discussions and comments. Discussions with A.~Altland, A.~Clerk, and W.~Zurek are greatly acknowledged. The author also acknowledges collaboration with S. Girvin and S. Sachdev on earlier work~\cite{psg}, which stimulated the author's interest to this topic. The author thanks Zhigang Wu for pointing out a missing factor of $1/2^M$ in Eqs.~(\ref{weyl_coherent_def})-(\ref{wig_coherent}). This work was supported by AFOSR YIP, Sloan Foundation and US NSF DMR-0907039.

\appendix

\section{Implementation of quantum corrections for the sine-Gordon and Bose Hubbard models.}
\label{sec:app_implement}

In this section we give some details of how one can implement quantum corrections in practice. First let us consider the sine-Gordon model (see Sec.~\ref{sec:sg}), which is described by the Hamiltonian (\ref{FK}). At the semiclassical level one has to solve the equations of motion (\ref{eq:sg}) subject to random initial conditions distributed according to the Wigner function~(\ref{wig_sg}). In the chosen setup the Wigner function is gaussian both at zero and at finite temperatures and there is no problem in sampling over the initial conditions.

In order to implement the quantum correction in this case we used quantum jumps given by the second term in the brackets of Eq.~(\ref{diff_6}) proportional to $\hbar^2$. The function $F_3(\xi)$ is given by Eq.~(\ref{f_xi}). To implement this jump in practice at each step of the Monte-Carlo sampling over the initial conditions we also select a random time $\tau$ and a random position $j$ of the jump. Then at this time we evaluate $V_{3,j}=-V(\tau)\beta^3\sin(\beta\phi_j(\tau))$, randomly choose $\xi$ from a Gaussian distribution with zero mean and unit variance and the rest of the function $F_3(\xi)$, namely $1/2(\xi^3/3-\xi)$, we multiply by $V_{3,j}$ and $\hbar^2/2$. And finally we transform $\delta n_j\to \delta n_j + \xi_j \sqrt[3]{\Delta\tau}$. Here $\Delta \tau$ is a sufficiently small interval which can be adjusted to improve the efficiency of the simulation: larger $\Delta\tau$ corresponds to faster convergence but worse accuracy. We checked that the results for $\Delta\tau=0.1$ and $\Delta\tau=0.2$ are practically indistinguishable.

Then the quantum correction term to the expectation value of say $\hat\Omega=\cos(\beta\hat\phi_0(t))$ is evaluated as the average over many runs of the following quantity:
\be
\left<-{t  M V(\tau)\beta^3\over 8}\sin(\beta\phi_j(\tau))(\xi^3/3-\xi)\cos(\beta \phi_0(t)) \right>,
\ee
where $\tau$ is a random time uniformly distributed in $[0,t)$, $j$ is the random position of the jump: $j\in[0,M-1)$, $M$ is the spatial size of the system ($M=20$ for our particular example), $\xi$ is the randomly distributed Gaussian variable with zero mean and unit variance. At the moment $\tau$ we shift the density $n_j(\tau)\to n_j(\tau)+\xi \sqrt[3]{\Delta \tau}$. At each Monte Carlo run we also randomly choose initial conditions distributed according to the Wigner function.

Next let us describe implementation of the quantum corrections in terms of the nonlinear response, which we used in the problem of turning on interactions in the system of bosons in optical lattice (see Sec.~\ref{sec:ramp}). In principle, the two implementations of corrections in terms of quantum jumps and nonlinear response are equivalent. We arbitrarily decided to use jumps for the sine-Gordon model and nonlinear response for the Hubbard model to illustrate that both methods work. Which one is more numerically efficient in which situation still remains to be investigated.

As in the previous case TWA is obtained by solving the classical discrete Gross-Pitaveskii equations with the initial conditions distributed to the Gaussian distribution~(\ref{wig_bos_coh}). The Weyl symbol for the Hamiltonian can be obtained e.g. by using Eqs.~(\ref{eq:pa}) and (\ref{eq:pa_dag}):
\be
\mathcal H_{W}(t)=-J\sum_{\langle j,k\rangle } (\psi_j^\star(t)\psi_k(t)+\psi_k^\star(t)\psi_j(t))+{U(t)\over 2}\sum_j \left[(|\psi_j(t)|^2-1)^2-{1\over 2}\right],
\ee
where the first sum is taken over all nearest neighbor pairs.

To find the first quantum correction to the TWA result we use Eq.~(\ref{corr4}), which in the case of Hubbard model takes the form
\beq
&&\langle \hat{\mathcal H}(t)\rangle_1=-\int D\psi^\star_{0,j}D\psi_{0,j}
W(\psi_{0,j}^\star,\psi_{0,j})\nonumber\\
&&\sum_i \int_0^t dt^\prime {U(t^\prime)\over
16}\left[\Im\psi_i(t^\prime)
{\partial\over\partial\epsilon_1}-\Re\psi_i(t^\prime)
{\partial\over\partial\epsilon_2 }\right]\left[{\partial^2\over
\partial\epsilon_1^2}+{\partial^2\over
\partial\epsilon_2^2}\right]\mathcal H_{W}(t,\epsilon_1,\epsilon_2),
\label{quant_corr}
\eeq
where $\epsilon_1$ and $\epsilon_2$ represent the real and imaginary parts of an infinitesimal jump of $\psi_i$ at the moment $t'$:
\be
\psi_i(t')\to \psi_i(t')+\epsilon_1+i\epsilon_2;
\ee
$\mathcal H_{W}(t,\epsilon_1,\epsilon_2)$ depends on $\epsilon_1$ and $\epsilon_2$ through the fields $\psi_j(t)$. To get Eq.~(\ref{quant_corr}) from Eq.~(\ref{corr4}) we used relations (\ref{compl_der}) because in practice it is easier to manipulate with real rather than comlex numbers. Evaluation of the correction according to Eq.~(\ref{quant_corr}) requires computing third order derivatives. This can be done using finite differences. E.g. for an arbitrary function $\Omega(\epsilon_1,\epsilon_2)$ we have
\beq
&&{\partial^3 \Omega(\epsilon_1,\epsilon_2)\over\partial \epsilon_1^3}\approx
{\Omega(2\epsilon_1,0)-\Omega(-2\epsilon_1,0)-2\Omega(\epsilon_1,0)+2\Omega(-\epsilon_1,0)\over
2\epsilon_1^3}\label{partial1}\\
&& {\partial^3 \Omega(\epsilon_1,\epsilon_2)\over\partial \epsilon_1\partial\epsilon_2^2}\approx {\Omega(\epsilon_1,\epsilon_2)+\Omega(\epsilon_1,-\epsilon_2)-\Omega(-\epsilon_1,\epsilon_2)
-\Omega(-\epsilon_1,-\epsilon_2)-2\Omega(\epsilon_1,0)+2\Omega(-\epsilon_1,0)\over 2\epsilon_1\epsilon_2^2},\phantom{XX}\label{partial2}
\eeq
where $\epsilon_1$ and $\epsilon_2$ are small. As in the previous example the actual values of $\epsilon_1$ and $\epsilon_2$ should be determined by the compromise: larger values give faster convergence but smaller accuracy. Similar expressions can be used to evaluate the remaining two derivatives. In practice one can evaluate this differences by time-evolving the original fields $\psi_j$ as well as twelve additional perturbed fields which obtained by shifting $\psi_i(t)$ in different ways:
\beq
&&\psi_{i,1}(t)=\psi_i(t)+\epsilon_1,\; \psi_{i,2}(t)=\psi_i(t)-\epsilon_1,\; \nonumber\\
&& \psi_{i,3}(t)=\psi_i(t)+2\epsilon_1,\; \psi_{i,4}(t)=\psi_i(t)-2\epsilon_1,\nonumber\\
&&\psi_{i,5}(t)=\psi_i(t)+i\epsilon_2,\; \psi_{i,6}(t)=\psi_i(t)-i\epsilon_2,\;\nonumber\\
&&\psi_{i,7}(t)=\psi_i(t)+2i\epsilon_2,\; \psi_{i,8}(t)=\psi_i(t)-2i\epsilon_2,\;\nonumber\\
&&\psi_{i,9}(t)=\psi_i(t)+\epsilon_1+i\epsilon_2,\; \psi_{i,10}(t)=\psi_i(t)-\epsilon_1+i\epsilon_2,\;\nonumber\\
&&\psi_{i,11}(t)=\psi_i(t)+\epsilon_1-i\epsilon_2,\; \psi_{i,12}(t)=\psi_i(t)-\epsilon_1-i\epsilon_2,\;\nonumber
\eeq
Then one can obtain desired finite differences using appropriately displaced solutions of the Gross-Pitaevskii equations (according to Eqs.~(\ref{quant_corr}), (\ref{partial1}), and (\ref{partial2})). As in the sine-Gordon case we can evaluate time integral and the sum by randomly choosing time $t'\in [0,t)$ and position at each Monte-Carlo run and then multiply the result by the product of $t$ and the number of sites.

\section{Emergence of the Weyl ordering of the Hamiltonian in the path integral representation of the evolution operator.}
\label{appendix_hamilt_ordering}

The emergence of the Weyl form of the Hamiltonian both in Eq.~(\ref{eq:omega1}) and (\ref{eq:omega4}) is somewhat subtle because usually path integrals are based on the normal ordering of operators~\cite{negele_orland}. In the coherent state picture the normal ordering corresponds to creation operators appearing on the left of annihilation operators. In the coordinate momentum representation by the normal ordering we will understand the one corresponding to the coordinate operators appearing on the left of the momentum operators. Note that usually the Hamiltonian is the sum of kinetic and potential energies the former being a function of momenta only and the latter being the function of the coordinates. Then the issue of ordering never arises because normal and Weyl orderings are  identical. This issue will become important if the Hamiltonian has cross terms like $\hat x^2\hat p$ which will emerge e.g. in the systems with nontrivial vector potential. In the coherent state picture the choice of the correct ordering can be very important. Even for the Hamiltonians with two-body interactions like~(\ref{hamilt_bos}) the Weyl compared to normal ordering introduces additional terms (see Eq.~(\ref{tilde_h})). As we argued in Sec.~\ref{sec:main_coherent} for spatially uniform two-body density-density interactions the Weyl ordering introduces an additional term proportional to the total number of particles, which is conserved in time and only affects an overall phase in the density matrix (and can be removed by a global gauge transformation). Such phase is usually unimportant unless we are interested in various interference phenomena between different systems. However, if the interaction between particles is spatially dependent due to some external potential, the additional term in the Hamiltonian can not be eliminated by a simple gauge transformation and have to be taken into account. At the end of this Appendix we will illustrate this point with an explicit example.

Let us discuss the emergence of the Weyl ordering in the coherent state representation. The discussion in the coordinate-momentum representation would be completely analogous. In the path integral representation of the Hamiltonian we have to deal with the evolution operator written as an infinite product (see Sec.~\ref{sec:der_coherent}):
\be
\prod_\alpha \exp[\pm i\hat{\mathcal H}(\psi^\dagger,\psi,\tau_\alpha)\Delta \tau].
\ee
For simplicity we suppressed all spatial indices. The two signs correspond to the forward and backward evolution. It is convenient to proceed with the Hamiltonian written in the normal-ordered form. Then by introducing the resolution of the identity as $|\psi(\tau_\alpha)\rangle\langle \psi(\tau_\alpha)\exp[-|\psi_\alpha|^2]|$ at each time step we end up with the products of the type:
\be
\prod_\alpha \exp\left[\pm \sum_\alpha \psi(\tau_\alpha)\biggl(\psi^\star(\tau_{\alpha+1}) - \psi^\star(\tau_\alpha)+i\mathcal H_n(\psi^\star(\tau_\alpha),\psi(\tau_{\alpha+1}),\tau_\alpha)\Delta\tau\biggr)\right].
\label{normal_prod}
\ee
Here the subindex $n$ emphasizes that the substitution $\hat\psi^\dagger\to\psi^\star$ and $\hat\psi\to \psi$ is taken in the normal ordered Hamiltonian. The key observation we have to make is that the field $\psi^\star$ appears in the earlier time than the field $\psi$. As we argued earlier, when we discussed non-equal time correlation functions, the path integral automatically produces Bopp operators for the fields $\hat\psi$ and $\hat\psi^\dagger$ in any observable. The same situation happens in the Hamiltonian, which can be formally seen by treating the term $\exp\left[i\mathcal H_n(\psi^\star(\tau_\alpha),\psi(\tau_{\alpha+1}),\tau_\alpha)\Delta\tau\right]$ at each time step as an observable.
Then we have to use the substitution in the Hamiltonian, $\hat \psi^\dagger\to \psi^\star-{i\over 2}\partial_\psi$. We remind that the derivative term comes from integrating over the quantum field $\eta$ (see Sec.~\ref{sec:der_coherent}). for the details. As we discussed in that section the derivative terms acting on the observables appearing at later times give raise to various quantum jumps. However, there is also an additional contribution from the jumps in $\psi$ on the Hamiltonian itself coming from the fact that in the normal ordered form $\psi$ appears at an infinitesimally later time than $\psi^\star$ (see Eq.~(\ref{normal_prod})). As we discussed in Sec.~\ref{sec:main_coherent} these derivative terms produce the Weyl symbol of the Hamiltonian. In other words the Weyl symbol of the Hamiltonian is produced by the self-action of the quantum jumps on the Hamiltonian. The same argument shows that the Weyl ordering should be used in the coordinate-momentum representation.

\begin{figure}[ht]
\includegraphics[width=12cm]{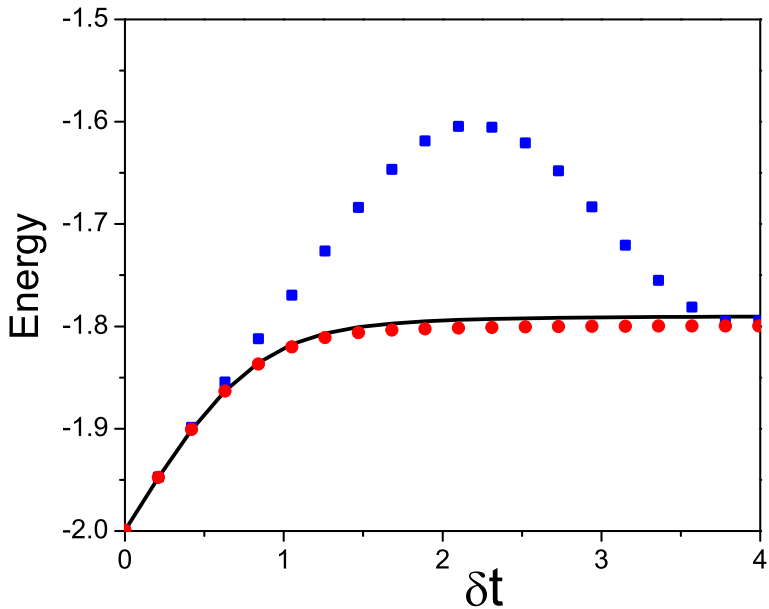}
\caption{Energy (per site) dependence on the product $\delta t$ for a periodic Bose-Hubbard chain consisting of eight sites with the modulated interaction potential (see the text for details). The black solid curve is the result of the exact calculation. Red circles correspond to TWA with the Weyl form of the Hamiltonian used in the time evolution and and blue squares show the result of TWA where the normal ordered Hamiltonian is used instead. }
\label{fig:hubbard_cubic}
\end{figure}

\begin{figure}[ht]
\includegraphics[width=12cm]{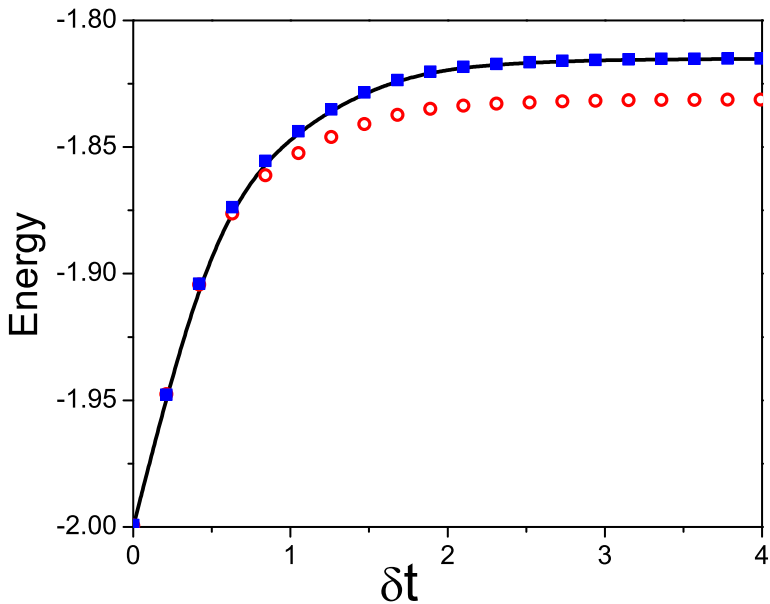}
\caption{Same as in Fig.~\ref{fig:hubbard_cubic} except for $\delta=1$. The black line is the exact result. The red circles and blue squares represent TWA and TWA + first quantum correction evaluated using the Weyl form of the Hamiltonian.}
\label{fig:hubbard_cubic_1}
\end{figure}

To illustrate the importance of the Weyl ordering we will consider an explicit example as in Sec.~\ref{sec:ramp} but with slight modification that the atoms are only interacting in off cites:
\be
\hat{\mathcal H}=-J\sum_{j} (\hat\psi_j^\dagger\hat\psi_{j+1}+\hat\psi_{j+1}^\dagger\hat\psi_j)+{U(t)\over 2}\sum_{j= {\rm even}} \hat\psi_j^\dagger\hat\psi_j^\dagger\hat\psi_j \hat\psi_j.
\ee
The classical normal-ordered counterpart of this Hamiltonian appears by simply substituting $\hat\psi^\dagger\to\psi^\star$ and $\hat\psi\to \psi$ in the Hamiltonian above. The Weyl symbol for this Hamiltonian contains an additional term:
\be
\mathcal H_W=-J\sum_{j} (\psi_j^\star\psi_{j+1}+\psi_{j+1}^\star\psi_j)+{U(t)\over 2}\sum_{j= {\rm even}} |\psi_j^\star|^4-U(t)\sum_{j={\rm even}} |\psi_j|^2.
\ee
The extra term now does not commute with the Hamiltonian and gives raise to the physical difference in the time evolution.

As in the example considered in Sec.~\ref{sec:ramp} we will start in the noninteracting case and assume that $U(t)=U_0\tanh\delta t$. We will use the same parameters as in Fig.~\ref{fig:hubbard_ramp}, i.e. eight sites, on average one particle per site, $J=1$, $U_0=1$ and $\delta=2.5$. In Fig.~\ref{fig:hubbard_cubic} we show the comparison of the exact calculation and TWA with the Hamiltonian in the Weyl and normal-ordered forms. Clearly the Weyl representation (red circles) gives much better accuracy. In addition to a larger mistake using normal-ordered Hamiltonian also yields an unphysical result where the energy is not conserved even when the interaction almost stops changing in time. In Fig.~\ref{fig:hubbard_cubic_1} we show that as in Fig.~\ref{fig:hubbard_ramp} the first quantum correction significantly improves the accuracy of TWA. We use a slightly smaller value of $\delta=1$ in Fig.~\ref{fig:hubbard_cubic_1} compared to Fig.~\ref{fig:hubbard_cubic} to introduce a larger mistake in TWA so that the effect of the correction is more visible.

\bibliography{wigner}

\end{document}